\documentclass[11pt]{iopart}
\pdfoutput=1
\usepackage[top=1.2 in, bottom=0.75 in, left=1 in, right=1 in]{geometry}

\usepackage[perpage]{footmisc}
\usepackage{iopams}
\expandafter\let\csname equation*\endcsname\relax
\expandafter\let\csname endequation*\endcsname\relax
\usepackage{amsmath,color,bbold,bbm,array,verbatim,harvard,ulem,tabularx}
\usepackage[colorlinks,bookmarks=false,citecolor=blue,linkcolor=blue,hyperfootnotes=true,urlcolor=blue]{hyperref}

\usepackage{tikz}
\usetikzlibrary{calc,arrows,decorations.pathreplacing}

\newcommand{\la}{\langle}
\newcommand{\ra}{\rangle}
\newcommand{\dg}{^\dagger}
\newcommand{\p}{\partial}
\newcommand{\up}{\uparrow}
\newcommand{\dn}{\downarrow}
\newcommand{\s}{\sigma}
\def\nn{\nonumber\\}

\usepackage{etoolbox}
\makeatletter
\def\@mkboth#1#2{}
\newlength\appendixwidth
\preto\appendix{\addtocontents{toc}{\protect\patchl@section}}
\newcommand{\patchl@section}{%
  \settowidth{\appendixwidth}{\textbf{Appendix }}%
  \addtolength{\appendixwidth}{1.5em}%
  \patchcmd{\l@section}{1.5em}{\appendixwidth}{}{\ddt}%
}
\makeatother

\makeatletter
\def\footnoterule{\kern-3\p@
  \hrule \@width 2in \kern 2.6\p@} 
\makeatother

\begin{document}

\review[Anomalies in the pseudogap phase of the cuprates]{Anomalies in the pseudogap phase of the cuprates:\\ Competing ground states and the role of umklapp scattering}

\author{Neil J Robinson}
\address{Institute for Theoretical Physics, University of Amsterdam,\\ 
Science Park 904, Postbus 94485, 1098 XH Amsterdam, The Netherlands}
\vspace{-2mm}
\ead{n.j.robinson@uva.nl}

\vspace{-3mm}
\author{Peter D Johnson}
\address{Condensed Matter Physics \& Materials Science Division, Brookhaven National Laboratory, Upton, NY 11973-5000, USA}
\vspace{-2mm}
\ead{pdj@bnl.gov}

\vspace{-3mm}
\author{T Maurice Rice}
\address{Theoretische Physik, ETH Zurich, 8093 Zurich, Switzerland and}
\address{Condensed Matter Physics \& Materials Science Division, Brookhaven National Laboratory, Upton, NY 11973-5000, USA}
\vspace{-2mm}
\ead{rice@phys.ethz.ch}

\vspace{-3mm}
\author{Alexei M Tsvelik}
\address{Condensed Matter Physics \& Materials Science Division, Brookhaven National Laboratory, Upton, NY 11973-5000, USA}
\vspace{-2mm}
\ead{atsvelik@bnl.gov}

\begin{abstract}
Over the past two decades, advances in computational algorithms have revealed a curious property of the two-dimensional Hubbard model (and related theories) with hole doping: the presence of close-in-energy competing ground states that display very different physical properties. On the one hand, there is a complicated state exhibiting intertwined spin, charge, and pair density wave orders. We call this \textit{`type A'}. On the other hand, there is a uniform $d$-wave superconducting state that we denote as \textit{`type B'}. We advocate, with the support of both microscopic theoretical calculations and experimental data, dividing the high-temperature cuprate superconductors into two corresponding families, whose properties reflect either the type A or type B ground states at low temperatures. We review the anomalous properties of the pseudogap phase that led us to this picture, and present a modern perspective on the role that umklapp scattering plays in these phenomena in the type B materials. This reflects a consistent framework that has emerged over the last decade, in which Mott correlations at weak coupling drive the formation of the pseudogap. We discuss this development, recent theory and experiments, and open issues. 

\vspace{2mm}

\end{abstract}

\maketitle

\hrule height 1.5pt
\tableofcontents
\vspace{0.8cm}
\hrule height 1.5pt

\section{Motivation and outline}

Although more than three decades have passed since the discovery of high temperature superconductivity in the cuprates~\cite{bednorz1986possible,wu1987superconductivity,maeda1988new,sheng1988bulk,schilling1993superconductivity}, the relevant microscopic description remains controversial. There is, however, general agreement that the appropriate starting point to describe the microscopic physics of a single CuO$_2$ layer is a single band Hubbard model, doped away from half-filling by holes.  Despite the physics of strongly interacting fermions presenting a continued challenge, there is a consensus about a key and unusual property of this model. This goes back to a set of variational Monte Carlo calculations by Himeda, Kato and Ogata~\cite{himeda2002stripe} early in this century. Starting from a strong coupling $t$-$J$ model and including a next-nearest-neighbour $t'$ hopping amplitude, they reported an almost degeneracy between two possible ground states. The first of these exhibits uniform $d$-wave superconductivity. On the other hand, the second is a complex state with hole stripes, $(\pi,\pi)$ antiferromagnetism with $\pi$ phase slips at the hole stripes, and $d$-wave superconductivity, also with $\pi$ phase slips between the local maxima located at the hole stripes~\cite{fradkin2015colloquium,keimer2015quantum}. 

\begin{figure}
  \begin{center}
    \includegraphics[width=0.6\textwidth]{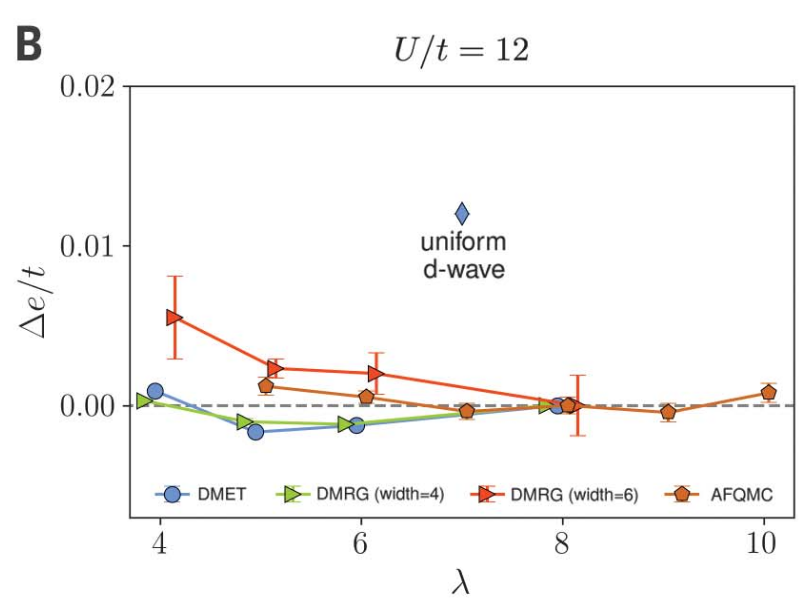}
  \end{center}
  \caption{Competition between stripe states with period of $\lambda$ lattice sites and the uniform d-wave superconducting state in the two-dimensional Hubbard model with $U/t = 12$. The energy difference per site between a plethora of stripe states and the uniform d-wave state is just $\Delta e/t \sim 0.01$. Results are shown for a number of different computational techniques, discussed in detail later. Figure reproduced from (Zheng et al. 2017).}
  \label{zheng6}
\end{figure}

In the intervening years there has been considerable progress by many groups on the numerical calculations using different algorithms~\cite{himeda2002stripe,capone2006competition,raczkowski2007unidirectional,aichhorn2007phase,chou2008clusterglass,yang2009nature,chou2010mechanism,corboz2011stripes,corboz2014competing,leblanc2015solutions,zheng2017stripe,iso2018competition}. Remarkably, the almost degeneracy between these two different ground states appears as a universal feature in such calculations, with the uniform $d$-wave superconducting state only slightly higher in energy by $\sim 0.01t$ per site~\cite{zheng2017stripe}, see figure~\ref{zheng6}. These microscopic calculations were all for a planar, square lattice in the absence of an external magnetic field, and we will also restrict our discussion to the zero field case. Clearly the two-dimensional Hubbard model is a simplification of the underlying lattice description, and so the choice of the ground state between these two very different states is impossible to predict \textit{a priori} for each cuprate. In view of this, we argue that the decision between these two ground states should be taken \textit{after} an examination of the experimental results on the individual cuprate. 

A recent review by Keimer and coworkers~\cite{keimer2015quantum} favoured a spatially modulated ground state in \textit{all} cuprates. Here we emphasise that the alternative uniform $d$-wave superconducting state can also capture the behaviour of the cuprates as the hole density is lowered into the pseudogap region. We will show how the anomalous changes that accompany the pseudogap~\cite{timusk1999pseudogap,norman2005pseudogap,lee2008from,hufner2008two} can arise from the uniform $d$-wave superconducting state at overdoping. With this splitting of the cuprates into two sets, we shall refer to the set possessing a spatially modulated ground state with intertwined orders as `\textit{type A}', while those with a spatially uniform ground state are called `\textit{type B}'. Let us now briefly address how families of cuprate materials can be placed into these two types, with more detailed discussion of experiments postponed until later.

In the first instance, neutron scattering can be used to identify the intertwined orders prevalent in the type~A cuprates. Early on, Tranquada and coworkers first identified both stripe charge density wave (CDW) and spin density wave (SDW) order, with periods $4a$ and $8a$ respectively, in La$_{2-x}$Ba$_x$CuO$_4$ (LBCO) at $x=1/8$ doping~\cite{tranquada1995evidence}. Pair density wave (PDW) order with period $8a$, associated with phase slips in the $d$-wave superconductivity, was also proposed for this material as it explains the very weak $c$-axis Josephson coupling between adjacent CuO$_2$ planes~\cite{li2007twodimensional,berg2007dynamical}. LBCO at $x = 1/8$ doping clearly belongs to the type A cuprates. 

Neutron scattering can also serve to distinguish long range magnetic order from short range magnetic order. In the first case, long range order leads to the static magnetic peaks and gapless $S=1$ magnon excitations observed in the type A cuprates. In the latter case of short range order, the magnon excitation is a gapped, finite frequency, mode centred on $(\pi,\pi)$, as seen in the type B cuprates. We will discuss these features further below.
  
Nuclear magnetic resonance (NMR) and nuclear quadrupolar resonance (NQR) also serve as useful probes that distinguish between the two types of cuprate. Bulk properties are probed by NMR and NQR, and the method has the advantage of being applicable to even small samples. CDW and SDW order can be probed separately when NMR is applied to a higher spin nuclei, such as the spin $J=5/2$ nucleus of $^{17}$O~\cite{warren1989cu,alloul1989nmr,walstedt1990nmr,takigawa1991cu,walstedt2010nmr}. Recently, Imai and coworkers~\cite{imai2017revisiting} reported a careful series of NMR experiments on the central transition of $^{63}$Cu nuclei, leading them to conclude that La$_{2-x}$Sr$_x$CuO$_4$ (LSCO) exhibits both CDW and SDW ordering patterns, consistent with those observed in LBCO. These experiments also show that Sr cuprates are much more disordered than Ba cuprates, which leads to imperfect superlattice ordering. The presence of PDW order has also been confirmed via non-linear optical response measurements on LSCO~\cite{rajasekaran2018probing}. These experiments lead us to also assign LSCO to the type A cuprates.

\begin{figure}
\begin{center}
\includegraphics[width=0.6\textwidth]{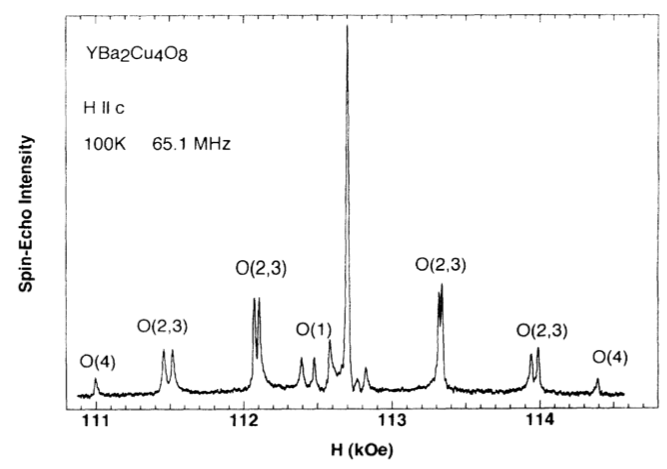}
\end{center}
\vspace{-4mm}
\caption{The $^{17}$O NMR spectra for YBa$_2$Cu$_4$O$_8$ with magnetic field applied parallel to the $c$-axis at $T=100$\,K. Notice that the O sites show well-resolved narrow lines, making this material particularly useful for probing charge density wave order. O(2) and O(3) sites are located within the CuO$_2$ planes, whilst O(1) and O(4) are the apical and double chain sites. Note that the electric field gradients on the O sites agree well with first principle local density approximation estimates from (Abrosch-Draxl et al. 1991). Figure reproduced from (Tomena et al. 1994).}
\label{fig:nmr}
\end{figure}

The type B cuprates behave quite differently in NMR, as can be seen in experiments on underdoped YBa$_x$Cu$_y$O$_{7-z}$ (YBCO) samples. The stoichiometric underdoped double chain compound YBa$_2$Cu$_4$O$_8$ (Y248) is particularly interesting, as it is an ordered compound with fractional hole doping of the CuO$_2$ layers. In figure~\ref{fig:nmr} we show the NMR spectrum for the CuO$_2$ plane. This has well-resolved narrow lines (for both the O and Cu nuclei, Cu not shown), with no signs of superlattice order at all temperatures~\cite{tomeno1994nmr}. We note that the measured electric field gradients on the O sites agree with first principle results~\cite{ambroschdraxl1991electronic}. The much studied YBa$_2$Cu$_3$O$_{7-x}$ cuprates have broadly similar NMR spectra, but show a small broadening of some lines as temperature is lowered~\cite{wu2015incipient}. The onset temperature, $T^\text{ons}$, for this broadening in YBa$_2$Cu$_3$O$_{7-x}$ with $T_c \sim 50$\,K coincides with the appearance of a new mode in the c-axis optical reflectivity and also with phonon broadening observed in inelastic X-ray scattering measurements, as will be discussed later. The application of a strong magnetic field, $H\sim17$\,T, does introduce a clear CDW pattern~\cite{wu2011magneticfieldinduced}, but the behaviour of these materials in high magnetic fields is outside the scope of this review; we instead refer the reader to the very recent review article~\cite{proust2018remarkable}.

Another widely studied cuprate is the single layer HgBa$_2$CuO$_{4+\delta}$ (Hg1201), where a range of hole densities stretching into the underdoped regime have been examined~\cite{itoh1998pseudo}. There the authors made extensive comparisons between NMR measurements on Hg1201 and those on LSCO. They found strong differences between the two cuprates, e.g., in the temperature dependence of the spin relaxation rate $1/T_1T$ of Cu nuclei, concluding that a magnetic pseudogap appears in the former material but not the latter. The Hg cuprates can also be grown as multilayer materials, with HgBa$_2$Ca$_{n-1}$Cu$_n$O$_{2n+2+d}$ ($n=1\to5$ layers) having been studied extensively in NMR and NQR, as summarised in work by the Kitaoka group~\cite{mukuda2012high}. In those Hg cuprates with wider unit cells, $n > 2$, the hole doping and superconductivity is strongest in the outermost layers, $j=1$ and $j=n$, which are closest to the acceptors, while the inner lays are more lightly doped and have very narrow NMR lines, indicating spatial homogeneity. We shall not discuss their many experiments in detail here, referring instead the reader to their review~\cite{mukuda2012high}. The main conclusion of their work, however, is that the phase diagram shows universal spatially homogeneous behaviour with long-range antiferromagnetic order at low densities $x<0.1$, which crosses over to d-wave superconductivity at higher dopings with a pseudogap at $T>T_c$. At higher hole densities, $x\sim0.16$, $T_c$ goes through a maximum and the pseudogap vanishes. In summary, these various results lead us to assign YBCO, Y248 and Hg cuprates to type B.

One of the limitations of the NMR technique is the requirement of well-ordered samples. This limits NMR on Bi$_2$Sr$_2$CaCu$_2$O$_{8+x}$ (BSCCO), the cuprate with a well defined surface suitable for surface sensitive techniques, such as angle resolved photoemission spectroscopy (ARPES) and scanning tunnelling microscopy (STM). Ishida   and coworkers~\cite{ishida1998pseudogap} probed the temperature dependence of the spin relaxation rate $1/T_1T$ and Knight shift of $^{63}$Cu nuclei in a underdoped sample with $T_\text{c} = 79$\,K. They found behaviour consistent with a normal state pseudogap at $T = 170$\,K in the density of states, in agreement with ARPES measurements of a pseudogap opening at $T^\ast = 170$\,K. The value of the spin gap in $1/T_1T$ coincides with the pseudogap in the quasiparticle density of states. The crystalline disorder shows up clearly in the NMR on $^{17}$O nuclei experiments by~\cite{crocker2011nmr} as a strong temperature independent quadrupolar and magnetic broadening that Does not change at $T^\ast$ or $T_\text{c}$. They attribute this broadening to disorder in the O spatial distribution. Taken together these experiments point towards type B behaviour in underdoped BSCCO.

In this review, our primary focus will be on the type~B cuprates (although we will often discuss similar or contrasting anomalous behaviour displayed by type~A materials). These do not display the complex intertwined order (CDW, SDW and PDW) found in the type~A cuprates, which have been the subject of several recent review articles to which we refer the reader~\cite{vojta2009lattice,taillefer2010scattering,fradkin2015colloquium,keimer2015quantum,kloss2016charge}. In our view, it is remarkable that the assignment of underdoped cuprates into two types, corresponding to the two almost degenerate ground states appearing in microscopic calculations of simple lattice models, is corroborated by experiments. We will discuss many of the anomalous properties displayed by the type~B cuprates, and theoretical approaches to describing them. Let us now describe how this review will proceed. 

In section~\ref{sec:history} we begin with an overview of the historical development of theoretical proposals to explain the anomalous properties associated with the pseudogap in type B cuprates. Perhaps the biggest challenge for theory is to explain the opening of the pseudogap in the antinodal region of the low-energy spectrum -- a clear deviation from standard Landau theory of Fermi liquids. The opening of the pseudogap does not appear to be accompanied by a clearly identifiable Landau symmetry breaking order parameter. Mott's original proposal for insulating behaviour in systems with a half-filled Bloch band~\cite{mott1949basis} is not a consequence of broken translation symmetry, but caused instead by strong onsite Coulomb repulsion. Mott insulating behaviour is, however, known to occur already at weak coupling in certain low-dimensional systems~\cite{lieb1968absence,balents1996weakcoupling} and we discuss two important examples of this, first reminding the reader of exact results on the one-dimensional Hubbard model in section~\ref{sec:onedhubbard}, before discussing the $d$-Mott ground state of the half-filled two-leg Hubbard ladder in section~\ref{sec:twoleghubbard}. In the latter case, we review the work of Balents and Fisher, who obtained the $d$-Mott state through a one-loop renormalization group (RG) analysis~\cite{balents1996weakcoupling}, and later with collaborators studied its properties using low-energy emergent integrability~\cite{lin1998exact}. Umklapp scattering plays a key role in both of these examples; in particular, in the two-leg ladder it leads to insulating behaviour associated with strictly short-range order in the $d$-wave Cooper and commensurate $(\pi,\pi)$ antiferromagnetic channels. These scattering processes connect the Fermi points $\pm k_\text{F}$, allowing one to scattering pairs of electrons from the right-to-left (or left-to-right) Fermi points with a net transfer of $4k_\text{F} = 2\pi/a$ momentum, which is absorbed by the lattice. 

Having seen how Mott insulating behaviour can emerge in one-dimensional systems, we move further towards the cuprates and consider the two-dimensional Hubbard model near half-filling in section~\ref{sec:twodhubbard}. Here umklapp scattering processes also appear prominently in one-loop functional RG calculations~\cite{honerkamp2001breakdown}, as discussed in section~\ref{sec:frg}. They also serve to enhance the nodal/antinodal dichotomy, a key feature of the cuprates, as shown by recent diagrammatic quantum Monte Carlo studies~\cite{wu2017controlling}, which we discuss in section~\ref{sec:diagqmc}. The role umklapp scattering plays is not entirely surprising: in Mott's original argument for an insulating state, there is an interplay between half-filling and strong interactions. When the interactions are translated into reciprocal space, the half-filling condition is clearly related to umklapp scattering and points towards its special role in generating Mott insulating behaviour (even away from strong coupling). In section~\ref{sec:cdmft} we briefly discuss recent results from cluster dynamical mean field theory, focussing on the nodal/antinodal dichotomy.

Having discussed the role umklapp scattering plays in various numerical results of the two-dimensional Hubbard model, we then review a number of analytical approaches to extending the (umklapp dominated) physics of half-filled two-leg Hubbard ladders (as discussed in section~\ref{sec:twoleghubbard}) to two dimensions. In section~\ref{sec:yrzprop} we first consider the phenomenological approach of Yang, Rice. and Zhang~\cite{yang2006phenomenological,rice2012phenomenological}, who proposed an ansatz for the single electron propagator. We discuss some of the microscopic analytical routes supporting this ansatz in the appendix. In the following section,~\ref{sec:ossadnik}, we provide a brief review of Ossadnik's wave packet approach~\cite{ossadnik2016wave} for describing short range ordered states in two dimensions.  Next our attention turns to recent work by Tsvelik~\cite{tsvelik2017ladder}, which links the physics of ladders to that of the spin-fermion model~\cite{abanov2003quantumcritical}, one of the cuprate theoretical paradigms. 

The history of theoretical developments finishes in section~\ref{sec:competinggs} with a discussion of one of the complications of the two-dimensional Hubbard and $t$-$J$ models: the presence of competing ground states (see, e.g.,~\cite{zheng2017stripe}). As we have already mentioned, these results  strongly motivated our partitioning of the cuprates into two types, and we discuss the various theoretical works showing this dichotomy.   

With the historical developments for theory of the type B cuprates covered, in section~\ref{sec:exp} we present a detailed discussion of the various anomalies that appear in the `normal states' of the cuprates. In section~\ref{sec:nmr} we first return to NMR/NQR studies, discussing these in more detail. Following this, in section~\ref{sec:arpes} we discuss the numerous ARPES experiments on BSCCO, and the observation of the now infamous pocket Fermi surface. This is followed by a discussion of STM measurements of the single particle energy gap (which persists in the antinodal region of the Fermi surface to temperatures far above the superconducting critical temperature $T_\text{c}$) in section~\ref{sec:stm}. Both ARPES and STM show clear signatures of the opening of the pseudogap, which is also accompanied by a rapid change in the carrier density, as measured by the Hall effect, as is discussed in section~\ref{sec:carrierdens}. Indeed, the carrier density changes from that of the standard Fermi surface at overdoping to a much smaller value associated with doping the Mott insulator at stoichiometry, see e.g.~\cite{badoux2016change}. We then discuss the presence of antiferromagnetic correlations in the pseudogap phase in section~\ref{sec:afmcorr}.

Our discussion then moves to understanding phonon features and short-range charge correlations. We first discuss optical properties of the pseudogap phase in section~\ref{sec:optical}, with a particular emphasis on the anomalous high temperature ($T \sim 3T_\text{c}$) onset of an additional peak in the $c$-axis infrared absorption~\cite{homes1993optical,schutzmann1995doping,berhard2000farinfrared,dubroka2011evidence}. This has been interpreted as a signature of intra-bilayer tunnelling of Cooper pairs that are present already at temperatures well above the superconducting critical temperature. In section~\ref{sec:xray} we review recent X-ray scattering experiments that reveal giant phonon anomalies starting also at $T \sim 3T_\text{c}$, and which grow as temperature is lowered towards $T_\text{c}$.

Finally, we turn to the well-known temperature-dependence of transport in the cuprates in section~\ref{sec:transport}.  At temperatures above the onset of the pseudogap, $T > T^\ast$, the d.c. resistivity shows linear-in-temperature behaviour, in contrast to the quadratic-in-temperature behaviour of a Fermi liquid~\cite{gurvitch1987resistivity,orenstein1992frequency,takagi1992systematic,mandrus1992resitivity}. Below the pseudogap transition temperature, there is increasing evidence for conventional $T^2$ behaviour~\cite{barisic2013universal}. We discuss this in section~\ref{sec:linear}. Following this, we return to the Hall effect, and consider its behaviour as a function of temperature in section~\ref{sec:hallT}. As is now well-known, transverse transport for the cuprates in a magnetic field is consistent with the predictions of Fermi liquid theory, in striking contrast to the d.c. resistivity~\cite{harris1995violation,terasaki1995normalstate,kimura1996inplane,barisic2015hidden}.  
The hole density, coinciding with the value from the doping the Mott insulator at stoichiometry, as measured in the Hall effect persists at $T > T^\ast$, even though the pseudogap has already dissolved in ARPES measurements~\cite{hwang1994scaling,wuyts1996resistivity,matthey2001hall,balakriev2003signature,vishik2018photoemission}.

Taken together in section~\ref{sec:exp}, these various anomalies indicate that highly unusual and puzzling behaviour abounds in the normal states of the cuprates. This presents a grand challenge for any microscopic theory of high temperature superconductivity. Finally, in section~\ref{sec:open} we discuss open issues. In particular, we focus on the question of the relation of the two very different ground states (type A and type B) found in theoretical studies.

\section{History of theoretical developments}
\label{sec:history}

At the beginning of the 1930s, Wilson presented his description of metals and insulators~\cite{wilson1931theory} in terms of non-interacting electrons. This was successfully applied to many materials, explained much of the phenomenology, and allowed many quantities to be computed. However, towards the end of that decade it was realised that Wilson's theory did not always work: de Boer and Verwey found that NiO is a transparent insulator, despite theory predicting it to be a metal with half-filled band~\cite{deboer1937semiconductors}. These experiments were greeted with a great deal of surprise and generated much discussion~\cite{mott1937discussion}, leading Peierls to propose (in passing) that electron-electron interactions could be responsible~\cite{mott1937discussion,mott1980memories}. This was a drastic modification of Wilson's band theory~\cite{wilson1931theory}, argued in detail later in the well-known 1949 work of Neville Mott~\cite{mott1949basis}, see also~\cite{mott1968metalinsulator,imada1998metalinsulator}~\footnote{We note that Eugene Wigner in 1938 proposed that electron-electron interactions could lead to a `crystallisation' of the electrons at low densities~\cite{wigner1938effects}.}.

The correlations-driven metal-insulator transition (herein the Mott transition) highlights the essential interplay between electron-electron interactions and localisation (i.e. the bandwidth). Mott's basic argument is a simple one~\cite{mott1968metalinsulator}; consider a cubic lattice formed from one-electron atoms with lattice spacing $d$. A metallic phase will have free charge carriers, so one needs to consider moving an electron from a given atom to another, i.e. creating an electron-hole pair. For large (but finite) $d$, an electron-hole pair with separation $r$ will experience an attractive potential $V(r) \sim -e^2/r$. This potential supports bound states; the electron-hole pair are bound together and there can be no charge transport. On the other hand, at small $d$ the attractive potential becomes screened by carriers on neighbouring atoms, $V(r) \to -e^2 \exp(-q r)/r$. For sufficiently strong screening ($q$ large) this potential no longer supports bound state solutions, so the electron is a mobile charge carrier and the crystal exhibits metallic behaviour. In between these two limiting cases, at some critical separation $d=d^\ast$, there is thus a metal-insulator transition. 

The existence of such a transition was rigorously proven in Kohn's seminal work on insulators~\cite{kohn1964theory}, with examples rapidly following in the mid-1960s from~\cite{hubbard1964electron}, \cite{kemeny1965model} and \cite{gutzwiller1965correlation}. The original scenario of Mott, the metal-insulator transition for one electron atoms, was later revisited by~\cite{brinkman1970application} using the Gutzwiller approximation, with the metal-insulator transition being revealed through a divergence of the electron effective mass at a critical interaction strength. 

For fixed lattice parameter $d$ (i.e., in the absence of a structural transition) and a half-filled Bloch band in arbitrary dimensions, the Mott transition occurs for sufficiently strong values of the interaction strength. It is known, however, that the Mott transition can already occur at weak coupling in low-dimensional systems, in particular for the case of a single spatial dimension. Here the Mott transition is driven by umklapp scattering, and we discuss two examples of this in Secs.~\ref{sec:onedhubbard} and~\ref{sec:twoleghubbard}. We then discuss how umklapp-driven physics carries through to two spatial dimensions, with the Mott insulator emerging at intermediate couplings and for a range of dopings in the 2D Hubbard model in section~\ref{sec:twodhubbard}. The challenge of generalising the physics of half-filled two-leg ladders (motivated by similarities between the physics there and that of the pseudogap) is discussed in section~\ref{sec:approaches}, in particular we consider the Yang-Rice-Zhang (YRZ) ansatz for the propagator in section~\ref{sec:yrzprop}, Ossadnik's wave packet approach in section~\ref{sec:ossadnik} and a recently revealed correspondence between ladder and spin-fermion model physics in section~\ref{sec:spinfermion}. We then return to the much discussed issue of competing ground states in the two-dimension Hubbard and $t$-$J$ models in section~\ref{sec:competinggs}. 

\subsection{Mott physics at weak coupling: the one dimensional Hubbard chain}
\label{sec:onedhubbard}

Restricting particles to move in only a single spatial dimension has profound implications~\cite{bethe1931zur,haldane1981luttinger}. No longer can one think separately of bosons and fermions, with an exchange of quasiparticles necessitating a scattering event (and thus mingling scattering and statistical phases). As a result, conventional approaches and techniques breakdown, and non-perturbative phenomena abound. Fortunately, there are also a panoply of techniques peculiar to one spatial dimension that allow one to tackle these problems and provide exact (or non-perturbative approximate) solutions, see~\cite{giamarchi2004quantum,gogolin2004bosonization,tsvelik2007quantum,imambekov2012onedimensional,james2018nonperturbative} to name but a few. Here we will discuss Mott physics in two closely-related one-dimensional models of interest. 

As a starting point, let us consider a purely one-dimensional chain of electrons, with hopping allowed between nearest neighbours and electrons only interacting when they reside on the same site of the chain. We thus realise the Hubbard model~\cite{hubbard1964electron} on a one-dimensional chain, whose Hamiltonian reads
\begin{equation}
H_\text{1D} = -t \sum_j \sum_{\s = \up,\dn}  \Big( c\dg_{j,\s} c_{j+1,\s} + \text{H.c.} \Big) + U \sum_j n_{j,\up} n_{j,\dn}.
\label{1dHubbard}
\end{equation}
Here $t$ is the hopping amplitude, $U$ is the Hubbard onsite interaction strength, and $c\dg_{j,\s}$ is the creation operator for an electron with spin $\s$ on the $j$th site of a chain with lattice spacing $a$. The number operators on each site are $n_{j,\s} = c\dg_{j,\s}c_{j,\s}$. This model is exactly solvable by the Bethe ansatz, as shown by Lieb and Wu~\cite{lieb1968absence} (see also \cite{essler1991complete} with regards to issues surrounding completeness of the solution). Many other details of the solution of the one-dimensional Hubbard model, and its properties, can be found in~\cite{essler2005one}. 

At half-filling (one electron per site), the model has a gap for charge excitations~\cite{lieb1968absence}
\begin{equation}
\Delta_\text{c} =  - 2 + 2u + 2 \int_0^\infty \frac{\rmd \omega}{\omega} \frac{J_1(\omega) \rme^{-\omega u}}{\cosh(\omega u)} , 
\qquad u = \frac{U}{4t}, \label{eq:chargegap}
\end{equation}
for any non-zero value of $U$, and thus is an insulator. In this sense, the Mott transition occurs at weak coupling in the Hubbard chain~\footnote{Note that Lieb and Wu's interpretation of insulating behaviour for any non-zero value of the interaction is that the Mott transition is \textit{absent}, in the sense that the interacting half-filled Hubbard model is \textit{always} a Mott insulator~\cite{lieb1968absence}.}. In~\eqref{eq:chargegap} we have introduced $J_1(\omega)$, a Bessel function of the first kind. 

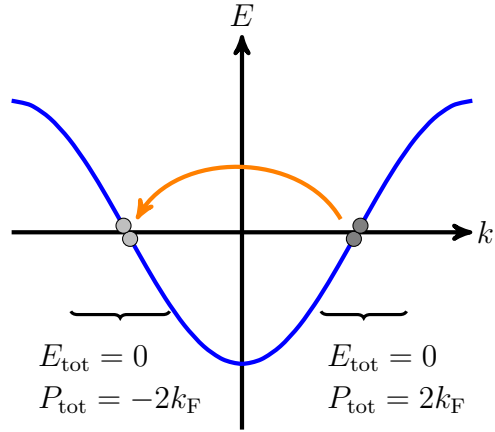
\begin{figure}
\begin{center}
\begin{tikzpicture}[scale=1.75]
		\draw[ultra thick,->,>=stealth'] (0cm,0cm) -- (3.5cm,0cm);
		\draw[ultra thick,->,>=stealth'] (1.75cm,-1.5cm) -- (1.75cm,1.5cm); 
		\node at (3.6cm,0cm) {$k$};
		\node at (1.75cm,1.65cm) {$E$};
		\draw[xscale=1,domain=0:3.5,smooth,variable=\x,blue,ultra thick] plot ({\x},{cos(2*3.14159/3.5*\x r)});
		\node [draw, circle,inner sep = 0pt,minimum size=0.2cm,fill=gray] (Rh) at (2.6cm,-0.05cm) {};
		\node [draw, circle,inner sep = 0pt,minimum size=0.2cm,fill=gray] (Rp) at (2.65cm,0.05cm) {};
		\node [draw, circle,inner sep = 0pt,minimum size=0.2cm,fill=lightgray] (Lh) at (0.9cm,-0.05cm) {};
		\node [draw, circle,inner sep = 0pt,minimum size=0.2cm,fill=lightgray] (Lp) at (0.85cm,0.05cm) {};
		\path[->,>=stealth',orange,ultra thick] (2.5cm,0.1cm) edge[bend right = 60] node [left] {} (0.95cm, 0.1cm); 
		\draw[decoration={brace,mirror,raise=5pt},decorate,very thick] 
		(2.35cm,-0.5cm) -- node[below=10pt] {
		\qquad\begin{tabular}{l} 
		$E_\mathrm{tot}=0$ \\ 
		$P_\mathrm{tot} = 2k_\text{F}$ 
		\end{tabular}} (3.0cm,-0.5cm);
		\draw[decoration={brace,mirror,raise=5pt},decorate,very thick] 
		(0.45cm,-0.5cm) -- node[below=10pt] {
		\begin{tabular}{l} 
		$E_\mathrm{tot}=0$ \\ 
		$P_\mathrm{tot} = -2k_\text{F}$ 
		\end{tabular}}(1.2cm,-0.5cm);
\end{tikzpicture}
\end{center}
\caption{A schematic example of umklapp scattering. Two right-moving electrons (dark grey) scatter to two left-moving (light grey) electrons, with a resulting change in momentum of $4k_\text{F}$. At half-filling this change in momentum, $4k_\text{F} = 2\pi/a$, is commensurate with the reciprocal lattice vector and hence is allowed by conservation of crystal momentum.}
\label{fig:umklapp1d}
\end{figure}

The opening of a spectral gap for charge excitations at half-filling is driven by umklapp scattering processes (we illustrate one such process in figure~\ref{fig:umklapp1d}, where two right-moving electrons scatter, resulting in two left-moving electrons). Such processes  are allowed at half-filling as the change in momentum $\Delta k = 4k_\text{F}$ is commensurate with the reciprocal lattice vector, $4k_\text{F} = 2\pi/a$, and hence conserve crystal momentum. Thus pairs of electrons at the Fermi points can scatter across the Brillouin zone (with the released momentum being absorbed by the lattice) and a spectral gap opens. As is well known, at low energies the problem can be described in terms of an antiferromagnetic insulator: the Heisenberg model with exchange parameter $J = 4t^2/U$.  

\subsection{Mott physics at weak coupling: the two-leg Hubbard ladder}
\label{sec:twoleghubbard}

Another example of the Mott transition occurring at weak coupling is the two-leg Hubbard \textit{ladder}. We will discuss this model in quite some detail, as the Mott insulating ground state at half-filling contains a precursor to $d$-wave superconductivity. This so-called $d$-Mott state will motivate many of the further works that we discuss and is, of course, pertinent to our discussion of the cuprates. We also note that ladder models have found a number of applications in the cuprate problems~\cite{dagotto1996suprises}.

The two-leg Hubbard ladder at weak coupling was studied in detail by~\cite{balents1996weakcoupling} using the one-loop RG. Subsequently, the model garnered much attention and there have been numerous other studies~\cite{lin1997nchain,lin1998exact,schulz1998son,konik2000twoleg,konik2001exact,konik2002interplay,controzzi2005excitation,essler2005application}, as well as various extensions, e.g.~\cite{jaefari2012paridensitywave,robinson2012finite,carr2013spinful}.  

The most general nearest-neighbour Hamiltonian for the two-leg ladder with U(1)$\times$SU(2)$\times\mathbb{Z}_2$ symmetry  is
\begin{eqnarray}
H_\text{ladd} &=& H_1 + H_2 + V_{12}, \label{hubbardladder}\\
 H_\ell &=& -t \sum_j \sum_{\s = \up,\dn}  \Big( c\dg_{j,\ell,\s} c_{j+1,\ell,\s} + \text{H.c.} \Big) \nn 
        && + \sum_j \Big( U  n_{j,\ell,\up} n_{j,\ell,\dn} + V n_{j,\ell}n_{j+1,\ell} + J{\bi S}_{j,\ell} \cdot {\bi S}_{j+1,\ell} \Big)  \label{singlechain}\\
  V_{12} &=& - t_\perp \sum_j \sum_{\s = \up,\dn} \Big( c\dg_{j,1,\s} c_{j,2,\s} + \text{H.c.} \Big) \nn
         & &+ \sum_j \Big(J'{\bi S}_{j,1}\cdot{\bi S}_{j,2} + V'n_{j,1}n_{j,2}\Big) \label{V12}
\end{eqnarray}
where $c\dg_{j,\ell,\s}$ is the creation operator for an electron of spin $\s$ on site $j$ of leg $\ell$ of the ladder, ${\bi S}_{j,\ell} = c\dg_{j,\ell.\alpha} (\vec{\s}_{\alpha\beta}/2) c_{j,\ell,\beta}$ is the local spin operator, and $n_{j,\ell} = n_{j,\ell,\up}+n_{j,\ell\dn}$ is the electron number operator on each site. $t$ is the along-leg hopping amplitude, while $t_\perp$ describes hopping between the legs (across the rungs). Moving from the chain to the ladder, the Hamiltonian~\eqref{1dHubbard} is no longer exactly solvable. Instead, for weak interactions ($U,V,J,V',J'\ll t,t_\perp$) one can harness the power of the RG to derive and study a low-energy effective theory~\cite{balents1996weakcoupling}. We note that some other interactions can be added without altering the essential physics. 

\subsubsection{Continuum limit and renormalization group.}
The theoretical approach taken to ladder models~\eqref{hubbardladder} depends on the value of the interchain coupling $t_\perp$ in~\eqref{V12}. Most of the literature discusses ladder models in real space, where such tunnelling is always present. In this section, we follow this tradition and assume that the interchain tunnelling dominates the interactions, $t_\perp \gg U,V,J,V',J'$. However, as we will see, the ladder model of most relevance to two-dimensional cuprates has an effective $t_\perp = 0$. We will discuss this in section~\ref{sec:spinfermion}. The choice between $t_\perp \gg U,V,J,V',J'$ and $t_\perp \ll U,V,J,V',J'$ can be reflected in the choice of basis for our fields and affects the expressions for the order parameter fields. The corresponding expressions will be given below in section~\ref{sec:spinfermion}.

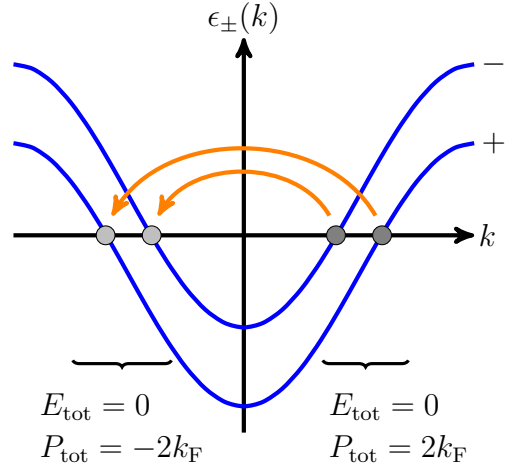
\begin{figure}
\begin{center}
\begin{tikzpicture}[scale=1.75]
		\draw[ultra thick,->,>=stealth'] (0cm,0cm) -- (3.5cm,0cm);
		\draw[ultra thick,->,>=stealth'] (1.75cm,-1.5cm) -- (1.75cm,1.5cm); 
		\node at (3.6cm,0cm) {$k$};
		\node at (1.75cm,1.65cm) {$\epsilon_\pm(k)$};
		\draw[xscale=1,domain=0:3.5,smooth,variable=\x,blue,ultra thick] plot ({\x},{cos(2*3.14159/3.5*\x r)-0.3});
		\node at (3.65cm,0.7cm) {$+$};
		\draw[xscale=1,domain=0:3.5,smooth,variable=\x,blue,ultra thick] plot ({\x},{cos(2*3.14159/3.5*\x r)+0.3});
		\node at (3.65cm,1.3cm) {$-$};
		\node [draw, circle,inner sep = 0pt,minimum size=0.25cm,fill=gray] (Rh) at (2.45cm,0.0cm) {};
		\node [draw, circle,inner sep = 0pt,minimum size=0.25cm,fill=gray] (Rp) at (2.8cm,0.0cm) {};
		\node [draw, circle,inner sep = 0pt,minimum size=0.25cm,fill=lightgray] (Lh) at (1.05cm,0.0cm) {};
		\node [draw, circle,inner sep = 0pt,minimum size=0.25cm,fill=lightgray] (Lp) at (0.7cm,0.0cm) {};
		\path[->,>=stealth',orange,ultra thick] (2.4cm,0.15cm) edge[bend right = 60] node [left] {} (1.1cm, 0.15cm); 
		\path[->,>=stealth',orange,ultra thick] (2.75cm,0.15cm) edge[bend right = 60] node [left] {} (0.75cm, 0.15cm); 
		\draw[decoration={brace,mirror,raise=5pt},decorate,very thick] 
		(2.35cm,-0.85cm) -- node[below=10pt] {
		\qquad\begin{tabular}{l} 
		$E_\mathrm{tot}=0$ \\ 
		$P_\mathrm{tot} = 2k_\text{F}$ 
		\end{tabular}} (3.0cm,-0.85cm);
		\draw[decoration={brace,mirror,raise=5pt},decorate,very thick] 
		(0.45cm,-0.85cm) -- node[below=10pt] {
		\begin{tabular}{l} 
		$E_\mathrm{tot}=0$ \\ 
		$P_\mathrm{tot} = -2k_\text{F}$ 
		\end{tabular}}(1.2cm,-0.85cm);
\end{tikzpicture}
\end{center}
\caption{The non-interacting band structure of the two-leg ladder model~\eqref{hubbardladder}, $\epsilon_\pm(k)$ in blue. We illustrate one possible umklapp scattering process, with a change in momentum of $\Delta k = 4k_\text{F}$, where two right moving electrons are scattering to left-moving states where band indices are not changed. Other processes are allowed at half-filling.}
\label{fig:ladderbands}
\end{figure}

The first step of the weak-coupling renormalization group treatment of~\eqref{hubbardladder} is to diagonalize the single-particle part of the Hamiltonian, i.e. setting $U=V=J=V' =J'= 0$. To do so, on each rung of the ladder we form symmetric ($+$) and antisymmetric ($-$) combinations of the fermions living on each leg: 
\begin{equation}
c_{\pm,j,\s} = \frac{1}{\sqrt2} \Big( c_{j,1,\s} \pm c_{j,2,\s} \Big).
\end{equation}
Following a subsequent Fourier transformation, the non-interacting Hamiltonian becomes a diagonal two-band model
\begin{equation}
H_\text{ladd}(U=0) = \sum_{d=\pm} \sum_{k,\s} \epsilon_d(k) c\dg_{d,k,\s} c_{d,k,\s},
\end{equation}
with the dispersion relation $\epsilon_\pm(k) = -2t\cos(k) \mp t_\perp$, shown in figure~\ref{fig:ladderbands}. 

In deriving an effective low-energy theory, we linearise the dispersion about the Fermi points, denoted $\{- k_{\pm,F}, k_{\pm,F}\}$, and take the continuum limit through the identities
\begin{equation}
c_{\pm,j,\s} = \sqrt{a_0} \Big[ R_{\pm,\s}(x) \rme^{\rmi k_{\pm,\text{F}}x} + L_{\pm,\s}(x) \rme^{-\rmi k_{\pm,\text{F}}x} \Big],  
\end{equation}
where $a_0$ is the lattice spacing and $R_{\pm,\s}(x)$/$L_{\pm,\s}(x)$ is a right/left moving fermionic field in the $\pm$ band with spin $\s$. Considering the case of half-filling, the continuum Hamiltonian reads~\cite{lin1998exact}
\begin{eqnarray}
&& H = \int \rmd x \Big( {\cal H}_0 + {\cal H}_\text{int} + {\cal H}_\text{um} \Big), \label{laddercontinuumH}\\
&& {\cal H}_0 = -\rmi v_\pm \Big( R\dg_{\pm,\s}\p_x R_{\pm,\s} - L\dg_{\pm,\s} \p_x L_{\pm,\s} \Big),\label{Lorentz}\\
  && {\cal H}_\text{int} = \sum_{\alpha,\beta} \Big[ b^\rho_{\alpha\beta} J^R_{\alpha\beta} J^L_{\alpha\beta} - b^\s_{\alpha\beta} J^{aR}_{\alpha\beta} J^{aL}_{\alpha\beta} \Big] \nn
  && \qquad  + \sum_{\alpha\neq\beta} \Big[ f^\rho_{\alpha\beta} J^R_{\alpha\alpha} J^L_{\beta\beta} - f^\s_{\alpha\beta} J^{aR}_{\alpha\alpha} J^{aL}_{\beta\beta} \Big],\\
&& {\cal H}_\text{um} = u^\rho_{\alpha\beta} (I^R_{\alpha\beta})\dg I^L_{\bar\alpha \bar\beta} - u^\s_{\alpha\beta} (I^{aR}_{\alpha\beta})\dg I^{aL}_{\bar\alpha\bar\beta}+\text{H.c.}
\end{eqnarray}
where $v_\pm = 2t \sin(k_{\pm,F})$ are the Fermi velocities in the two bands, $\bar \alpha = - \alpha$, and we define the following currents
\begin{eqnarray}
J_{\alpha\beta}^R = R\dg_{\alpha,\s} R_{\beta,\s}, \quad 
J_{\alpha\beta}^{aR} = R\dg_{\alpha,\s} s^a_{\s,\s'} R_{\beta,\s'},  \nn
I_{\alpha\beta}^R = R_{\alpha,\s} \epsilon_{\s,\s'} R_{\beta,\s'}, \quad
I_{\alpha\beta}^{aR} = R_{\alpha,\s} (\epsilon s^a)_{\s,\s'} R_{\beta,\s'}, \nonumber
\end{eqnarray}
where $s^a = \frac12 \s^a$ with $\s^a$ being the Pauli matrices, and $\epsilon_{\s,\s'}$ the antisymmetric tensor. The couplings in the continuum Hamiltonian~\eqref{laddercontinuumH} are related to the microscopic parameters in a simple manner (for example, for $J=V=J'=V'=0$ we have $4 b_{\alpha\beta}^\rho = b_{\alpha\beta}^\s = 4 f_{\alpha\beta}^\rho = f_{\alpha\beta}^\s = U$), but the details do not matter for sufficiently weak coupling, as the RG flow exhibits universality~\cite{lin1998exact,konik2002interplay}. The final term, ${\cal H}_\text{um}$ describes umklapp scattering (an example of which is shown in figure~\ref{fig:ladderbands}), and is absent away from half-filling~\cite{balents1996weakcoupling}~\footnote{We note that for other commensurate fillings, for example when one of the bands is half-filled, other umklapp terms can arise. We do not consider this scenario here; some examples can be found in \cite{jaefari2012paridensitywave,robinson2012finite,carr2013spinful}.}. 

At half-filling, under the RG flow we find a low-energy effective field theory where two of the interaction parameters remain small ($b_{11}^\rho$ and $f_{12}^\s$), while the remaining ones flow to strong coupling with fixed ratios~\cite{lin1998exact}~\footnote{Note that hermiticity implies $b_{12} = b_{21}$, parity imposes $f_{12}=f_{21}$, and the anti/symmetric nature of currents also enforces $u_{12} = u_{21}$. Furthermore, $I^{aR}_{ii} = 0$ so we set $u_{ii}^\s = 0$. }
\begin{equation}
b_{12}^\rho = \frac14 b_{12}^\s = f_{12}^\rho = -\frac14 b_{11}^\s = 2u_{11}^\rho = 2u_{12}^\rho = \frac12 u_{12}^\s = g > 0.
\label{rgflow} 
\end{equation}
In the low-energy effective field theory, there is thus a single coupling constant $g$.

\subsubsection[Low-energy effective theories at half-filling.]{Low-energy effective theories at half-filling: SO(8) symmetry, trialities, and dualities.}
\label{sec:effectivetheory}
This remarkable result from the renormalization group for the low-energy physics is universal at one-loop~\cite{lin1998exact,konik2000twoleg,konik2001exact,konik2002interplay,essler2005application}, applying for general weak interactions. Either the interactions die out, or they increases such that the system scales towards the strongly correlated state with the maximum possible symmetry. In the case at hand, the two-leg ladder~\eqref{hubbardladder} has initial symmetry U(1)$\times$SU(2)$\times\mathbb{Z}_2$. At half-filling, particle-hole symmetry promotes the U(1) charge symmetry to SU(2), so the maximum possible symmetry of this model is SO(8). Moreover, for sufficiently weak bare couplings the renormalization of the velocities (originating from the interaction terms containing only right- or left-moving fermions) for different excitations can be neglected. Then the low-energy theory preserves the Lorentz symmetry of~\eqref{Lorentz} and, in the ideal case of sufficiently weak bare couplings, excitations are massive particles with the relativistic dispersion 
\begin{equation}
\epsilon_a(p) = \sqrt{(vp)^2 + m_a^2}, \label{dispersion}
\end{equation}
which transform according to irreducible representations (irreps) of the SO(8) group. Here $m_a$ is the mass of the corresponding excitation. This idealisation serves as a convenient starting point for the description of ladder physics~\cite{lin1998exact,konik2000twoleg,konik2001exact,konik2002interplay,essler2005application}, and it also allows us to consider perturbations that lead to deformations of this idealised picture. 

As mentioned, excitations can be characterised by irreps of the SO(8) group; there are three 8-dimensional and one 28-dimensional irreps. The 8-dimensional irreps consist of one vector and two spinor representations. This constitutes a remarkable \textit{triality}: the theory can be formulated in three different, but equivalent, manners~\cite{maldacena1997majorana}. The excitations in the spin representations are related to the bare electronic degrees of freedom, carrying the same quantum numbers of spin $\pm1/2$, band (or chain if $t_\perp=0$) index and charge $\pm e$. The two spinor irreps have different parities. On the other hand, the excitations in the vector representation are nonlocal in terms of the original fermions. The relation between these different representations can be understood in terms of the bosonization/refermionization toolbox, which we do not propose to cover in detail here, instead referring the reader to the extensive literature that is available, e.g., see~\cite{giamarchi2004quantum,gogolin2004bosonization,tsvelik2007quantum}. At heart, bosonization is a correspondence between fermions and bosons in one-dimensional quantum systems that arises from the intermingling of statistical and scattering phases~\footnote{In 1D the wave function transforms under exchange of neighbouring particles as $\Psi(x_1,x_2,x_3,\ldots,x_N) = \exp(\rmi \theta_\text{scat} + \rmi\theta_\text{stat}) \Psi(x_2,x_1,x_3,\ldots,x_N)$. For bosons  $\theta_\text{stat} = 2\pi$, while fermions have $\theta_\text{stat} = \pi$. The scattering phase $\theta_\text{scat}$ depends on details of the system. If we consider bosons with interactions such that $\theta_\text{scat} = \pi$, the wave function transforms as if the particles were fermions: $\Psi(x_1,x_2,x_3,\ldots,x_N) = - \Psi(x_2,x_1,x_3,\ldots,x_N)$.}.  For the problem at hand, this correspondence is summarised in the bosonization identities~\cite{lin1998exact}
\begin{equation}
R_{\alpha,\s} = \frac{\kappa_{\alpha,\s}}{\sqrt{2\pi a_0}} \rme^{\rmi \phi_{\alpha,\s}}, \quad L_{\alpha,\s} = \frac{\kappa_{\alpha,\s}}{\sqrt{2\pi a_0}} \rme^{\rmi \bar \phi_{\alpha,\s}},  
\end{equation}
where $\phi,\bar\phi$ are chiral bosonic fields obeying the commutation relations
\begin{equation}
\begin{split}
\Big[\phi_{\alpha,\s}(x), \phi_{\alpha',\s'}(x') \Big] &= \rmi \pi \delta_{\alpha,\alpha'} \delta_{\s,\s'} \text{sgn}(x-x'), \\
\Big[\bar\phi_{\alpha,\s}(x), \bar\phi_{\alpha',\s'}(x') \Big] &= - \rmi \pi \delta_{\alpha,\alpha'} \delta_{\s,\s'} \text{sgn}(x-x'), \\
\Big[\phi_{\alpha,\s}(x), \bar \phi_{\alpha',\s'}(x') \Big] &= \rmi \pi \delta_{\alpha,\alpha'} \delta_{\s,\s'},
\end{split}
\end{equation}
and $\kappa_{\alpha,\s}$ are Klein factors, which ensure fermions in different bands and with different spins anti-commute, and thus obey $\{ \kappa_{\alpha,\s}, \kappa_{\alpha',\s'}\} = 2\delta_{\alpha,\alpha'} \delta_{\s,\s'}$. There is a gauge freedom in the definitions of the Klein factors: we fix this by setting $\Gamma = \kappa_{+,\up}\kappa_{+,\dn} \kappa_{-,\up}\kappa_{-,\dn} = 1$ herein~\footnote{By construction $\Gamma^2 = 1$, so alternatively one could choose $\Gamma = -1$. This would modify details of the correspondence between the bosonic and fermionic theories, such as the signs of couplings. This choice does not modify any physics, with all physical quantities being gauge invariant.}.

Under such a transformation to bosonic fields, the interaction terms in the field theory~\eqref{laddercontinuumH} become non-linear functions of the fields. We will see that the problem becomes simpler under a change of basis of the bosonic fields:
\begin{equation}
\begin{split}
\phi_{\alpha,\s} = & \frac{1}{2}\Big(\phi_\text{c} + \alpha\phi_\text{f} + \s\phi_\text{s} + \alpha\s\phi_\text{sf}\Big), \qquad \alpha = \pm 1,~~\s = \pm 1,\\
 \bar\phi_{\alpha,\s} = & \frac{1}{2}\Big(\bar\phi_\text{c} - \alpha\bar\phi_\text{f} + \s\bar\phi_\text{s} + \alpha\s\bar\phi_\text{sf}\Big),
\end{split}
\label{basis}
\end{equation}
The new bosonic fields now describe symmetric (c/s) and antisymmetric (f/s) charge/spin excitations. Notice that the bosonic modes with different chiralities are transformed differently (for later convenience) through the signs of the f fields.

The change of basis~\eqref{basis} transforms between the spinor and vector representations of the SO(8) group. Refermionisation gives the excitations from the vector representation: 
\begin{equation}
 \chi_{a1} = \frac{\kappa_a}{\sqrt{\pi a_0}}\cos(\phi_a), \qquad \chi_{a2} = \frac{\kappa_a}{\sqrt{\pi a_0}}\sin(\phi_a), \qquad a= \text{c, f, s, sf}, \label{chi}
\end{equation}
where $\kappa_a$ are new Klein factors. These fields are real (Majorana) fermions $\chi\dg_a(x) = \chi_a(x)$, satisfying anti-commutation relations $\{\chi_a(x),\chi_b(y)\} = \delta_{ab}\delta(x-y)$. In terms of these fields, we have the simplest representation of the most general SO(8) symmetric two-leg ladder Hamiltonian~\footnote{This is the SO(8) Gross-Neveu model~\cite{lin1998exact}, which is an integrable theory where one can perform various exact calculations, see~\cite{lin1998exact,essler2005application}}: 
\begin{equation}
H_\text{SO(8)} = \frac{\rmi v}{2}\big(\bar\chi_\alpha\p_x\bar\chi_\alpha - \chi_\alpha\p_x\chi_\alpha\big) - g\sum_{\alpha>\beta}\big(\rmi r_\alpha\bar\chi_\alpha\chi_\alpha\big)\big(\rmi r_\beta\bar\chi_\beta\chi_\beta\big), \label{GN}
\end{equation}
where $\alpha,\beta = 1,2,\ldots,8$, and $r_\alpha =\pm 1$, $g$ are determined by the bare interactions, and the Majorana fields have been relabelled to solely numerical indices. 

For $g>0$ the Hamiltonian~\eqref{GN} has a gapped spectrum. This can easily be obtained by performing a Hubbard--Stratonovich transformation
\begin{equation}
- g\sum_{\alpha>\beta}(\rmi r_\alpha\bar\chi_\alpha\chi_\alpha)(\rmi r_\beta\bar\chi_\beta\chi_\beta) \rightarrow \frac{\Delta^2}{2g} + \rmi\Delta \sum_\alpha (r_\alpha\bar\chi_\alpha\chi_\alpha),
\end{equation}
and looking for the saddle point configuration of the Hubbard--Stratonovich field $\Delta$. As expected from the symmetry of the model, this yields identical (modulo signs) masses for each of the Majorana fermions. The signs, $r_\alpha$, do not affect the dispersion of the excitations~\eqref{dispersion}, but correlation functions do strongly depend on them.  This was demonstrated, for example, in~\cite{controzzi2005excitation} where it was found that there are several possible phases of the half-filled two-leg ladder (the phase of interest to the cuprates problem, the $d$-Mott state, is just one of them and we will discuss this phase further  in section~\ref{sec:dmott}). There are eight inequivalent sets of $r_\alpha$ (automorphisms of the SO(8) group), given by:
\begin{equation}
\begin{split}
&(r_{\text{c}1}r_{\text{c}2}; r_{\text{f}1}r_{\text{f}2}; r_{\text{sf}2}r_{\text{sf}1}; r_{\text{s}1}r_{\text{s}2}) ~~ : \\
 &(++;++;++;++), ~~(++;--;++;++), \\
 &(++;++;+-;--), ~~ (++;--;+-;--), \\
&(++;++;-+;++), ~~(++;--;-+;++),\\
&(++;++;--;--), ~~(++;--;--;--),
\end{split}
\label{grstates}
\end{equation}
Each of these sets preserves the original U(1) charge symmetries of the $c,f$ fields, as well as the SU(2) symmetry of each spin sector. As a result, the spin excitations fall into triplet and singlet multiplets, reflecting the SU(2)$\times$SU(2) on each rung of the ladder. The triplet contains the $\chi_{\text{s}1},\chi_{\text{s}2}, \chi_{\text{sf},1}$ Majorana fermions, whilst the singlet is realised through the $\chi_{\text{sf},2}$ Majorana fermion. 

Even though deviations from the weak interaction picture in realistic ladder materials deform the dispersion, as well as the precise symmetry, it seems reasonable that phase diagram will remain relatively robust. One can also understand what happens to the excitation spectrum under such deviations. The Majorana fermions are collective excitations of the two-leg ladder, and can be classified according to the original symmetries of the Hamiltonian. There are charge neutral $S=1$ and $S=0$ modes (spin excitons), as well as spin-less charge $2e$ Cooperon modes and a charge neutral mode carrying a dipole moment, each of which are formed from two Majorana fermions. For a case with precise SO(8) symmetry, all these excitations have the same mass; when the symmetry is not perfect (as presumably happens for stronger bare couplings) the degeneracy between these excitations is lifted. It is also likely that strong deviations from the SO(8) symmetry destroy the bound states belonging to the 28-dimensional irrep. 
 
Besides the Majorana fermions associated with the vector representation of the SO(8) group, which are collective excitations of the original fermion model~\eqref{hubbardladder}, the model~\eqref{GN} also has quasiparticle excitations. These correspond to kinks in the Hubbard--Stratonovich field $\Delta$, which interpolated between the two degenerate ground states $\Delta(x) = \pm \Delta_0$. These kinks are dressed by Majorana zero energy bound states, which always exist for such a potential~\cite{jackiw1976solitons}. As a result of this dressing, the kinks carry non-trivial quantum numbers. 
 
The Majorana bound states $\gamma_a$ comprise an eight-dimensional Clifford algebra
\begin{equation}
\big\{\gamma_a, \gamma_b\big\} = \delta_{ab},
\end{equation}
whose representations are the spinor irreps of the SO(8) group, according to which the original fermions transform. The topological nature of these `kink plus Majorana bound state' quasiparticles means they must be stable under deviations from the precise SO(8) symmetry~\cite{tsvelik2011field}. This is in clear contrast to the single fermionic chain  which has incoherent  single particle excitations. The coherence  is reflected in the single electron Green's function; in the SO(8)-symmetric case, this reads~\cite{konik2000twoleg}:
\begin{equation}
G(\rmi\omega,k) = -Z\frac{\rmi\omega + \epsilon(k)}{\omega^2 +\epsilon(k)^2 + \Delta^2}.
\end{equation}
This has the characteristic form of the BCS Green's function, but the off-diagonal part is zero. This is due to the fact that left- and right-movers having  different parities belong to different spinor irreps. This is a very important property of the SO(8) Gross-Neveu model which will be exploited below in the YRZ ansatz. 

The web of dualities in the SO(8) symmetric model is in fact even more remarkable: one can rewrite the Majorana fermion model~\eqref{GN} in terms of bosonic Ising variables, which describe the continuum limit of \textit{eight} coupled Ising chains (here labelled by $a,b$): 
 \begin{equation}
 H = \sum_{n}\bigg[J_\text{I} \sum_{a}\Big(-\s^x_{n,a}\s^x_{n+1,a} + \s^z_{n,a}\Big) -g \sum_{a>b}(r_a\s^z_{n,a})(r_b\s^z_{n,b})\bigg]. \label{Ising}
\end{equation}
Here $\s^x_{n,a},\s^z_{n,a}$ are Pauli matrices acting on the $n$th site of the $a$th chain, and $J_\text{I} \sim v$ is the Ising exchange interaction. In this case a Hubbard--Stratonovich transformation leads to 
\begin{equation}
H = \frac{\Delta^2}{2g}  + J_\text{I} \sum_n \sum_{a}\Big(-\s^x_{n,a}\s^x_{n+1,a} + h_a\s^z_{n,a}\Big),
\end{equation}
where $h_a = 1 + r_a\Delta/J$ and $\Delta$ is the Hubbard--Stratonovich field. The role that the signs $r_a$ play in determining the ground state behaviour is now clear, with this varying depending on whether the transverse field $h_a$ is smaller or large than 1 in absolute value.  For small values of the transverse field, the Ising spins order along the $x$-axis, taking a finite expectation value $\la\s^x\ra \neq 0$. On the other hand, at large $h^a$ the spins point along the $z$ axis and $\la\s^x\ra =0$. The quantum Ising model possesses a self duality (yet one more duality in this wonderful SO(8) universe!); namely, by introducing the operators
\begin{equation}
\mu^z_{n+1/2} = \s^x_{n}\s^x_{n+1}, \quad \mu^x_{n+1/2} = \prod_{k <n}\s^z_k,
\end{equation}
the Ising model is transformed into itself, but with different parameters. This is known as the Kramers--Wannier duality~\cite{kramers1941statistics}, under which the Hamiltonian transforms as: 
\begin{equation}
\sum_n\Big(-\s^x_{n}\s^x_{n+1} + h\s^z_{n}\Big) \to \sum_n\Big(-\mu^z_{n+1/2} +h\mu^x_{n+1/2}\mu^x_{n-1/2}\Big).
\end{equation}
Thus when $h > 1$ we have $\la\mu^x\ra \neq 0$ (and hence $\la\s^x\ra = 0$). It also turns out that the superconducting and CDW order parameters of the original fermions are related to products of the Ising fields $\s^x$ and their dual counterparts $\mu^x$ (the so-called disorder parameters). We will discuss this correspondence further in section~\ref{sec:spinfermion}.

Sometimes it is convenient to work with the Hamiltonian written in a different form to~\eqref{GN}, instead leaving the charge sector in terms of the bosonic fields $\phi_\text{c},\phi_\text{f}$. Then the Hamiltonian reads~\cite{lin1998exact}~\footnote{If one did not refermionise the spin sector, the non-linear interaction term within the bosonic theory reads $-4g \sum_{a\neq b} \cos\Theta_a \cos \Theta_b$, which is suggestive of a high symmetry.}:
\begin{equation}
\begin{split}
{\cal H} = &
\frac{v}{8\pi} \sum_{a=\text{c,f}} \Big[ (\p_x \Theta_a)^2 + (\p_x \Phi_a)^2 \Big]  - \frac{g}{2\pi^2} \sum_{a=\text{c,f}} \p_x \phi_a \p_x \bar \phi_a \\
& -4g  \cos\Theta_\text{c} \cos\Theta_\text{f} - 2\rmi g' \big(\!\cos\Theta_\text{c} + \cos\Theta_\text{f}\big) \sum_{a=\text{s,sf}}\sum_{b=1,2} \bar\chi_{ab}\chi_{ab}  \\
& + \frac{\rmi v}{2} \sum_{a=\text{s,sf}}\sum_{b=1,2} \big( \bar\chi_{ab} \p_x \bar\chi_{ab} - \chi_{ab} \p_x \chi_{ab} \big) 
\end{split}
\label{ladderbosonH} 
\end{equation}
where we have written some terms in terms of $\Phi_a = \phi_a +\bar\phi_a$ and $\Theta_a = \phi_a -\bar\phi_a$.  The coupling $g$ in the Hamiltonian~\eqref{ladderbosonH} is the effective coupling after running the RG procedure, see~\eqref{rgflow}. 

\subsubsection{The $d$-Mott state.}
\label{sec:dmott}
It would be fair to call the $d$-Mott phase a failed  superconductor due to the presence of umklapp scattering. The corresponding order parameter field can be represented as 
\begin{equation}
\Delta = \rme^{\rmi\Phi_c/2}A, \label{Delta}
\end{equation} 
where the amplitude $A$ is a product of the Ising order and disorder parameters and has a finite vacuum average only for the particular combination of $r_\alpha$'s characterising the $d$-Mott phase. This combination is naturally realised in the spin-fermion model, the particulars of which we will discuss in section~\ref{sec:spinfermion}. At half filling the charge field $\Phi_c$ fluctuates wildly and cannot condense, but at finite doping the fluctuations become critical and  quasi-long-range order is established. In other words, at finite doping the system develops a strongly singular pair susceptibility $\chi_P \sim T^{-\alpha}$. At half filling the dynamical pair susceptibility has a pole on the real axis in the $\omega$ plane, corresponding to the emission of a gapped excitation with charge $2e$ (known as a Cooperon). Another coherent collective mode is the $S=1$ spin exciton centred on the commensurate wave vector $(\pi,\pi)$.

With a completely gapped ground state established at half-filling~\cite{balents1996weakcoupling,lin1998exact}, let us consider the correlations within this state. These will all be short-ranged (i.e., exponentially decaying) due to all excitations being massive. Nevertheless, we can consider the pairing operator within a given band $\alpha$:
\begin{equation}
\Delta_\alpha(x) = R_{\alpha,\up}(x) L_{\alpha,\dn}(x). 
\end{equation}
To find the relative phase of pairing between the bands, we compute 
\begin{equation}
\la \Delta_+ \Delta_-\dg \ra= - \la \rme^{\rmi (\Theta_\text{sf} + \Theta_\text{f} )}\ra .
\end{equation}
In the ground state, the bosons in this expression become pinned to $2N\pi$ with $N$ an integer~\footnote{This is easy to see in the purely bosonic language, as the interaction term $-4g \cos \Theta_\text{sf} \cos\Theta_\text{f}$ at strong coupling pins the bosons to $(\Theta_\text{sf},\,\Theta_\text{f}) = (2m\pi,\,2n\pi\big)$ or $(\Theta_\text{sf},\,\Theta_\text{f}) =\big( [2m+1]\pi,\,[2n+1]\pi\big)$ with $n,m\in\mathbb{Z}$.}. The relative phase of the pairing between the two bands is thus $\Delta_+ \Delta_-\dg < 0$ and hence has \textit{$d$-wave symmetry!}

And so we return to the statement at the beginning of this section: the Mott insulating ground state of the two-leg Hubbard ladder contains the precursor to $d$-wave pairing. To see this, one can consider doping away from half-filling. Then the total charge boson $\Phi_{c}$ will no longer be gapped, and quasi-long-range order  (power law decaying two-point functions) can arise in the pairing channels, with the relative phase between bands having $d$-wave symmetry~\cite{balents1996weakcoupling,lin1998exact}. The analogy with the high-temperature $d$-wave superconductivity in the cuprates, which emerges from the undoped Mott insulator, is evident. 

In either the doped or undoped cases, the short range antiferromagnetic order means that the lowest energy $S=1$ excitations are gapped (at finite frequency) and centred about momentum $(\pi,\pi)$. Later in this review, we will see that finite frequency $(\pi,\pi)$ $S=1$ excitations are one of the distinguishing features of the type B cuprates, in contrast to the gapless $S=1$ excitations of the type A materials. This clear difference can be seen in neutron scattering experiments~\cite{chan2016commensurate}.

\subsubsection{Other phases of the ladder model.}
\label{sec:otherphases}
As has been mentioned above, the $d$-Mott state (and its finite doping analogue that displays $d$-wave pairing fluctuations) is just one of many ground states that exists within the phase diagram of the two-leg Hubbard ladder~\eqref{hubbardladder}. According to \eqref{grstates} such models can describe a rich variety of physics~\cite{carr2013spinful}, from `standard' CDW, SDW and PDW phases and the ubiquitous 1D Tomanaga-Luttinger liquid, to phases exhibiting superconductivity (of various symmetries) and other more exotic orders (including composite order that mixes spin and charge degrees of freedom on different legs, etc). A discussion of the full phase diagram and the properties of these phases is far beyond the scope of this work; we refer the interested reader to existing reviews and literature, such as~\cite{carr2013spinful} and references therein for details. 

It is, at this point, perhaps worth briefly mentioning that the plethora of phases in ladder models can be seen as a weakness in view of applications to the cuprates: how does one guarantee access to the appropriate physics? We will discuss herein various approaches and pieces of evidence to suggest that this is quite possible. However, one of the beauties of alternative approaches, such as the spin-fermion model~\cite{abanov2003quantumcritical}, is that they can give the phase that supports $d$-wave superconductivity \textit{and nothing else!} 

\subsection{Mott physics at moderate couplings: the two dimensional Hubbard model}
\label{sec:twodhubbard}

In the previous section, we have seen that one-dimensional quantum systems can exhibit the Mott transition at weak coupling, and can show some precursors to $d$-wave superconductivity. It is tempting to make an analogy with the cuprates, but we have already seen that physics in one spatial dimension is rather special. It is thus natural to ask what happens when we move away from one dimension: how much of this physics carries over? In this section, we turn our attention to the Hubbard model in two dimensions and Mott physics that arises there at moderate couplings. 

\subsubsection{Functional renormalization group approaches.}
\label{sec:frg}
Renormalization group techniques have been used to study the weak-to-intermediate coupling regime of models of cuprates since the discovery of high temperature superconductivity~\cite{dzyaloshinskii1987superconducting,schulz1987superconductivity,lederer1987antiferromagnetism}. For a comprehensive and insightful review of the functional RG method and its application to the cuprates, and related topics, see~\cite{metzner2012functional}.  The main focus of study is the single band Hubbard model in two dimensions, thought to be the minimal model of the CuO$_2$ planes. The Hamiltonian reads
\begin{equation}
  \begin{split}
  H_{2d}=& - t \sum_{\la i,j\ra} \sum_{\s=\up,\dn} \Big( c\dg_{\s,i} c_{\s,j} + \text{H.c.} \Big) + U \sum_{j} n_{j,\up} n_{j,\dn} \\
  &  - t' \sum_{\la\la i,j\ra\ra}\sum_{\s=\up,\dn}  \Big( c\dg_{\s,i} c_{\s,j} + \text{H.c.} \Big).
  \end{split}
\label{2dHubbard}
\end{equation}
Here $\la i,j\ra$ denotes a sum over nearest neighbour sites, and $\la\la i,j\ra\ra$ next-nearest neighbours. As emphasised by~\cite{metzner2012functional}, the functional RG is well suited for the analysis of models with scale-dependent behaviour and competing instabilities, both of which are present in the two-dimensional Hubbard model with nearest- and next-nearest-neighbour hopping. 

In the cases when a single instability diverges at low energy scales, the divergence can be understood using mean field theory. However, in the present case divergences appear in \textit{two competing channels simultaneously}, $d$-wave pairing in the particle-particle channel and antiferromagnetism in the particle-hole channel at similar energy scales. This suggests that one looks for a solution which reconciles these competing divergences as happens in the two-leg Hubbard ladder at half-filling~\cite{furukawa1998instability,honerkamp2001breakdown,honerkamp2002flow,lauchli2004dmott}. Indeed, functional RG calculations show clearly that umklapp scattering flows to strong coupling for finite values of the doping in~\eqref{2dHubbard}~\cite{halboth2000renormalizationgroup,honerkamp2001breakdown,honerkamp2002flow,lauchli2004dmott}. 

\begin{figure}
\begin{center}
\begin{tabular}{ll}
(a) & (b) \\
\includegraphics[width=0.4\textwidth]{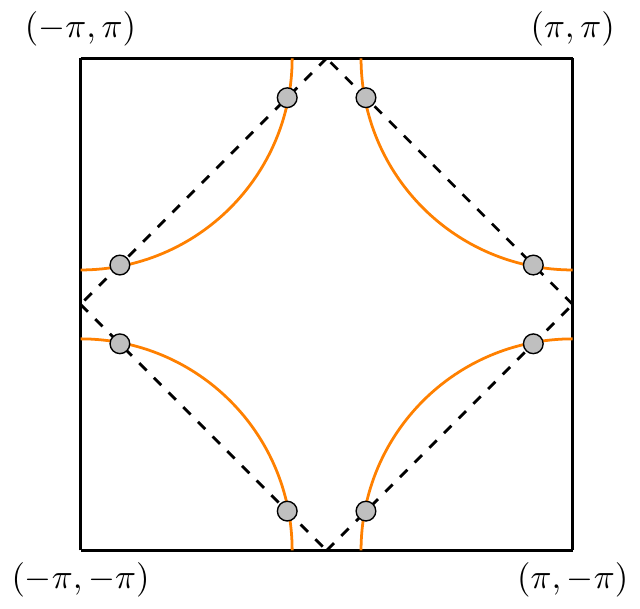} &
\includegraphics[width=0.4\textwidth]{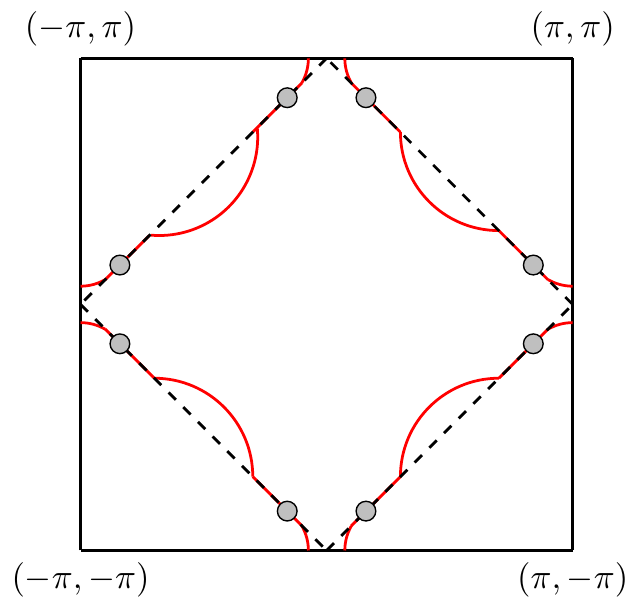} 
\end{tabular}
\end{center}
\vspace{-4mm}
\caption{(a) A schematic illustration of the non-interacting Fermi surface of the two dimensional Hubbard model (orange solid lines) at finite doping, with the umklapp surface shown as dashed black lines. Solid black lines denote the Brillouin zone boundary (with momenta $(k_x,k_y)$ denoted in units of the inverse lattice spacing). We remind that the umklapp surface is \textit{not a surface of constant energy}, unlike the Fermi surface. Shaded grey circles highlight the so-called `hot spots' where the non-interacting Fermi surface and umklapp surface coincide. (b) A (vastly exaggerated) schematic of the  deformed Fermi surface (red lines) due to strong umklapp scattering.}
\label{fig:umklappsurface}
\end{figure}

The presence of strong scattering can lead to the opening of an energy gap and the lowering of the total energy, provided the associated nesting condition of the scattering is satisfied (i.e. for scattering that transfers momentum $\bi{Q}$, there exist electrons at the Fermi surface that differ by momentum $\bi{Q}$). In the present case, the gap will open along the umklapp surface, which is illustrated by the dashed lines in figure~\ref{fig:umklappsurface}(a) (it coincides with the antiferromagnetic Brillouin zone boundary). The nesting condition can be satisfied in two scenarios: firstly, when the hole density is low, such that the non-interacting Fermi surface overlaps approximately with the umklapp surface in the antinodal region. Alternatively, the ground state in the presence of strong interactions and umklapp scattering may be regarded as being connected to an \textit{excited} state of the non-interacting band structure, where the energy of this excitement is more-than-compensated by the opening of the gap. This is illustrated schematically in figure~\ref{fig:umklappsurface}(b)~\footnote{Interestingly, recent gauge theory calculations that are proposed to apply to the pseudogap phase of the cuprates find such a deformation of the Fermi surface. See figure~7(a) of~\cite{sachdev2018gauge}.}. Such a scenario is similar to that proposed for Cr alloys to explain the observation of commensurate antiferromagnetic order in a wide range of doping with the N\'eel temperature double that of the incommensurate order in pure Cr. This illustrates the substantial energy gained by commensurate antiferromagnetic order in itinerant antiferromagnetism~\cite{rice1970bandstructure}.

\subsubsection[Importance of umklapp scattering in the 2D Hubbard model.]{Further evidence for the importance of umklapp scattering in the 2D Hubbard model.}
\label{sec:diagqmc}

\begin{figure}
\begin{center}
\includegraphics[trim=0 0 135 0,clip,width=0.55\textwidth]{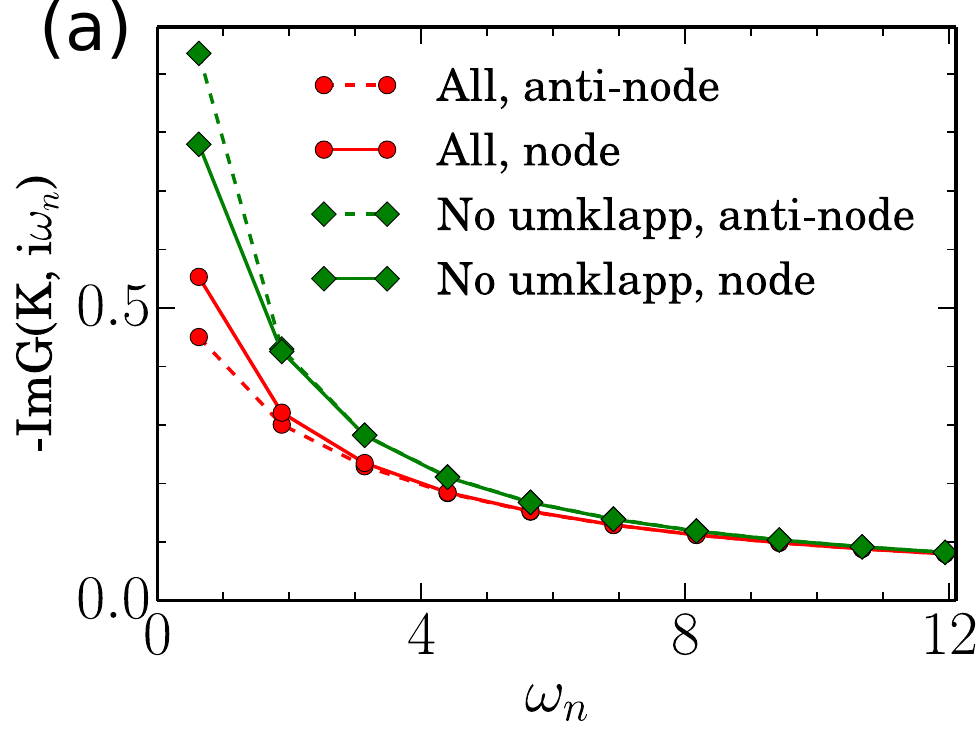}
\end{center}
\vspace{-7mm}
\caption{The imaginary part of the Green's function as a function of Matsubara frequency, plotted in the nodal (points with solid lines) and antinodal (points with dashed lines) regions of the Brillouin zone, computed via diagrammatic quantum Monte Carlo. Results are for the two-dimensional Hubbard model~\eqref{2dHubbard} with $U = 5.6$, $t' = -0.3$ with hole doping $x=0.04$ and temperature $T = 0.2$ (with energies measured in units of the nearest neighbour hopping amplitude $t$) . The circular points show results including all interaction vertices, while diamonds show results with umklapp interactions forbidden. Note that the zero frequency spectral weight at the node/antinode swaps in magnitude when umklapp scattering is included. Figure reproduced from (Wu et al. 2017).}
\label{fig:wu2017fig3}
\end{figure}

The influence umklapp scattering has on the finite doping, moderate coupling physics of the two-dimensional Hubbard model~\eqref{2dHubbard} can further be revealed by other numerical techniques. For example, recent state-of-the-art diagrammatic quantum Monte Carlo simulations have shown that umklapp scattering plays a crucial role in suppressing spectral weight (especially in the antinodal region)~\cite{wu2017controlling}. This is highlighted in figure~\ref{fig:wu2017fig3}, which shows the imaginary part of the Green's function in both the nodal (solid lines) and antinodal (dashed lines) region of the Brillouin zone. With direct control of the diagrammatic summation, Wu and collaborators are able to forbid umklapp scattering (green data in figure~\ref{fig:wu2017fig3}) revealing both that spectral weight is suppressed by umklapp scattering \textit{and} the coherence of the antinodal region is reduced (both in overall magnitude and in comparison to the nodal region). Furthermore, the authors found that the importance of umklapp scattering \textit{grows as the perturbation order of the diagram is increased}. 

Umklapp scattering can also help explain the temperature dependence of the d.c. resistivity in the overdoped phase (which will not be the subject of much discussion herein). In this part of the phase diagram, the cuprates have a large Fermi surface and the high temperature d.c. resistivity can take on a distinctly non-Fermi liquid form, $\rho(T) = \rho_0 + a T^x$ with $x\neq 2$ ($x=2$ is the expectation from Fermi liquid theory). Theoretical calculations~\cite{buhmann2013unconventional} on a single band Hubbard model, with scattering from non-magnetic impurities, shows that umklapp scattering combined with conventional Fermi liquid impurity scattering, can lead to deviations away from $x=2$.

\subsubsection{Cluster dynamical mean field theory.}
\label{sec:cdmft}
 
\begin{figure}
  \begin{center}
    \includegraphics[width=0.5\textwidth]{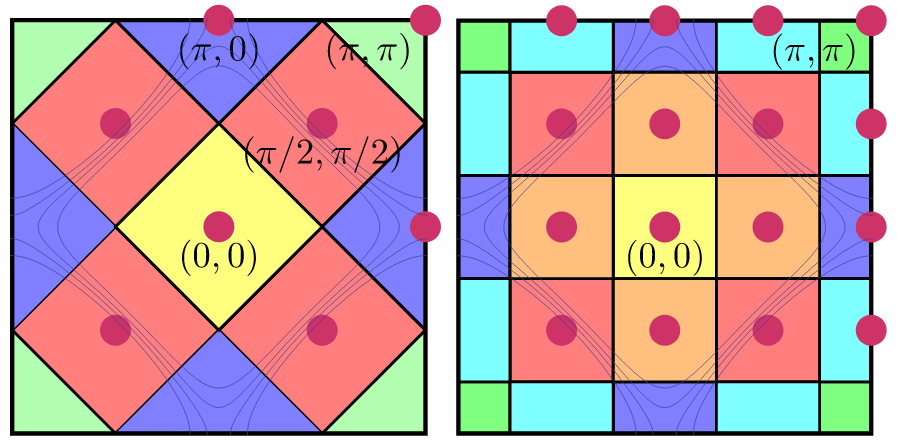}
  \end{center}
  \caption{Examples of momentum space tilings corresponding to (left) $N_c = 8$ and (right) $N_c=16$ cluster approximations (below the tiling non-interacting Fermi surfaces are shown). Red dots show the momenta corresponding to the patches that appear in Eq.~\eqref{cluster}. Figure adapted from (Gull et al. 2010). }
  \label{fig:cluster}
\end{figure}

The approach to the Mott insulator from the hole-doped side, and the opening of the pseudogap in the antinodal, has also beeen studied using cluster dynamical mean field theory in the two dimensional Hubbard model. This approach improves on the original dynamical mean field theory, where the self energy is treated as momentum independent. In the cluster approach, the momentum dependence is partially restored by approximating the self energy as 
\begin{equation}
\Sigma(\omega_n, {\bf k}) \approx \sum_j^{N_c}\Sigma_j(\omega_n)\phi_j({\bf k}), \label{cluster}
\end{equation}
where $N_c$ is the number of clusters and $\phi_j({\bf k})$ constitute a basis in momentum space. In the limit $N_c \rightarrow \infty$ the procedure becomes exact, but in practice the number of clusters is limited. Despite limitation to small $N_c$, the results obtained so far can distinguish between nodal and antinodal parts of momentum space (see figure~\ref{fig:cluster}), and so capture differences in behaviour of the self-energy in these regions. 

In the range of parameters of relevance to the cuprates, the approach to the Mott insulator from finite hole-doping proceeds in two steps. Firstly, in the high-doping regime that properties of the two-dimensional Hubbard model are those of a Fermi liquid. On the other hand, as doping decreases a dramatic difference emerges for the nodal and antinodal regions, with the antinodal debeloping a pseudogap in self energy~\cite{lin2010physics}. No long range order is required for this, such that the pseudogap is a feature of the Hubbard model at intermediate correlation lengths. Furthermore, it was established that at higher temperatures the nodal and antinodal scattering rates have different temperature dependences~\cite{gull2010momentumspace}: $\Gamma \sim T^2$ and $\Gamma \sim T$, respectively. We will see that the analytical approaches, discussed in the next subsection, reproduce this dichotomy in behaviour.

Cluster dynamical mean field theory has also been used to determine the collective modes that give the largest contributions to the self-energy, and so give rise to the pseudogap in the antinodal. Using fluctuation diagnostics, \cite{gunnarsson2015fluctuation} unambiguously identified a well-defined collective spin mode at $(\pi,\pi)$ as being responsible for the pseudogap behaviour in the antinodal, as well as majority of the differentiation between the nodal and antinodal regions. 

\subsection[Extending half-filled two-leg Hubbard ladders to two dimensions]{Analytical approaches to extending the physics of the short range ordered ground state of half-filled two-leg Hubbard ladders to two dimensions}
\label{sec:approaches}

The so-called $d$-Mott ground state of the half-filled two-leg Hubbard ladder is special and raises the question whether an extension to two spatial dimensions is possible. Its fully gapped spectrum, accompanied by strictly short range order, is similar to the features of the pseudogap state in the antinodal region of the Brillouin zone. Several methods to generalise the short range order physics of two-leg ladders to two dimensions have been proposed. In this section we will discuss three approaches in particular. Firstly, a phenomenological ansatz for the single particle propagator was proposed by Yang, Rice and Zhang in 2006~\cite{yang2006phenomenological}. Later, Ossadnik proposed a wave packet approach to describe short range ordered states with momentum space anisotropy in two dimensions~\cite{ossadnik2016wave}. Finally, we will discuss recent results from Tsvelik, who approached the problem using a strong coupling spin-fermion model formulation of the two dimensional Hubbard model, and showed that some of the physics maps to ladder physics~\cite{tsvelik2017ladder}.

\subsubsection{The Yang-Rice-Zhang ansatz for the single particle propagator.}
\label{sec:yrzprop}

The extension of results for the two-leg Hubbard ladder (at and near to half-filling) to the case of two spatial dimensions is difficult. Instead Yang, Rice and Zhang (YRZ)~\cite{yang2006phenomenological} looked for an approximate closed form for the single electron propagator in two dimensions by examining the evolution of the single electron propagators on $N$-leg ladders as $N$ is increased. Here we will briefly summarise their approach, referring the reader to a review for more details~\cite{rice2012phenomenological}. In the appendix we discuss some attempted microscopic approaches to deriving the YRZ ansatz for the single electron propagator.

The simplest case is the four-leg ladder with nearest neighbour hopping; the model contains four bands crossing the Fermi level. At half-filling these bands can be grouped into inner and outer pairs, whose Fermi velocities differ with the larger velocity being for the outermost bands. Turning on interactions, larger energy gaps open on the outer pair of bands. As a result, when hole-doped away from half-filling, initially all the holes enter the inner bands, leading to a truncation of the original four-band Fermi surface into just two bands. We see that energy is gained by splitting the Fermi surface in momentum space into gapped and open parts. This naturally led to the proposal to extend this distortion of the original band structure Fermi surface in two dimensions. It is then straightforward to write down a single electron propagator under this approximation scheme, 
\begin{equation}
\begin{split}
G_\text{YRZ}(\bi{k},\omega) &= \frac{g_t(x)}{\omega - \xi(\bi{k}) - \Sigma_\text{YRZ}(\bi{k},\omega)}, \\
\Sigma_\text{YRZ}(\bi{k},\omega) &= \frac{\Delta_\text{R}^2(\bi{k})}{\omega + \xi_0(\bi{k})}, \qquad g_t(x) = \frac{2x}{1+x}, 
\end{split}
\label{eq:yrz}
\end{equation}
where 
\begin{equation}
\begin{split}
  \xi(\bi{k}) &= \xi_0(\bi{k}) - 4t'(x) \cos k_x \cos k_y \\
  &~~ - 2t''(x) (\cos 2k_x + \cos 2k_y) - \mu, \\
\xi_0(\bi{k}) &= - 2 t(x) (\cos k_x + \cos k_y), \\
\Delta_\text{R}(\bi{k}) &= \Delta_0 (\cos k_x - \cos k_y).
\end{split}
\end{equation}
Here the effective hopping parameters $t,t',t''$ are renormalised by the interactions, e.g. by Gutzwiller factors, and thus are dependent on the doping, $x$.

The key feature in the YRZ ansatz for the single electron propagator~\eqref{eq:yrz} is the placing of the pole in the self-energy on the surface $\xi_0(\bi{k})=0$. This surface, often called the umklapp surface, is spanned by wave vectors $(\pi,\pm\pi)$ that coincide with the Brillouin zone in the presence of antiferromagnetic order. This leads to zeroes of the propagator, $G_\text{YRZ}(\bi{k},\omega=0)$ near the antinodal region. Closed Fermi pockets are present in the nodal regions with spectral weight that varies strongly around the pocket. Exact values for the parameters $t,t',t'',J,\Delta_0,\mu$ are not specified within the YRZ scheme and can be chosen as fits to experimental data or, alternatively, to fit ab initio calculations. 

\begin{figure}
\begin{center}
\includegraphics[width=0.8\textwidth]{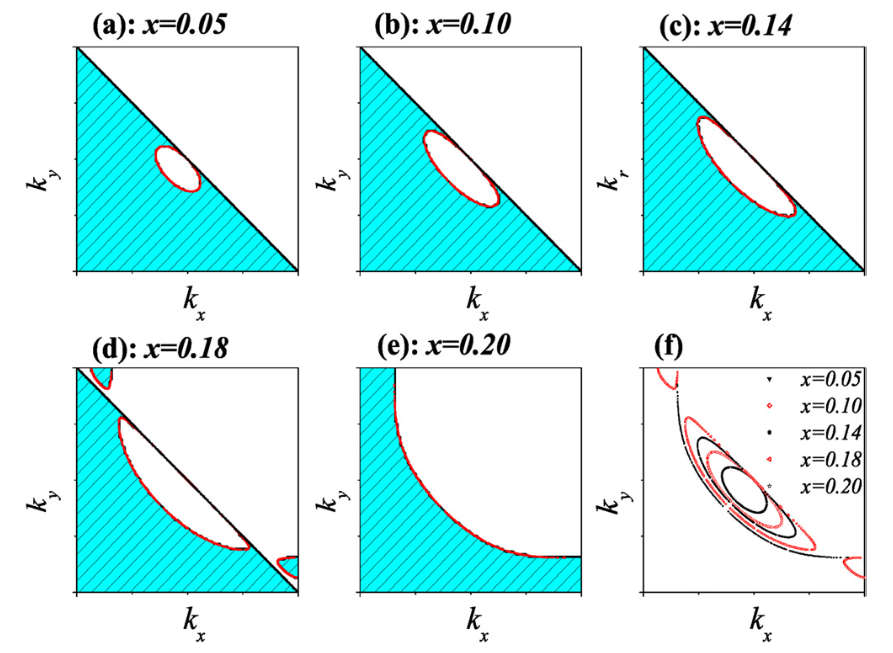}
\end{center}
\vspace{-4mm}
\caption{Evolution of the Fermi surface as a function of hole doping $x$ for the YRZ single electron propagator, showing quarter of the Brillouin zone. The Fermi pockets (formed from red lines in (a)--(c)) are defined through poles of the Green's function, enclose an area of $x$ in the Brillouin zone. The black lines denote zeros of the propagator, rather than poles, which run along the umklapp surface. At higher dopings, (e), a large Fermi surface is recovered. Figure reproduced from (Rice et al. 2012).}
\label{fig:rice2012}
\end{figure}

The YRZ single electron propagator~\eqref{eq:yrz} captures much of the phenomenology of the cuprates~\cite{rice2012phenomenological}. As illustrated in figure~\ref{fig:rice2012}, it has hole pockets at small doping [see panels (a)-(c)], whilst at large doping it has a single large Fermi surface [see panel (e)]. The antinodal points $(\pm\pi,0)$ and $(0,\pm\pi)$ are connected via lines of zeroes of the propagator along the antiferromagnetic Brillouin zone boundaries. This coincides with the umklapp surface that appears in functional RG calculations on the weak coupling 2D Hubbard model; as discussed in~\cite{honerkamp2001breakdown,honerkamp2002flow,lauchli2004dmott}, at weak to moderate couplings umklapp scattering in both particle-hole and particle-particle channels grows strongly under the RG. As a result an energy gap opens on this umklapp surface below a critical scale (and hence there is only short range, rather than long range, ordering). The hole pockets shown by the YRZ propagator, see figure~\ref{fig:rice2012}, are defined through the Fermi surface where $G_\text{YRZ}(\bi{k},\omega=0)$ changes signs through infinities. We note that the total area enclosed by the four hole pockets in units of the Brillouin zone volume is equal to the hole density $x$, and a generalised Luttinger sum rule is satisfied [see, e.g.,~\cite{konik2006doped}].  We note that poles in the self energy of the single particle Green's function in the two-dimensional Hubbard model have been observed in cluster dynamical mean field theory calculations~\cite{stanescu2006fermi,haule2007strongly,sakai2009evolution,lin2010physics,sakai2018direct}, although these studies do not relate the poles to umklapp processes.

To summarize the main picture: functional RG shows a flow to strong coupling of umklapp scattering at weak to moderate values of the Hubbard interaction. Two instabilities, singlet d-wave superconductivity and $(\pi,\pi)$ antiferromagnetism, are incompatible as long range order, and the flow to strong coupling of umklapp scattering avoids these, imposing short range order in both channels. Furthermore, the strong umklapp scattering changes the self energy, in turn introducing singularities along the umklapp surface in momentum space. It is worth emphasizing that this is \textit{not} equivalent to the formation of charge density wave order and subsequent reconstruction of the Fermi surface as seen, for example, in the presence of a large magnetic field~\cite{sebastian2011quantum,sebastian2012towards,vignolle2013from,sebastian2015quantum}.

The YRZ ansatz has provided successful fits for various experimental data in the pseudogap phase. The work of Carbotte, Nicol and coworkers shows this for the in-plane optical conductivity~\cite{iles2009nonbcs}, the $c$-axis optical conductivity~\cite{ashby2013caxis}, the specific heat~\cite{leblanc2009specific}, the London penetration depth~\cite{carbotte2010effect} and Raman scattering~\cite{leblanc2010signatures}, with each being considered as a function of doping and temperature. Other authors have also considered applying the YRZ ansatz to describe magnetic response~\cite{james2012magnetic,dean2013highenergy}. \cite{valenzuela2007phenomenological} had previously analysed the Raman scattering results under a `two-gap scenario', where they used difference dependences for the antinodal pseudogap and nodal superconducting gap as compared to the YRZ ansatz. This alternative approach has been discussed by many authors, see e.g.~\cite{hufner2008two}.

\subsubsection{Ossadnik's wave packet approach.}
\label{sec:ossadnik}

The short range ordered ground state of the half-filled two-leg Hubbard ladder cannot be described in terms of a mean field theory plus one-dimensional fluctuations, which rules out applying such an approximation in two dimensions. The challenge of constructing a microscopic theory for systems with strictly short range order and no broken symmetries inspired Matthias Ossadnik~\cite{ossadnik2016wave} to propose that the standard extended Bloch basis, as used in long range ordered systems, should be replaced by a more localised wave packet basis. He introduced an orthonormal Wilson-Wannier basis~\cite{sullivan2010phase}, formed from even and odd combinations of Wannier functions, and rewrote the many-body Hamiltonian is this wave packet basis. This led him to a fermionic formulation of the short range ordered ground state of the two-leg Hubbard ladder at half-filling, which agreed with the standard bosonization analysis discussed above. It was then demonstrated that this wave packet approach could be generalised, in a straightforward manner, to two dimensions (unlike the bosonization approach). We note, however, that explicit calculations were not carried out for this case.

In recent work,~\cite{liu2017transformation} used this approach to obtain qualitative results, in the simplest possible approximation scheme. The great advantage of this formalism, in comparison to a real space approach, is that it allows one to examine behaviour in momentum space near the Fermi energy \textit{and} that in other regions of the Brillouin zone. Thus it is suitable for describing the opening of an energy gap in the two-particle spectrum due to umklapp scattering in the antinodal region of the Brillouin, where low-energy effective interactions are strongest, while still retaining gapless degrees of freedom in the nodal region. Analogously to the Hubbard ladder, the presence of umklapp scattering is key to turning the superconducting gap at overdoping into an insulating pseudogap at underdoping. 

Earlier work by~\cite{honerkamp2001breakdown} noted that the Hubbard model on a two-dimensional square lattice near half-filling has a Fermi surface that is close to the umklapp surface (the surface defining the Brillouin zone on the square lattice with antiferromagnetic order). In cuprates with next-nearest-neighbour hopping, this umklapp surface is not a surface of constant energy, but each $\bi{k}$-point on the surface is part of a degenerate set of eight points, each connected via umklapp scattering processes. \cite{liu2017transformation} divided these eight points into two subsets of four points, with velocities parallel to the $(1,1)$ and $(1,-1)$ directions, respectively. Within each subset, the four points are connected by umklapp scattering processes that transfer momentum $(\pi,\pi)$, similar to the case of two-leg ladders at half-filling. $\bi{k}$-points within the two different subsets are also connected by umklapp scattering (but not with $(\pi,\pi)$ momentum transfer), which were neglected in~\cite{liu2017transformation} to approximate the ground state of the eight $\bi{k}$-point ``star'' as the product of two independent four $\bi{k}$-point sets. The point of this approximation is that calculations are significantly simplified, with a four $\bi{k}$-point set being mapped onto a two-leg Hubbard ladder. 

\subsubsection{Mapping between the spin-fermion model and two-leg ladders.}
\label{sec:spinfermion}
Recent work by Tsvelik has also shed light on the intimate relation between the physics of the cuprates and ladder models~\cite{tsvelik2017ladder}. The starting point is a strong-coupling formulation of the two-dimensional Hubbard model known as the spin-fermion model~\cite{abanov2003quantumcritical}. It is described by the Hamiltonian 
\begin{equation}
\begin{split}
H_\text{sf} =& \sum_\bi{k} \epsilon(\bi{k}) \psi\dg_\s(\bi{k}) \psi_\s(\bi{k}) + \frac{1}{2}\sum_\bi{q} \chi^{-1}(\bi{q}) \bi{S}(-\bi{q}) \cdot \bi{S}(\bi{q})\\
& + g \sum_{\bi{k},\bi{q}} \psi\dg_\alpha(\bi{k}+\bi{q}) \vec{\s}_{\alpha\beta} \psi_\beta(\bi{k}) \cdot \bi{S}(\bi{q}). 
\end{split} 
\label{Hsf}
\end{equation}
where $\psi\dg_\s(\bi{k})$ is the creation operator for an electron of spin $\s$ with momentum $\bi{k}$, $\bi{S}(\bi{q})$ describe spin fluctuations with momentum $\bi{q}$, $\epsilon(\bi{k})$ is the electronic band structure, $g$ is the spin-fermion coupling strength, and $\chi(\bi{q})$ is the spin susceptibility
\begin{equation}
\chi(\bi{q}) = \frac{\chi_0}{ \xi^{-2} +  \big( \bi{Q}-\bi{q}\big)^2},
\label{spinsus}
\end{equation}
with $\bi{Q} = (\pi, \pm \pi)$ being the wave vectors for antiferromagnetic scattering and $\xi$ the spin correlation length. In the context of the cuprates, we consider the case where the electronic band structure is such that there exist hotspots: points on the Fermi surface that are connected by wave vector $\bi{Q}$. As $\bi{Q}$ is the antiferromagnetic wave vector, this is equivalent to the statement that the Fermi surface crosses the umklapp surface, see for example figure~\ref{fig:umklappsurface}. 

The spin-fermion model is a phenomenological model which has been extensively used to describe the physics of the cuprates~\cite{abanov2003quantumcritical}. It assumes that the important low-energy degrees of freedom in a strongly correlated doped Mott insulator consist of quasiparticles with an extended Fermi surface and collective magnetic excitations, known as paramagnons. This is consistent with the results from cluster dynamical mean field theory discussed in section~\ref{sec:cdmft}. The paramagnons have a dispersion relation that is centred on the wave vector $\bi{Q}$. The starting point for considerations is thus a strong Hubbard repulsion $U\gg t$, and it is therefore remarkable that we will find the same physics here as in the weak-coupling limit of the two-leg ladder.

The strongly peaked nature of the spin susceptibility~\eqref{spinsus} about $\bi{Q}$ means that strong correlations only emerge in the vicinity of the hotspots. As is the case in section~\ref{sec:frg} (and see (b) of figure~\ref{fig:umklappsurface}), one can speculate that strong spin-fermion coupling will lead to a modification of the Fermi surface in order to stabilise nesting. In order for this, and the subsequent formation of a gap, to happen the coupling $\gamma \propto \chi_0 g^2$ must exceed some critical value (the self-consistent solution of the spin-fermion model at small bare coupling~\cite{schlief2017exact} shows a critical behaviour instead). The hotspots can be grouped into two quartets, one in which hotspots on opposite sides of the Fermi surface are connected by $\bi{Q} = (\pi,\pi)$, and the other in which $\bi{Q} = (\pi,-\pi)$. To first approximation these two sets can be treated as decoupled; we label the hotspots by $\bi{k}^a_{R,L}$ with $a=\pm$ and focus on electrons in their vicinity through the tomographic projection
\begin{equation}
\begin{split}
R_{a\s}(x) &= \frac{1}{\sqrt{2\pi}} \int \rmd k_\parallel \, \psi_\s(\bi{k}^a_R + k_\parallel \bi{e})\rme^{\rmi k_\parallel x},\\
L_{a\s}(x) &= \frac{1}{\sqrt{2\pi}} \int \rmd k_\parallel \, \psi_\s(\bi{k}^a_L + k_\parallel \bi{e}) \rme^{\rmi k_\parallel x}, 
\end{split}
\end{equation}
where $\bi{e} = (1,1)/\sqrt{2}$ and the above defines the right/left directions. The low-energy effective description for these degrees of freedom now reads $H = \int \rmd x ({\cal H}_0 + {\cal V})$ with
\begin{equation}
\begin{split}
{\cal H}_0 &= -\rmi v_F \sum_{a,\s} \Big( R\dg_{a\s} \p_x R_{a\s} - L\dg_{a\s} \p_x L_{a\s} \Big), \\
{\cal V} &= - \gamma \sum_{a,b=\pm} \Big( R\dg_{a \alpha} \vec{\s}_{\alpha\beta} L_{a\beta} + L\leftrightarrow R \Big)\\
& \qquad\qquad \times \Big( R\dg_{b\gamma} \vec{\s}_{\gamma\delta} L_{b\delta} + L\leftrightarrow R \Big), 
\end{split}
\label{sfladder}
\end{equation} 
where $\gamma \propto \chi_0 g^2$. Here we emphasise the importance of commensurability of the magnetic excitations, which are responsible for umklapp terms in this low-energy effective Hamiltonian.

We now see that the spin-fermion model has now been reduced to an effective ladder model~\eqref{sfladder}. This is not entirely surprising -- we have focused on the physics in the vicinity of a set of four Fermi points. The ladder model~\eqref{sfladder} is of a similar form to that of the half-filled ladders discussed in section~\ref{sec:twoleghubbard}. Such a correspondence is powerful: non-perturbative methodologies specific to one-dimensional quantum systems can now be turned to study the physics of this model~\cite{james2018nonperturbative}.

The model~\eqref{sfladder} has a number of distinctive features, including an absence of interchain tunnelling $t_\perp = 0$ and enhanced bare symmetry. As we have explained in section~\ref{sec:twoleghubbard}, the general ladder model possesses U(1)$\times$U(1)$\times$SU(2)$\times\mathbb{Z}_2$ symmetry, but at half-filling~\eqref{sfladder} has an even higher O(5)$\times$O(3) symmetry due to the special structure of the interactions. This symmetry enhancement simply strengthens the general statements on two-leg ladders made in section~\ref{sec:twoleghubbard}. Namely, at small values of $\gamma$ one-loop perturbative renormalization group calculations show an enhancement to a gigantic SO(8) symmetry in the low-energy limit~\cite{lin1998exact,konik2000twoleg,konik2001exact,konik2002interplay,essler2005application}. The low-energy effective theory is a Gross-Neveu model~\cite{gross1974dynamical}, which is exactly solvable and allows many quantities to be computed~\cite{lin1998exact,konik2000twoleg,konik2001exact,konik2002interplay,essler2005application}. When $\gamma$ is not small, and perturbative calculations are not valid, this SO(8) symmetry becomes approximate but many physical features of the model remain qualitatively similar to the exactly solvable case~\cite{tsvelik2011field}. 

As we also previously mentioned, generically half-filled ladder models can describe a number of phases (see section~\ref{sec:otherphases}). The most important feature of the ladder formulation of the spin-fermion model, discussed above, is that it puts the systems squarely into the $d$-Mott phase~\cite{tsvelik2017ladder}, which contains precursors to $d$-wave superconductivity. This is described by the order parameter
\begin{equation}
\Delta_d = R_{+\up}L_{-\dn} - R_{-\up}L_{+\dn} -  R_{+\dn}L_{-\up} + R_{-\dn}L_{+\up}, \label{orderparam}
\end{equation}
which can be expressed in terms of the Ising order/disorder parameters fields, discussed in section~\ref{sec:effectivetheory}, as 
\begin{equation}
\Delta_d \sim \rme^{\rmi\Theta_c/2}\Big(\s_3\sigma_4\sigma_5\mu_6\mu_7\mu_8 + \rmi \mu_3\mu_4\mu_5\s_6\s_7\s_8\Big). \label{Deltad}
\end{equation}
Here the fields are numbered according to the classification of~\eqref{grstates}. As was already pointed out, since the phase field $\Theta_c$ remains strongly fluctuating, the $d$-wave superconducting order parameter $\Delta_d$ cannot condense. Nevertheless, the phase can contain, so to speak, ``seeds of superconductivity'' if the amplitude of the operators in the bracket has non-zero vacuum average. For this to happen, the Majorana fermions labelled $3,4,5$ and $6,7,8$ must have different signs, corresponding to the ground state configuration $(++;++;+-;--)$ in~\eqref{grstates}~\footnote{There is a subtle difference between the band and the chain representations of the two-leg ladder. In the case of large $t_{\perp}$ discussed in section~\ref{sec:twoleghubbard} the $d$-Mott state corresponds to all signs of $r_a$ being equal~\cite{controzzi2005excitation}}.

In the $d$-Mott phase, the Cooperon is a coherent excitation with a finite energy gap that leads to insulating, not superconducting, behaviour. It is also interesting that the existence of such an excitation in the $d$-Mott phase also requires the existence of a coherent $S=1$ neutral exciton centred at the commensurate wave vector $(\pi,\pi)$, thereby enhancing the gain in magnetic energy. Below we will discuss the relation of this to the cuprates. 
  
\begin{figure}
  \begin{center}
    \includegraphics[width=0.6\textwidth]{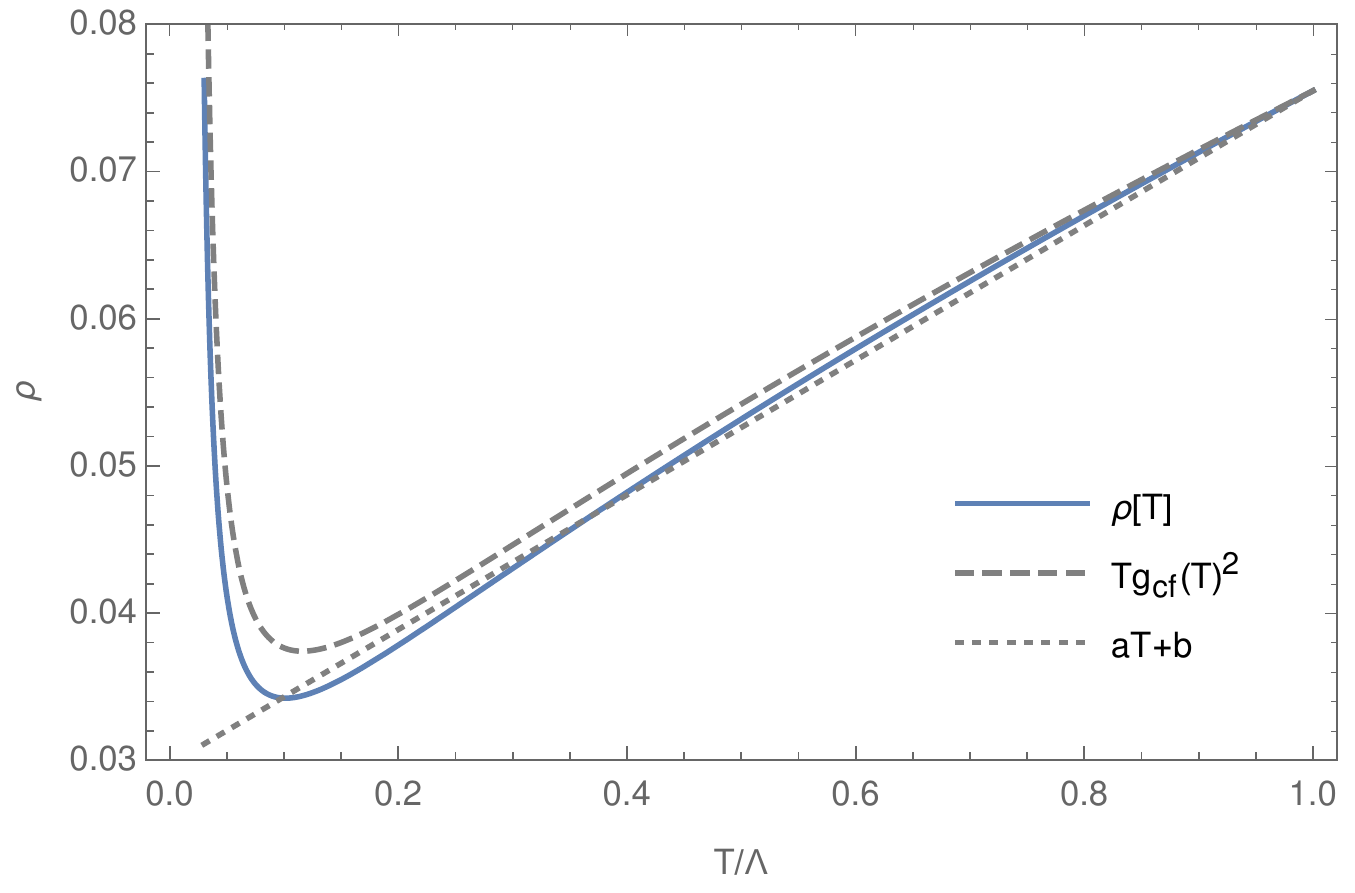}
  \end{center}
  \caption{The d.c. resistivity $\rho(T)$ as a function of temperature $T$, obtained from the optical conductivity $\sigma(\omega,T)$ computed within the memory function formalism for the ladder formulation of the spin-fermion model (Tsvelik 2017). At high temperatures the resistivity is approximately linear in temperature (dotted line), as expected within the strange metal phase. The low-temperature divergence signals the formation of the pseudogap in the antinodal region described by the ladder formulation. In the full model, this diverging resistivity is short-circuited by the nodal Fermi-liquid contributions, which dominate at temperatures below $T^\ast$ and are neglected in the ladder calculation. Figure reproduced from (Classen et al. 2018).}
  \label{classen}
\end{figure}

The described ladder formulation of the spin-fermion model was recently used to compute the antinodal contributions to the optical conductivity, $\sigma(\omega,T)$, as a function of frequency $\omega$ and temperature $T$~\cite{classen2018ladderlike}. The calculation made use of the memory function formalism, which was first applied to Luttinger liquids with umklapp scattering by~\cite{giamarchi1991umklapp}. At finite temperatures $T$, the extrapolation of the optical conductivity to zero frequency yields the d.c. resistivity $\rho(T)$ shown in figure~\ref{classen}. At high temperatures, this is consistent with the linear-in-$T$ behaviour of the strange metal phase, while at low-temperature (below $T^\ast$) the resistivity diverges due to the formation of the pseudogap in the antinodal region. In practice, this divergence is regulated by the Fermi-liquid-like contribution from the nodal quasiparticles, which is neglected in the ladder formulation and dominates the transport upon the opening of the pseudogap. This dichotomy in scattering rates between the nodal and antinodal is the same as that observed in cluster dynamical mean field theory, see section~\ref{sec:cdmft}. At high frequencies the optical conductivity was found to behave as a power law $\text{Re}\,\sigma(\omega) \sim \omega^{-\alpha}$ with a non-universal power law $\alpha$ that depends on the bare interaction strength.

\subsubsection{The two fluid model.}
\label{sec:twofluid}
A common feature of the works described in the previous sections,~\cite{ossadnik2016wave,liu2017transformation,tsvelik2017ladder}, is the partitioning of the Fermi surface into two separate regions, the nodal and the antinodal. This naturally sets the stage for a two-fluid model of transport~\footnote{We note that two-fluid models have been used to fit experimental data, such as in~\cite{clayhold2010constraints}, but without microscopic justification. A separate two-fluid model, based on the upper and lower branches of the YRZ formalism is described in~\cite{storey2012twocomponent}.}. Using a deformed Fermi surface, much like that shown in figure~\ref{fig:umklappsurface}, a two-fluid model of the d.c. conductivity was proposed in \cite{rice2017umklapp}
\begin{equation}
\s(T) \approx \s_\text{nodes}(T) + \s_\text{antinodes}(T). \label{twofluid}
\end{equation}
The antinodal regions, corresponding to the nested parts of the Brillouin zone, were treated non-perturbatively within a ladder formalism in a similar manner to the previous section~\cite{tsvelik2017ladder}. The nodal regions were instead assumed to be a Fermi liquid. Combining these in the two fluid model~\eqref{twofluid}, the authors computed both the temperature-dependence of the d.c. resistivity and the Hall angle, finding results consistent with experiments on the cuprate superconductors.

To be precise, following a bosonization analysis in the ladder formalism, the temperature dependence of the d.c. resistivity was found to be~\cite{rice2017umklapp}
\begin{equation}
\rho(T) = A T \left[ \exp\left( -\frac{\alpha T^\ast}{T}\right) + \frac{B T^\ast}{T} \right]^{-1} .
\label{twofluidresistivity}
\end{equation}
At high temperatures, $T \gg T^\ast$, this is dominated by umklapp scattering from the nested antinodal regions, which leads to linear-in-temperature resistivity, $\rho(T) \propto T$. On the other hand, at temperatures below the pseudogap onset $T^\ast$ the antinodal region becomes gapped and the Fermi liquid nodal pockets dominate. This leads to a conventional $\rho(T) \propto T^2$ behaviour at low temperatures. The constants $A,B,\alpha$ can be treated as fit parameters, and the form~\eqref{twofluidresistivity} fits well to experimental data~\cite{rice2017umklapp}, and reproduces the behaviour observed in \textit{a priori} calculations, such as cluster dynamical mean field theory~\cite{gunnarsson2015fluctuation}. 

The temperature dependence of the Hall angle, describing transverse transport in a magnetic field, also follows straightforwardly from the form of the deformed Fermi surface and the two fluid separation. The ladder formulation of the antinodal region has two degenerate bands, meaning that there is no effective interchain tunnelling. Thus these regions cannot support transverse transport, and only the Fermi liquid nodal pockets contribute. This leads to the experimental observation of a Fermi liquid-like $T^2$ transverse transport coefficient in a magnetic field. This simple reasoning is supported by a detailed calculation, using Ong's geometric approach for the Hall effect in 2D materials~\cite{ong1991geometric}, as shown in the supplemental material of~\cite{rice2017umklapp} 

In the context of the two-fluid model the $d$-wave superconductivity  originates from coupling of the Cooperon to nodal quasiparticles. We remind the reader that Cooperon are bound states of two \textit{massive} electrons (or holes) with zero momentum---excitations in the vector representation of the SO(8) group. These excitations live within the single-particle gap, and their energy will vanish (allowing condensation) before the closing of this gap; this mechanism was already studied in previous work on ladder models~\cite{konik2006doped}, which also served as motivation for the YRZ ansatz. The coupling between Cooperons and nodal quasiparticles is described by
\begin{equation}
  \begin{split}
    \delta H &= \gamma \sum_{{\bi q},{\bi k}}\Delta_d({\bi q}) \big(\cos k_x - \cos k_y\big)\\
    &\qquad\quad \times \psi^+_{\s}\left({\bi k}+\frac{\bi q}2\right)\psi^+_{-\s}\left(-{\bi k} + \frac{\bi q}2\right) + \text{H.c.},
    \end{split}
\end{equation}
where $\Delta_d(\bi{q})$ is the Cooperon field. When $\gamma = 0$ the Cooperon is massive, but the presence of finite interactions $\gamma\neq0$ lowers the energy of both the Cooperon and the nodal quasiparticles, eventually leading to the formation of the quasiparticle $d$-wave order parameter and the condensation of Cooperon.

\subsection[Competing ground states in the two-dimensional Hubbard model]{Complications: competing ground states in the two-dimensional Hubbard model}
\label{sec:competinggs}

Finally, let us discuss in more detail the issue of competing ground states. With general agreement that the essential physics of the cuprates is captured by the Hubbard model at intermediate coupling $U/t \sim  8$ and the $t$-$J$ model, its strong coupling version, extensive numerical works have been undertaken~\cite{himeda2002stripe,maier2005quantum,capone2006competition,raczkowski2007unidirectional,aichhorn2007phase,chou2008clusterglass,yang2009nature,chou2010mechanism,corboz2011stripes,hu2012absence,gull2013superconductivity,corboz2014competing,leblanc2015solutions,corboz2016improved,zheng2017stripe,huang2017numerical,ehlers2017hybridspace,iso2018competition,huang2018stripe,jiang2018superconductivity,huang2018strange,darmawan2018stripe}. With a lack of controllable analytical tools to tackle the problem, such numerics provide essential (and non-perturbative) insights into the nature of ground states of the Hubbard model. Surprisingly, almost degenerate ground states have emerged as a universal feature of these studies. Split by just a small fraction of the bandwidth, $\sim 0.01t$ per site, these two state exhibit wildly different properties: (A) states containing intertwined spin and charge orders, with or without coexisting pair orders; (B) a uniform $d$-wave superconducting state. As already mentioned in the introduction, these results strongly motivate our classification of the cuprates in types A and B, respectively. 

\begin{figure}
\begin{center}
\begin{tabular}{ll}
(a) & (b) \\
\includegraphics[width=0.45\textwidth]{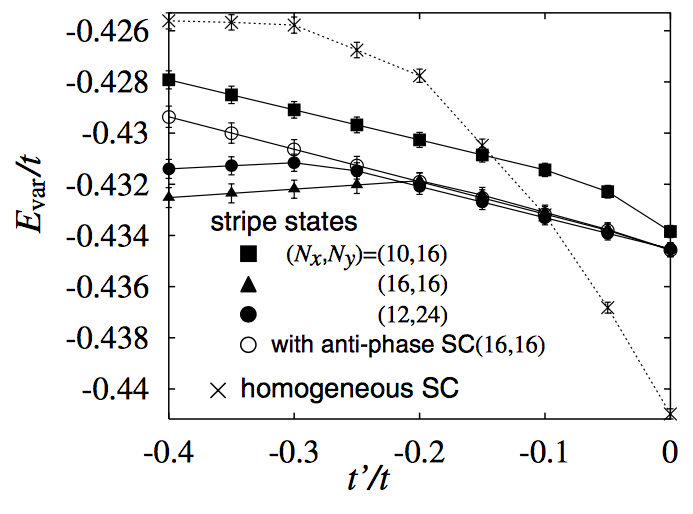} &
\includegraphics[width=0.45\textwidth]{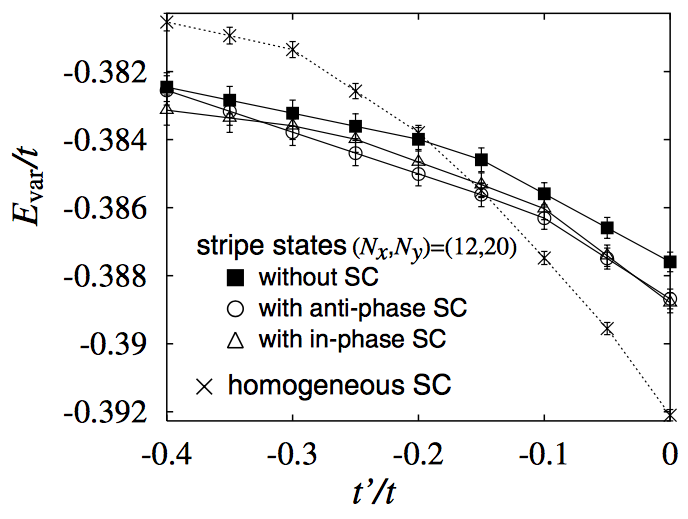} 
\end{tabular}
\end{center}
\vspace{-4mm}
\caption{Comparison of the variational energy densities of stripe states in the $t-t'-J$ model with $J/t = 0.3$, $t=1$ as a function of $t'$ for (a) $x=1/8$ doping; (b) $x=1/10$ doping. Stripe states, exhibiting intertwined charge, spin and pair order, are close in energy with the homogeneous superconducting state (cross symbols). $N_x, N_y$ denote the number of sites in the $x$,$y$ directions, respectively. Figure reproduced from (Himeda et al. 2002).}
\label{fig:himeda2002}
\end{figure}

One of the first studies to highlight the presence of competing ground states was the variational quantum Monte Carlo analysis of Himeda and coworkers~\cite{himeda2002stripe}. They considered the two dimensional $t$-$t'$-$J$ model, related to the Hubbard model through an exclusion of double occupied sites, with the Hamiltonian
\begin{eqnarray}
  H &=& -t \sum_{\la i,j\ra, \s} \mathbb{P} \Big( c\dg_{i,\s}c_{j,\s} + \text{H.c.} \Big) \mathbb{P} + J \sum_{\la i,j \ra} \bi{S}_i \cdot \bi{S}_j \nn
        && - t' \sum_{\la\la i,j \ra\ra,\s} \mathbb{P} \Big( c\dg_{i,\s}c_{j,\s} + \text{H.c.} \Big) \mathbb{P}, 
\end{eqnarray}
where $\mathbb{P} = \prod_i (1 - n_{i,\up}n_{i,\dn})$ is known as the Gutzwiller projection operator, which projects out doubly occupied sites. $t$ is the nearest-neighbour hopping amplitude, $t'$ is the next-nearest-neighbour hopping amplitude, and $J$ is the nearest neighbour spin exchange interaction. Trial states, including those with homogeneous $d$-wave superconductivity and those with intertwined stripe order, were variationally optimised to minimise their energy. Results of the variational quantum Monte Carlo studies for various system sizes (a lattice of $N_x\times N_y$ sites) are shown in figures~\ref{fig:himeda2002} for (a) $x=1/8$ doping and (b) $x=1/10$ doping~\cite{himeda2002stripe}. These show that in the regime applicable to the cuprates (approx $t'/t \sim -0.3$ to $t'/t \sim -0.1$) there is close competition between stripe states (both those with and without superconductivity) and the homogeneous superconducting state. Indeed, the difference in variational energy per site between the translational symmetry breaking stripe states and uniform superconductivity is less than $0.01t$ on average, a very small energy scale. 

Modern advances in numerical simulations have given further credence to the picture of~\cite{himeda2002stripe}. Tensor network methods, such as infinite projected-entangled pair states, allow the simulation of thermodynamic systems relevant to the cuprates to unprecedented levels of accuracy~\cite{corboz2011stripes,corboz2014competing,corboz2016improved,zheng2017stripe}. Other techniques have also been continuously pushed, using both increased computational resources and improvement in algorithms, see the recent works~\cite{zheng2017stripe,huang2017numerical} for further information. 

\begin{figure}
\begin{center}
\includegraphics[width=0.8\textwidth]{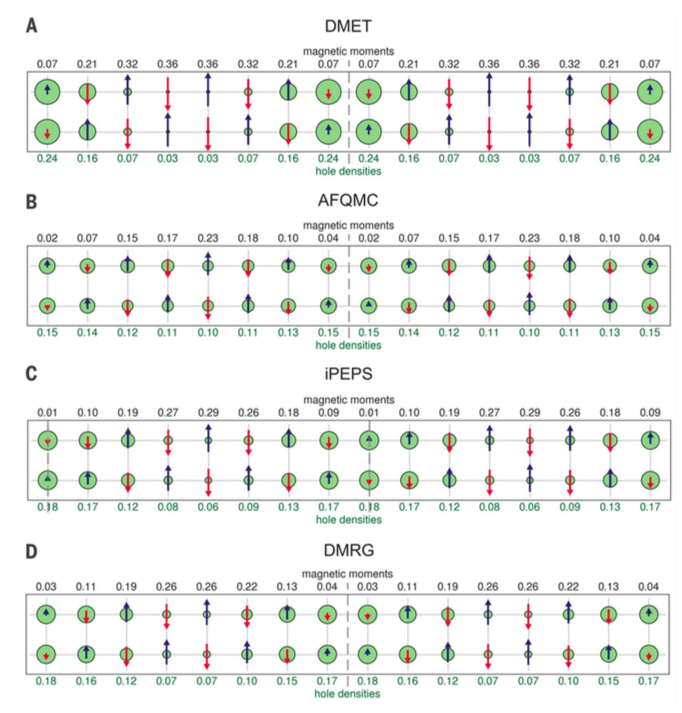}
\end{center}
\vspace{-4mm}
\caption{Charge and spin order in the period 8 stripe states from (A) density matrix embedding theory (DMET); (B) auxiliary field quantum Monte Carlo (AFQMC); (C) infinite system projected entangled pair states (iPEPS) ; (D) density matrix renormalization group (DMRG) simulations. Numerical values for the magnetic moment (hole density) on each site are shown above (below) each plot. Arrow size denotes the spin density and circle size the hole density. The position of the maximum hole density and the magnetic domain wall are shown by the grey dashed lines. Figure reproduced from (Zheng et al. 2017).
}
\label{fig:zheng2017}
\end{figure}

Figure~\ref{fig:zheng2017} shows a summary of results for the period-8 ground states of the two-dimensional Hubbard model at $x=1/8$ hole doping with $U/t =8$, as obtained by state-of-the-art numerical methods~\cite{zheng2017stripe}.  A vertical stripe ground state, see~\cite{zheng2017stripe}, is found to be the absolute ground state, but uniform $d$-wave and alternative stripe ordering are proximate in energy density $\sim 0.01t$. It is fascinating that this competition between almost-degenerate ground states, which exhibit very different physical properties, is borne out in experiments on the cuprates. Besides these qualitative similarities there \textit{are} differences between theory and the experimental observations. Theory predicts a longer wavelength for the stripe order and, more importantly, with a few notable exceptions it does not find PDW order. These discrepancies can be interpreted as an indication that the Hubbard model is not sufficient to provide a \textit{detailed} description of the physics of cuprates. One important feature omitted in the Hubbard model is the long-range Coulomb interaction. Estimates in~\cite{zheng2017stripe} show that this interaction reduces the inter-stripe distance. Another important neglected factor is the coupling between the electronic degrees of freedom and the lattice. Recent X-ray scattering experiments~\cite{chen2018persistent} have demonstrated that the stripe pattern in $x=1/8$ LBCO is determined by the domain pattern of the low temperature orthorhombic phase. These experiments found that the domain pattern of ordered stripes was reproduced after a heating cycle through the CDW transition provided the maximal temperature of the cycle did not exceed the low-temperature orthorhombic transition temperature. 
  
This small separation in energy between states (with very different properties) presents a challenge for numerical algorithms. Small biases in methods, which could result from finite size effects, incomplete treatments of certain physics (such as fluctuations), or systematic issues such as the sign problem, can easily favour one state over another~\cite{zheng2017stripe}. It also presents a challenge for numerical methods that intrinsically work at finite temperature, with the need to distinguish very small energy gaps. The height of the potential barrier that separates these two close-in-energy states is also not yet clear. Recent experimental works~\cite{miao2017hightemperature,miao2018incommensurate} have argued that it may not be particularly high, perhaps being of the order of $100\,$K in $x=1/8$ LBCO. There the temperature was raised above the CDW transition, and it was observed that charge and spin responses remained peaked around certain wave vectors ($Q_\text{CDW}$ and $Q_\text{SDW}$, respectively) with the locking relation $Q_\text{CDW} = 2Q_\text{SDW}$ being lost.

This small separation in energy between potential ground states also has important implications when realistic, additional, terms are added to our low-energy effective descriptions. For example, terms related to next-neighbour hopping and longer-range interactions may play an important role in choosing the absolute ground state. Indeed, recent works studying the addition of next-neighbour hopping to the 2D Hubbard model suggest that these may be important for capturing the stripe period at $x=1/8$ doping observed in experiments on type A materials~\cite{iso2018competition,jiang2018superconductivity}.

As was discussed in the introduction, these numerical results strongly motivate our partitioning of the cuprates into two sets, so-called `type A' and `type B' cuprates. The above theoretical developments review the path to describing the properties of the uniform type B ground state, which captures various anomalous features of the normal states. In the following section, we will now discuss the experimental signatures of these.

\section{Experimental overview: anomalies and signatures of the pseudogap} 
\label{sec:exp}

The study of high-temperature superconductivity in the cuprates has been driven by experiments since their discovery~\cite{bednorz1986possible,wu1987superconductivity,maeda1988new,sheng1988bulk,schilling1993superconductivity}. Numerous techniques have been brought to bear on the problem, analysing every available aspect of the material. The mysterious pseudogap phase in particular has received much attention~\cite{rigamonti1998basic,timusk1999pseudogap,hufner2008two,vishik2018photoemission}, motivated by a need to understand the `normal' state from which superconductivity emerges as temperature is decreased.

\subsection{Nuclear magnetic and quadrupolar resonance}
\label{sec:nmr}

NMR is a powerful, and widely used, tool to study electronic structure and it has been applied to assess many aspects of the superconductivity (and other properties) of cuprates~\cite{warren1989cu,alloul1989nmr,walstedt1990nmr,takigawa1991cu,machi1991nuclear,asayama1996nmr,rigamonti1998basic}. A detailed and comprehensive account can be found in the recent book by R.~E.~Walstedt~\cite{walstedt2010nmr}. 

We begin with LBCO, the best example of the type A underdoped cuprates. The pioneering neutron scattering studies by Tranquada and coworkers~\cite{tranquada1995evidence} established the interplay between CDW (in the form of hole stripes) and SDW ordering with phase slips at hole stripes. The presence of a very weak $c$-axis Josephson coupling in the same family of materials as ascribed to the presence of phase slip lines in the $d$-wave superconductivity, i.e. PDW order~\cite{berg2007dynamical}. These three different intertwined orders lead to a complex order ground state in LBCO. 

The isoelectronic cuprate LSCO is much more disordered that LBCO, which leads to an even more complex interplay between these translational symmetry breaking components. The disorder complicates the interpretation of the NMR spectra. Shortly after the experiments of Tranquada et al. evidence from NMR spectra suggested charge ordering within LSCO~\cite{hunt1999nqr}. This interpretation was hotly contested by X-ray scattering, which suggested an absence of charge ordering.  Only in the last few years, with advances in X-ray scattering techniques, has the presence of charge ordering been revealed~\cite{thampy2014rotated,croft2014charge}. Recent work by Imai and coworkers~\cite{imai2017revisiting} undertook a careful and detailed NMR study of LSCO, finding a similar interplay between CDW and SDW order as in LBCO, but with detailed patterns that are strongly modified by disorder. They also show that the amplitude of the CDW is very small, explaining why initial x-ray scattering experiments could not observe the associated Bragg peaks. We will not go into further detail here, instead assigning LSCO to be a type A cuprate and focusing our attention on those underdoped cuprates without broken translational symmetry.  

\begin{figure}
\begin{center}
\includegraphics[width=0.7\textwidth]{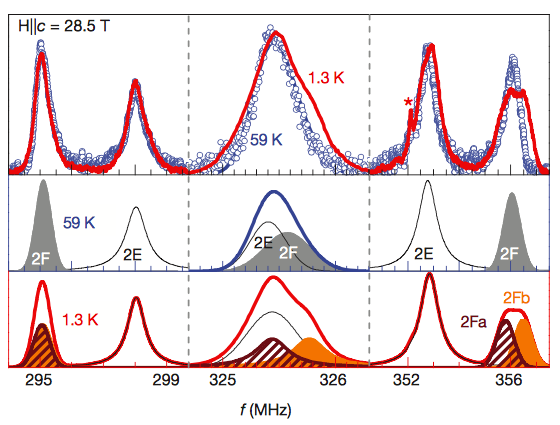}
\end{center}
\vspace{-4mm}
\caption{(Top panel) $^{63}$Cu NMR spectra of YBa$_2$Cu$_3$O$_{6.54}$ for an applied magnetic field parallel to the $c$-axis of strength $H = 28.5$\,T. Data is shown at high temperature $T = 59$\,K (blue points) and low temperature $T=1.3$\,K. The NMR spectra can be decomposed into contributions from different sites, as shown in the lower two panels for $T= 59$\,K and $T = 1.3$\,K, respectively. Charge density wave formation below $T_\text{CDW} = 50$\,K causes the Cu2F peak to split into Cu2Fa and Cu2Fb, as seen on the right of the lower plot. Figure adapted from (Wu et al. 2011).}
\label{fig:nmrH}
\end{figure}

The type B underdoped cuprates, for example Y248, YBa$_2$Cu$_3$O$_{7-x}$, and Hg1201, show quite different behaviour in NMR to LBCO and other type A underdoped cuprates. There are no signs of static long range CDW or SDW order, except under very strong magnetic fields~\cite{wu2011magneticfieldinduced,wu2013emergence,jang2016ideal}. This is clearly seen in the NMR spectra of figure~\ref{fig:nmrH}: at low temperatures and high magnetic fields, CDW order is seen through the splitting of the Cu2F peak into two sub-peaks, Cu2Fa and Cu2Fb. The case of the stoichiometric double chain cuprate Y248, is especially interesting because the mobile hole per formula unit doping of the CuO$_2$ layers is spread over the extra full CuO$_2$ chain when compared with YBa$_2$Cu$_3$O$_{7-x}$. Thus in Y248 the planar hole density is reduced and as a result it is the cleanest and best ordered underdoped cuprate. It displays a simple NMR pattern at all temperatures~\cite{tomeno1994nmr}. Unfortunately, it is difficult to grow samples of this cuprate, which has ruled out many other investigations (but not NMR, where one can study small samples). 

The flexibility and wide range of applicability of NMR is illustrated by experiments of Mali, Keller and collaborators~\cite{strassle2011absence} to directly check for the presence of orbital currents around the planar Cu$_4$O$_4$ plaquettes. Such currents are present in a d-density wave state which has been discussed in the literature in connection with the pseudogap~\cite{chakravarty2001hidden,lee1998su2,varma2006theory}. A key feature of the Y248 crystal structure is the asymmetric position of the Ba nucleus along the $c$-axis between Cu$_4$O$_4$ plaquettes. As a result the Ba nucleus should experience a net magnetic field in a $d$-density wave state. However their experiments failed to observe a magnetic field, which led them to put a tight upper bound on the size of such a magnetic field~\cite{strassle2011absence}. This has been further supported by polarised neutron scattering in charge-ordered YBCO~\cite{croft2017no}. 

\subsection{Photoemission studies of the pseudogap regime}
\label{sec:arpes}

\begin{figure}
  \begin{center}
    \includegraphics[width=0.4\textwidth]{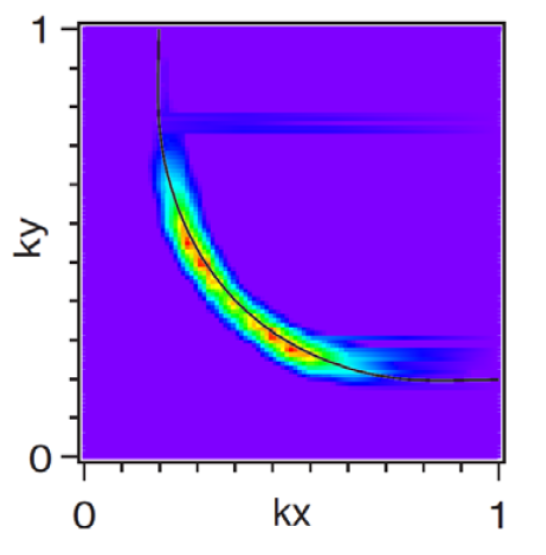}
  \end{center}
  \vspace{-4mm}
  \caption{An ARPES measurement of the low lying excitations around the Fermi surface, typical of the underdoped regime of the cuprate superconductors. The measurement is shown in one quadrant of the Fermi surface, with kx and ky expressed in reciprocal lattice units. The pseudogaps are observed at the end of the arcs in the $(\pi,0)$ and $(0,\pi)$ directions. Figure adapted from (Norman et al. 2007).}
  \label{arpes1}
\end{figure} 

Driven by technological advances and the need to understand the electronic structure of high-$T_c$ superconductors, ARPES has developed into one of the key probes of condensed matter systems today~\cite{damascelli2003angleresolved}. The technique allows the direct mapping of the dispersing electronic structure and, more importantly for this review, identifies the momentum dependence of any single particle gaps in the spectral response. Following the identification of the superconducting gap and its momentum dependence in optimally doped Bi$_2$Sr$_2$CaCuO$_{2+\delta}$ (Bi2212), ARPES studies of the underdoped regime in the normal state above $T_c$ quickly identified the presence of a spectral gap, the ``pseudogap'', in the antinodal (i.e. the copper-oxygen bond) directions~\cite{ding1996spectroscopic,loeser1996excitation}. The presence of the pseudogap in these directions leads to disconnected Fermi arcs in the nodal directions; a representative ARPES spectrum is shown in figure~\ref{arpes1}~\cite{norman2007modeling}. Photoemission studies of highly overdoped cuprate materials, on the other hand, found evidence of a full Fermi surface in the normal state with an area proportional to $1+p$, with $p$ the doping level~\cite{plate2005fermi}. The transition from arcs at underdoping to full Fermi surface at overdoping presumably reflects the crossover from the doped Mott insulator to something more akin to a Landau-Fermi liquid.  

\begin{figure}
  \begin{center}
    \includegraphics[width=0.95\textwidth]{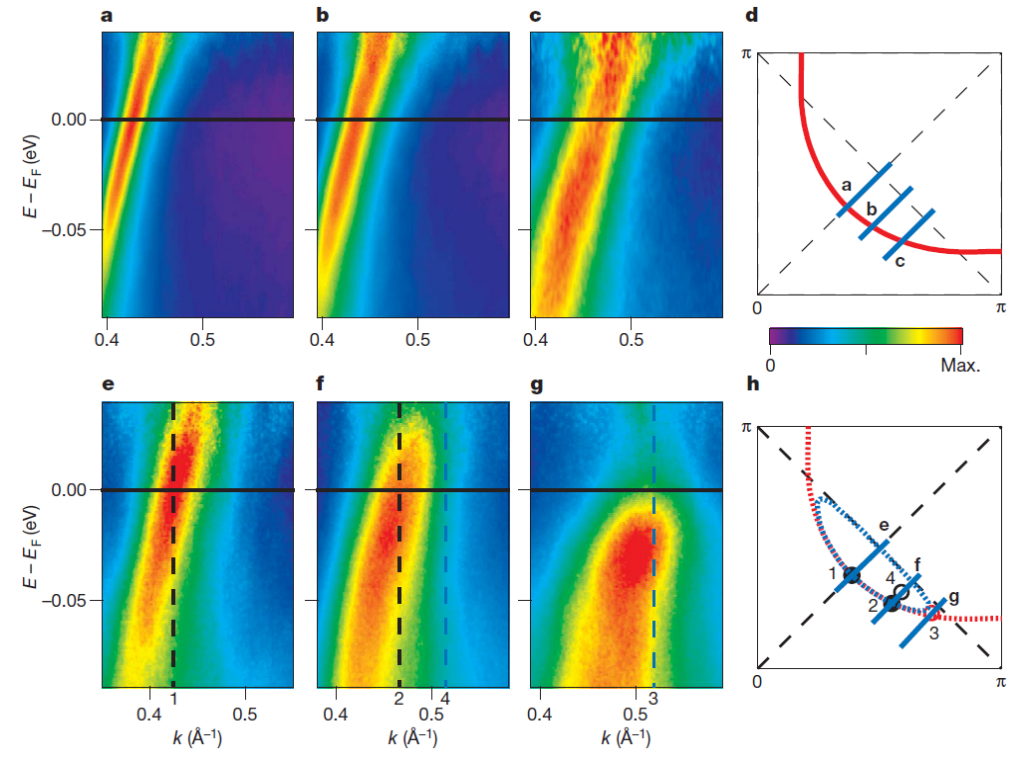}
  \end{center}
  \caption{(a)--(c) ARPES spectra recorded in optimally-doped Bi2212 ($T_c = 91$\,K) in both the nodal direction (a) and away from it, as indicated in (d). (e)--(g) are ARPES spectra for underdoped Bi2212 ($T_c = 65$\,K) for the cuts indicated in (h). The magnetic zone boundary would lie at $k = 0.58$ \AA$^{-1}$. In (e) and (f) the vertical black dashed line indicates the Fermi surface crossing. In (f) and (g), the vertical blue dashed line indicates the turning point at the top of the dispersion. These are indicated in (h) by the open circles, black indicating a turning point above the Fermi level and red a turning point below the Fermi level. The filled black circles indicate the position of Fermi crossings. Figure reproduced from (Yang et al. 2008). }
  \label{arpes2}
\end{figure}

In general, we certainly expect closed Fermi surfaces in condensed matter systems and, as such, the anomalous Fermi arcs in the underdoped materials have been the subject of considerable investigation. Several studies have been interpreted as indicating a temperature dependent arc length~\cite{kanigel2006evolution},  others a doping dependent length~\cite{yang2008emergence,kaminiski2015pairing}. Photoemission requires an electron in the initial state; recognising this Yang and collaborators~\cite{yang2008emergence} raised the sample temperature and normalised the measured spectral intensity to the appropriate Fermi distribution, so exploiting the ability of ARPES to also examine states above the chemical potential. As shown in figure~\ref{arpes2}, in a study contrasting optimally doped and underdoped Bi2212, they found a previously unidentified band gap that moves down and crosses the chemical potential when moving away from the nodal direction towards the antinodal direction~\cite{yang2008emergence}. As noted, near the nodal direction the crossing of the chemical potential defines a Fermi arc, one of the defining characteristics of the underdoped cuprates. In the antinodal directions, this Fermi arc gives way to a spectral gap. Yang and coworkers associated this antinodal gap with the pseudogap that is central to the YRZ ansatz discussed earlier (see section~\ref{sec:yrzprop}). It is also important to emphasise that the particle-hole asymmetry identified in the study of Yang et al. will not be evident in studies that simply employ symmetrisation in energy around the chemical potential to investigate potential gaps in the ARPES response. 

\begin{figure}
  \begin{center}
    \includegraphics[width=\textwidth]{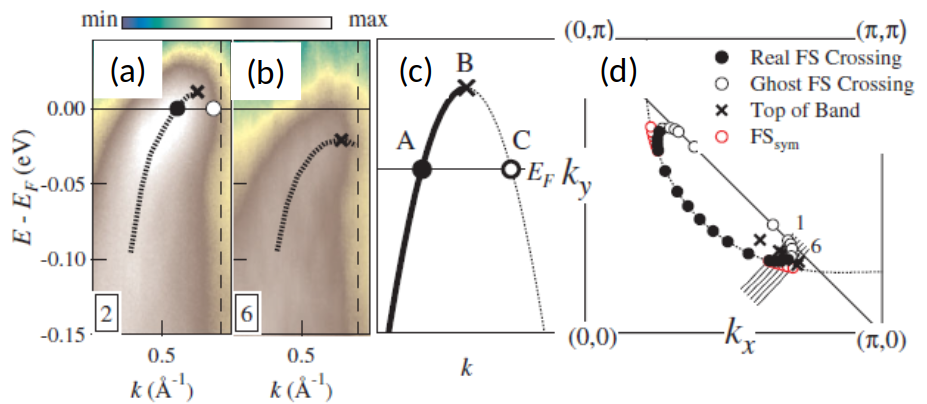}
  \end{center}
  \caption{(a) The ARPES spectra from a cut on the Fermi arc after `Fermi normalization'. The Fermi surface crossing (solid circle) and the top of the band (cross) are indicated. A ``ghost'' Fermi surface crossing (open circle), as deduced from symmetrisation in momentum is also indicated. The dashed line shows the position of the magnetic Brillouin zone boundary. (b) The same plot showing that there is no Fermi surface crossing beyond the arc. (c) A schematic depicting the discussed points (symbols the sames as in (a), (b)). (d) The pseudo-pocket as determined for the $T_c = 65$\,K sample. Here the black circles indicated the Fermi surface crossing corresponding to point A in the schematic. Similarly, the position of the top of the band, point B in the schematic, is shown by crosses. The open circles represent the ``ghost'' Fermi surface, point C in the schematic. The red circle are the Fermi surface crossing obtained from symmetrisation of the data. Finally, the dashed line in (d) indicated the large Fermi surface predicted by first principles local density approximation calculations. Figure reproduced from (Yang et al. 2011).}
  \label{arpes3i}
\end{figure}

\begin{figure}
  \begin{center}
    \includegraphics[width=0.8\textwidth]{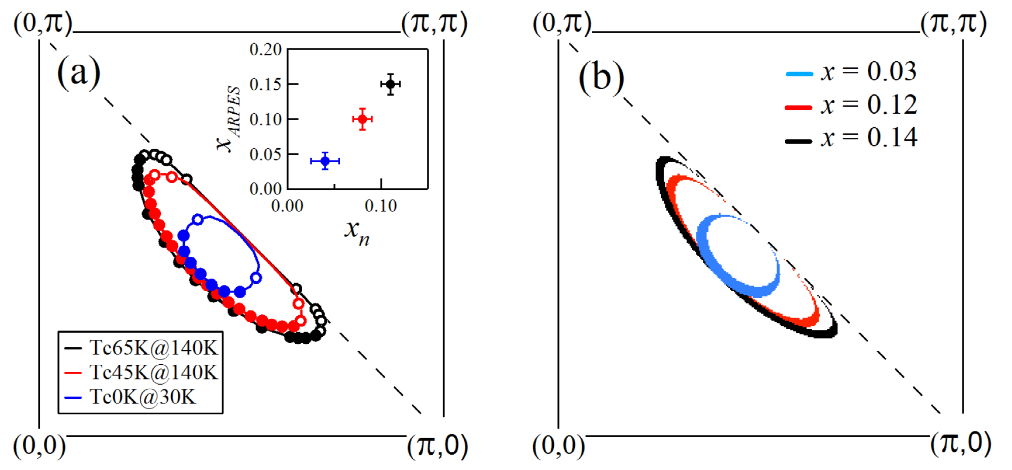}
  \end{center}
  \caption{(a) The pseudo-pockets determined for three different doping levels. The black data correspond to the $T_c = 65$\,K sample, the red data correspond to the $T_c = 45$\,K sample, and the blue data correspond to the non-superconducting $T_c = 0$\,K sample. The area of the pockets in ARPES scales with the nominal doping level $x_n$, as shown in the inset. (b) The Fermi pockets derived from the YRZ ansatz with equivalent doping levels. Figure reproduced from (Yang et al. 2011).}
  \label{arpes3}
\end{figure}

Assuming that the observed particle-hole asymmetry could be associated with the presence of the pseudogap, Yang and collaborators followed up their earlier study with a more detailed examination of the doping dependence of the asymmetry in a range of underdoped samples~\cite{yang2011reconstructed}. As illustrated in figure~\ref{arpes3i}, the measured dispersion was fitted to a parabolic form, as proposed in several theoretical models of the pseudogap region. This allowed the doping-dependence of the ``hole pockets'' to be mapped, and providing evidence that the ``pockets'' had an area proportional to the doping, as shown in figure~\ref{arpes3}.  

The photoemission spectral intensity, $A({\bf k},\omega)$, is determined by the imaginary part of the Green's function~\footnote{The ARPES intensity ${\cal I}({\bf k},\omega,T)$ at temperature $T$ is related to the imaginary part of the Green's function via  ${\cal I}({\bf k},\omega) = {\cal M}({\bf k},\omega) f(\omega,T) A({\bf k},\omega)$. Here ${\cal M}({\bf k},\omega)$ is the matrix elements of the dipole operator, and $f(\omega,T)$ is the Fermi-Dirac distribution.}:
\begin{equation}
  \begin{split}
    A({\bf k},\omega) &= {\rm Im}\, G({\bf k},\omega),\\
    &= \frac{\Sigma_2({\bf k},\omega)}{\left[ \omega - \epsilon_0 - \Sigma_1({\bf k},\omega) \right]^2 + \left[\Sigma_2({\bf k},\omega) \right]^2},
    \end{split}
\end{equation}
where $\epsilon_0$ represents the bare velocity and $\Sigma_{1,2}({\bf k},\omega)$ represent the real and imaginary components of the self-energy, which describes the interaction between the electron and its environment. For the ``pockets'' discussed above, it is important to note that these consist of the observed Fermi arc on one side (defined through infinities of the Green's function) and, on the backside, zeros of the Green's function as it switches from positive to negative along the umklapp surface. The latter coincides with the antiferromagnetic Brillouin zone boundary. The zeros of the Green's function on the backside of the ``pocket'', which will be invisible in any ARPES experiment, conform to the Luttinger sum rule~\cite{luttinger1960fermi}, as generalised by Dzyaloshinskii~\cite{dzyaloshinskii2003some}.

Interestingly, a study of La-Bi2201 did seemingly provide evidence of completely closed hole pockets in the nodal direction~\cite{meng2009coexistence}. However a subsequent study attributed the closure on the backside to superstructure replicas of the main and shadow bands~\cite{king2011structural}. Similar superstructures are also evident in detailed studies of Bi2212 [see, e.g.,~\cite{drozdov2018phase}]. However all of the authors of the latter studies note that the observed superstructures do not explain the earlier of observations of Yang and coworkers~\cite{yang2008emergence}. 

It is also interesting that a recent ARPES study of the Fermi arcs at temperatures below $T^\ast$ appears to indicate that the Fermi arc is indeed confined within the antiferromagnetic Brillouin zone, with the tip falling short of the boundary~\cite{kaminiski2015pairing}. This is consistent with the study of Yang and colleagues~\cite{yang2008emergence}. As noted earlier, both studies~\cite{yang2008emergence,kaminiski2015pairing} report a temperature-independent arc length, consistent with the concept of the small hole pockets having an area proportional to the doping.

\begin{figure}
  \begin{center}
    \includegraphics[width=\textwidth]{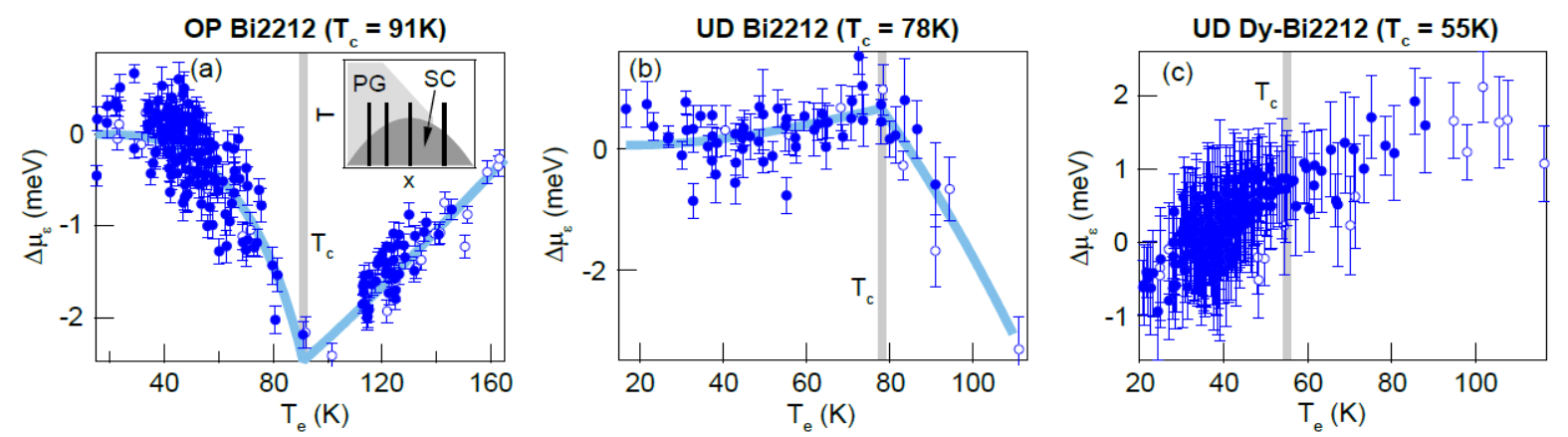}
  \end{center}
  \caption{The function $\Delta\mu_\epsilon(T_e)$ plotted for: (a) optimally-doped Bi2212 with $T_c = 91$\, K; (b) slightly under-doped Bi2212 with $T_c = 78$\,K; and (c) very underdoped Dy-Bi2212 with $T_c = 55$\,K. Data are collected over many delay times, pump ﬂuences, and initial temperatures. The inset of (a) shows the cuprate phase diagram, including the superconducting (SC) and pseudogap (PG) phases, and black lines indicating the doping of the four samples in this study. Empty circles denote data taken between $-0.3$ and $0.3$ ps delay, and grey bars indicate $T_c$. Figure reproduced from (Miller et al. 2017).}
  \label{arpes4}
\end{figure}

As an alternative to raising the temperature of the sample and using thermal excitation to populate states above the Fermi level, infrared pumping can also be used to examine these states. The particle-hole asymmetry, a signature of the pseudogap, has been investigated in this manner in two pump-probe photoemission studies of underdoped systems. In the first such study,~\cite{miller2017particlehole} explored the density of states in the vicinity of the chemical potential by pumping with infrared radiation to excite electrons to unoccupied states above the Fermi level. There it was noted that if the density of states is constant with respect to temperature then any change in chemical potential $\Delta \mu_\epsilon$ will be given by
\begin{equation} 
\Delta \mu_\epsilon \propto W^2 \frac{D'(E)}{D(E)}\Bigg\vert_{E\sim E_F}
\end{equation}
where $W$ is the bandwidth, $D(E)$ the density of states, and $D'(E)$ its first derivative with respect to energy. As shown in figure~\ref{arpes4} for the highly underdoped region, with pumping, the change in sign of $D'(E)$ in the vicinity of the Fermi energy $E_F$ is well-described by the particle-hole asymmetry characteristic of the YRZ model (see section~\ref{sec:yrzprop}).

In the second pump-probe study, \cite{freutel2019optical} measured the change in shape of the Fermi surface on ``photodoping'', the process by which the effective hole count is changed via infrared pumping. They found that on timescales of the order of $100$\,fs the Fermi arc would expand and then relax back to its equilibrium value. Comparing an optimally doped sample with an underdoped one, they found that the maximum change in the Fermi wave vector $k_F$ was near the end of the arc in the underdoped sample, again consistent with the concept of small pockets in the pseudogap regime, and different to the large Fermi surface interpretation.

\begin{figure}
  \begin{center}
    \includegraphics[width=\textwidth]{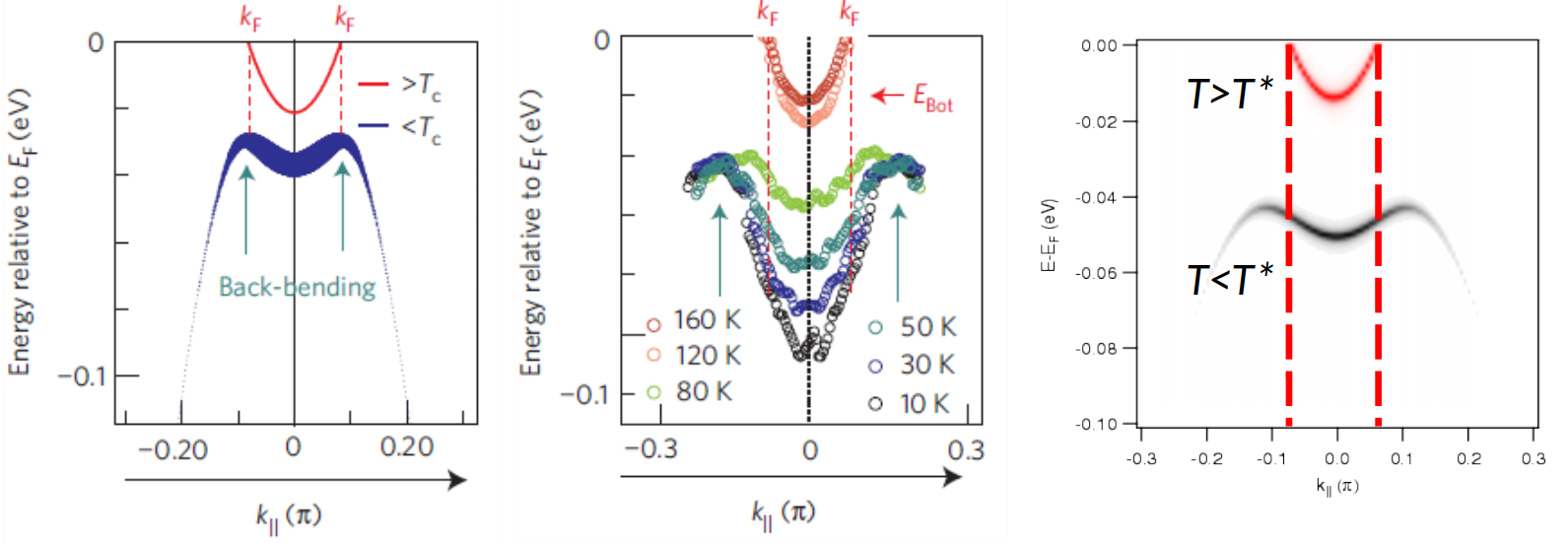}
  \end{center}
  \caption{(a) The simulated dispersion for d-wave homogeneous superconductivity with order parameter V = 30 meV. The quasiparticle energy for a given momentum state k is given by $E(k)=\sqrt{[\epsilon(k)]^2 + [\Delta(k)]^2}$. Cuts are along $(\pi,-\pi)-(\pi,0)-(\pi,\pi)$. The red (blue) curve is for the true normal (gapped) state. The spectral weight is indicated by the curve thickness. The back-bending (or saturation) of the dispersion and $k_F$ are indicated in the panel. Note that the back-bending momentum in the gapped state remains aligned with $k_F$. (b) Summary of the experimentally measured intensity maximum dispersions at different temperatures. (c) Simulation of the experimental dispersion with and without the pseudogap within the framework of the YRZ model with short-range spin correlations (Kidd et al. unpublished).}
  \label{arpes5}
\end{figure}

A related BSCCO ARPES study examined the development of the pseudogap in the antinodal region~\cite{hashimoto2010particlehole}. They reported a particle-hole asymmetry and attributed the momentum dependence of the gap to some form of symmetry-breaking charge order. In particular, they noted that the top and bottom of the gap were shifted from each other in momentum space, which is inconsistent with any form of pairing.  However, the same momentum dependence was subsequently shown to be potentially derived from short-range spin correlations through application of the YRZ model, as shown in figure~\ref{arpes5}~\cite{storey2015closing,kidd2018private}.

The discussion relating to experimental observations has thus far focused on the particle-hole asymmetry associated with the pseudogap. Several other studies have pointed to the possibility of a particle-hole symmetric gap with d-wave symmetry extending into the nodal gap.  This gap is often associated with preformed pairs and a temperature scale $T_\text{pair}$ for this pairing is often measured to be above the bulk superconducting temperature $T_c$, but below the pseudogap temperature $T^\ast$~\cite{kaminiski2015pairing,reber2015coordination}.  This same temperature scale $T_\text{pair}$ is usually coincident with a temperature defined by studies of the Nernst effect~\cite{xu2000vortexlike,li2010diamagnetism}. Zaki and coworkers \cite{zaki2017cuprate} have considered this phenomena and suggested that, in fact, the temperature scale $T_\text{pair}$ can be defined by local variations in the superconducting state associated with nanoscale inhomogeneities that are characteristic of BSCCO~\cite{gomes2007visualizing,pasupathy2008electronic}. This then provides a natural explanation of why ARPES often displays a gap in the spectral response in the antinodal region at doping levels above a critical doping level defined by $p=0.19$.

The fact that the two gaps, pseudogap and superconducting gap, can coexist suggests that one gap is not necessarily the precursor of the other. The pseudogap exists in the antinodal region, the superconducting gap has its initial development in the Fermi arc region. 

\subsection{Scanning tunnelling microscopy}
\label{sec:stm}

In scanning tunnelling microscopy (STM) a tip is scanned across a surface of interest and the current flowing to or from the tip is measured locally~\cite{binnig1982surface}. In its most straight forward application, such a study will provide a real space image of the local topography. However, it is a tunnelling spectroscopy and as such it is also providing information on the local electronic structure. Extracting this information is known as spectroscopic intensity scanning tunnelling microscopy (SISTM). It is a (conceptually) simple exercise to Fourier transform this real-space information, providing access to the electronic structure information in momentum space. This Fourier transform scanning tunnelling spectroscopy (FT-STS), often called quasiparticle interference spectroscopy, was first applied to metallic systems in the early 90s reflecting sharp edges and impurities, two-dimensional electron gas semiconductors, and absorbed impurities on conventional superconductors (see, e.g., \cite{simon2011fouriertransform} for a review). 

Quasiparticle interference spectroscopy was first applied to the cuprate problem by Davis and colleagues~\cite{hoffman2002imaging,mcelroy2003relating,wang2003quasiparticle}. It measures the local density of states as a function of energy at a given location $r$ in real space, LDOS$(E,r)$, which is related to the momentum space eigenstates $\Psi_k(r)$ through~\cite{hoffman2003search}
\begin{align}
  \text{LDOS}(E,r) \propto \sum_k |\Psi_k(r)|^2 \delta(E-\epsilon_k),
\end{align}
where $\epsilon_k$ is the underlying electronic dispersion. Impurity scattering mixes eigenstates with different momenta $k$ and the same energy $\epsilon_k$. Thus scattering that mixes states $k_1$ and $k_2$ results in a standing wave with wave function $\Psi_q$ having a wave vector $q=(k_1-k_2)/2$. The local density of states then represents an interference pattern with a wave vectors $q$ or wavelength $\lambda = 2\pi/q$. This is an important distinction between ARPES and SISTM, the former is a spectroscopy directly in momentum space $k$, while the latter is a spectroscopy reflecting scattering in momentum space given by $q$.

\begin{figure}
  \begin{center}
    \begin{tabular}{c}
      \includegraphics[width=0.475\textwidth]{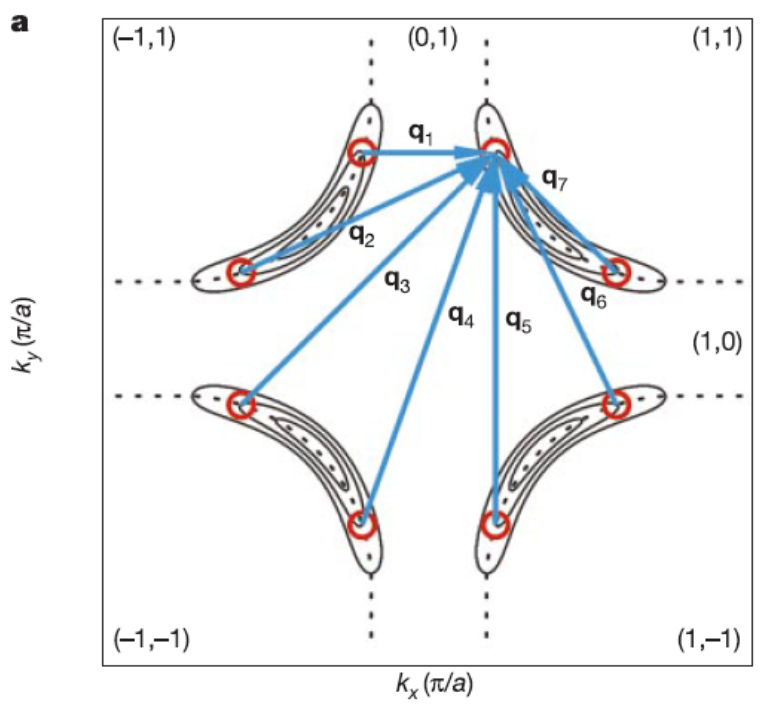} \\
       ~~\,\includegraphics[width=0.45\textwidth]{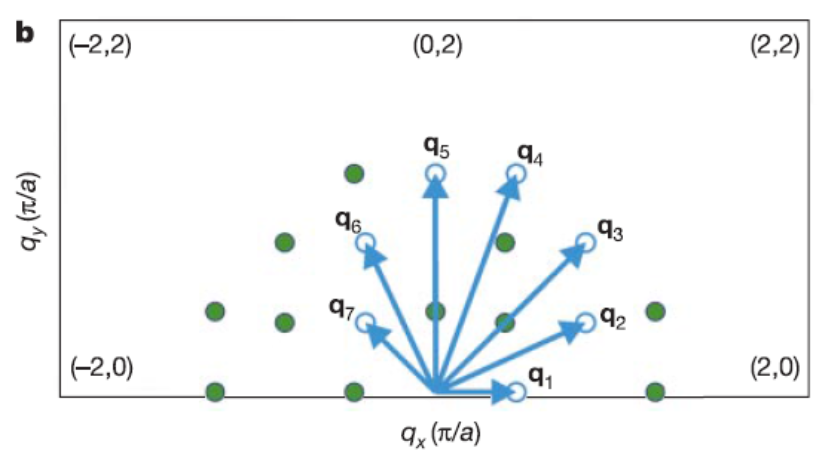}
    \end{tabular}
    \end{center}
    \caption{The quasiparticle interference patterns in a superconductor with the band structure of Bi2212 will be dominated by certain wave vectors. (a) The banana-shaped solid lines show the contours of constants quasi-particle energies, which increase in size with increasing energy. For a specific energy, the regions with highest density of states are shown by the octet of red circles. The seven dominant scattering ${\bf q}$-vectors, which connect the regions with highest density of states, are shown by the blue arrows. (b) Each ${\bf q}$-vector is shown as a blue arrow that originates from the origin, ending in an empty circle. All other inequivalent ${\bf q}$-vectors of the octet model (determined from the symmetry planes of the Brillouin zone) are denoted by filled circles. Thus it would be expected that Fourier transform scanning tunnelling spectroscopy can measure sixteen inequivalent maxima in the detectable region of ${\bf q}$-space. Figure reproduced from (McElroy et al. 2003).}
    \label{fig:stm1}
\end{figure}

\begin{figure}
  \begin{center}
    \includegraphics[width=0.475\textwidth]{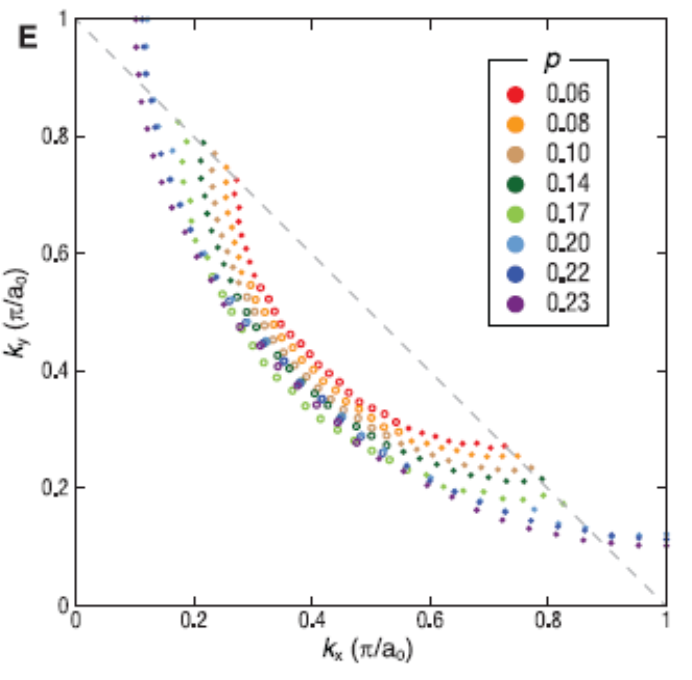}
  \end{center}
  \caption{The doping dependence of the Fermi surface topology as obtained from SISTM.  The arcs at underdoping terminate at the antiferromagnetic Brillouin zone boundary. The arcs transition to complete hole pockets surrounding $(\pm1,\pm1)\pi/a_0$ at $p \approx 0.19$. Figure reproduced from (Fujita, Kim, Lee, Lee, Hamidian, Firmo, Mukhopadhyay, Eisaki, Uchida, Lawler, Kim \& Davis 2014).}
  \label{fig:stm2}
\end{figure}

The relative contribution to the quantum interference pattern in $q$ space reflects a joint density of states. The most prominent features will reflect the largest density of states for that particular energy $E$. This density of states is given by
\begin{align}
  \text{DOS}(E) \propto \frac{1}{|\nabla_k E|}.
\end{align}
Thus $\text{DOS}(E)$ will be largest when $\nabla_k E$ is smallest. Consideration of the symmetry of the cuprates and the measured dispersion from ARPES studies led to the conclusion that the quantum interference pattern would be dominated by an octet of $q$-vectors as shown in figure~\ref{fig:stm1}~\cite{mcelroy2003relating}. Studies of this type resulted in a picture of the Fermi arcs that we have previously discussed in relationship to ARPES studies (see section~\ref{sec:arpes}). As figure~\ref{fig:stm2} shows, SISTM studies found that the Fermi arcs increased in length as a function of doping, up until some critical doping in the vicinity of $x = 0.19$, where the arcs switched to a full Fermi surface~\cite{fujita2014simultaneous}.

Aside from studies of the Fermi surface topology, SISTM has also been applied to studies of charge ordering. An early study identified a four unit cell periodic pattern of quasi-particle states surrounding vortex cores in Bi$_2$Sr$_2$CaCu$_2$O$_{8+\delta}$~\cite{hoffman2002four}. The states were found to exhibit a copper-oxygen bond-oriented ``checker board'' pattern, with four unit cell ($4a_0$) periodicity and a $\sim30$\,\AA$~$decay length.

\begin{figure}
  \begin{center}
    \includegraphics[width=0.5\textwidth]{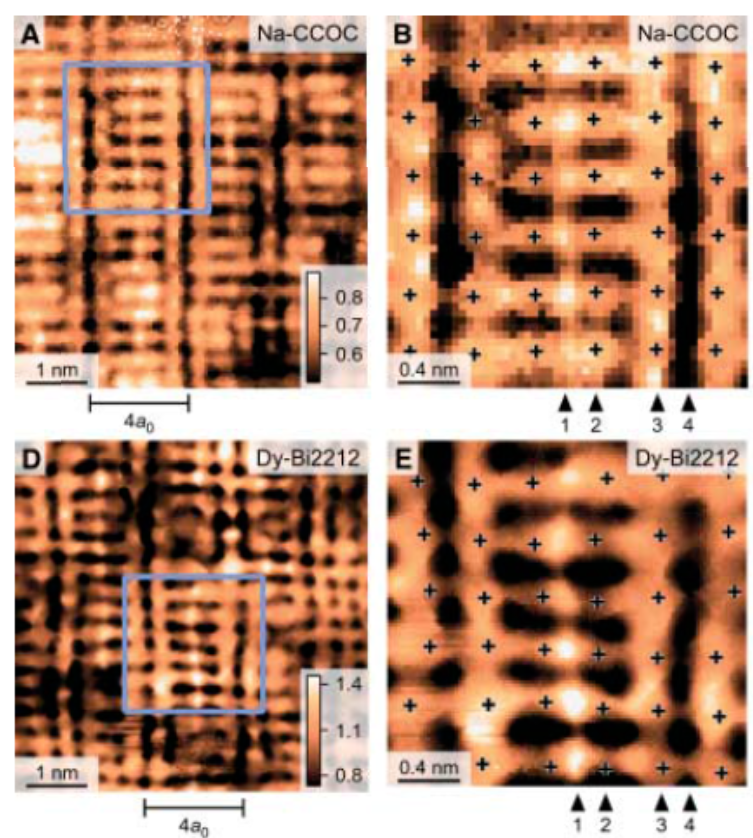}
  \end{center}
  \caption{(A,D) Maps of the tunnelling ratio $R(r,V)$ at $V=150$\,mV in Ca$_{1.88}$Na$_{0.12}$CuO$_2$Cl$_2$ (Na-CCOC) and Bi$_2$Sr$_2$Dy$_{0.2}$Ca$_{0.8}$Cu$_2$O$_{8+\delta}$ (Dy-Bi2212), respectively. The blue boxes in (A) and (D) are shown in higher resolution in (B) and (E), respectively, with the locations of the Cu atoms marked by black crosses. Figure adapted from (Kohsaka et al. 2007).}
  \label{fig:stm3}
\end{figure}

The recognition that in these complex systems there should be a detectable imbalance between the tunnelling rate for electron injection and extraction---a tunnelling asymmetry---led Kohsaka and colleagues to introduce the concept of atomic-resolution tunnelling-asymmetry imaging~\cite{kohsaka2007intrinsic}. As shown in figure~\ref{fig:stm3}, they were able to demonstrate the existence of a bond-centred electronic glass with unidirectional domains in the underdoped cuprates by plotting the ratio $R$ between the tunnelling into empty states and the tunnelling out of filled states:
\begin{align}
  R(r,V) = \frac{I(t,z,+V)}{I(r,z,-V)}. 
\end{align}
Here $z$ represents variations in local electron heterogeneity. The authors noted that the $R(r,V)$ plots shown in figure~\ref{fig:stm3} have the advantage of filtering out variations in intensity from ``unknown sources''.

Using the same site-specific $R(r,V)$-factor measurement (localised within a CuO$_2$ unit cell), and carrying out a phase sensitive Fourier transform analysis, Fujita and colleagues proposed a state with a d-density wave form factor in the underdoped cuprates~\cite{fujita2014direct}. In a later study, the same group argued that the energy scale associated with these modulations was in fact the characteristic pseudogap energy~\cite{hamidian2016atomicscale}. The d-density wave states suggested by Hamidian and coworkers were theoretically proposed by Chakravarty et al. in a Hartree-Fock analysis some years ago~\cite{chakravarty2001hidden}. They do not, however, show up in the more recent sophisticated numerical analyses of the two-dimensional Hubbard and $t$-$J$ models~\cite{corboz2014competing,zheng2017stripe}.

\begin{figure}
  \begin{center}
    \begin{tabular}{cc}
      \includegraphics[width=0.4\textwidth]{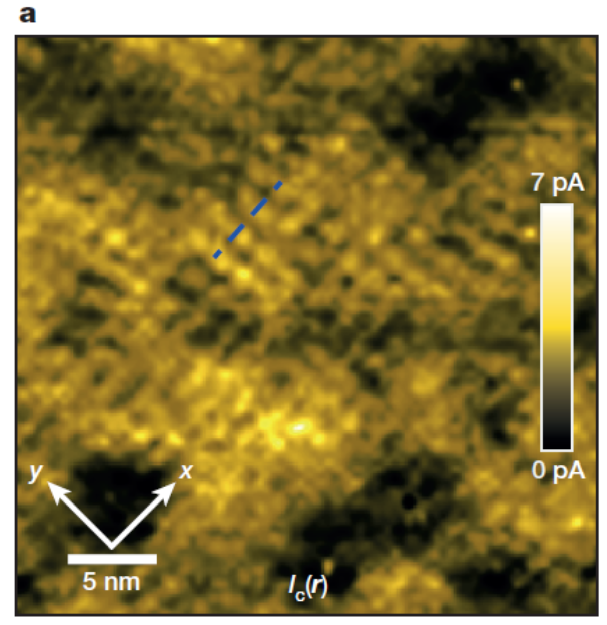}
      &
      \includegraphics[width=0.4\textwidth]{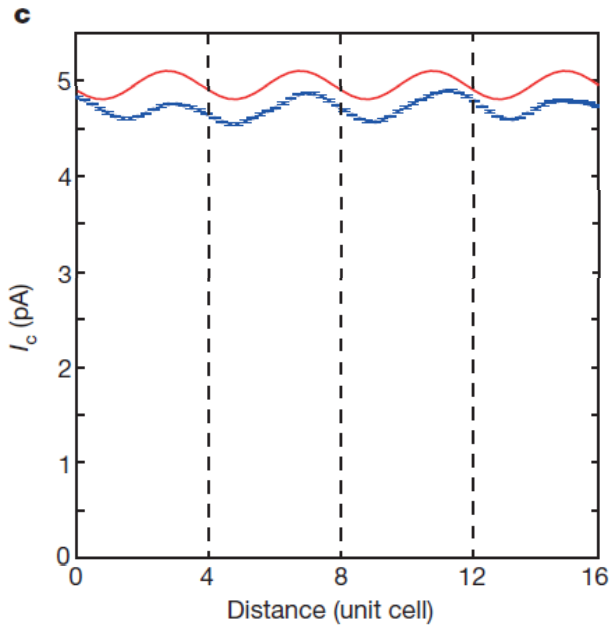}
    \end{tabular}
  \end{center}
  \caption{(a) A 35 nm $\times$ 35 nm field of view image showing the $I_c(r)$ modulations parallel to the CuO$_2$ $x$ and $y$ axes. (b) $I_c(r)$ along the dashed blue line in (a) (blue dotted line); the statistical error bars are the variance of $I_c(r)$ transverse to the dashed line. The red line shows the global amplitude and $Q_p$ of the $I_c(r)$ modulation obtained from the region around the maxima of the Fourier transform of $I_c(r)$. Figure reproduced from (Hamidian, Edkins, Joo, Kostin, Eisaki, Uchida, Lawler, Kim, Mackenzie, Fujita, Lee \& Davis 2016).}
  \label{fig:stm4}
\end{figure}

More recently activities in this area have turned to the search for pair density wave states, where the density of Cooper pairs modulates periodically in space with a wave vector $Q_p$, the finite centre of mass momentum. Such a periodic modulation is seen as a potential explanation of why intra-planar superconductivity is established at higher temperatures than inter-planar superconductivity. How might one see such a pair density wave? Two approaches have recently evolved. One possibility is through single particle scanning tunnelling microscopy whereby variations in the local energy gap $\Delta(r)$ are given by
\begin{align}
  \Delta(r) = \Delta_0 + \Delta_p \cos(Q\cdot r)
\end{align}
Here $\Delta_0$ is the gap associated with the homogeneous superconductivity and $\Delta_p \cos(Q\cdot r)$ is that associated with the PDW. A second possibility, developed recently, is that of Josephson tunnelling spectroscopy. Here a piece of the superconducting substrate is captured by the scanning tip, thereby allowing direct tunnelling of Cooper pairs between the sample and tip. The locally measured Josephson current, $I_j(r)$, will be given by
\begin{align}
I_j(r) = I_{j0} + I_{jp} \cos(Q\cdot r),
\end{align}
where $I_{j0}$ represents the homogeneous superconducting background and $I_{jp}\cos(Q\cdot r)$ reflects the contribution from pair density waves. Shown in figure~\ref{fig:stm4}, in an experiment of this type Hamidian and coworkers identified a modulation in the Cooper pair tunnelling with wavelength $4a_0$, the same periodicity as the charge density wave~\cite{hamidian2016detection}. In the figure, the maximum current is $I_c(r) \propto I_j(r)/2$. The small amplitude of the oscillations for the pair density wave, in light of previous STM works that observed rather large spatial anisotropy of the single-particle gap, requires further clarification.

\begin{figure}
  \begin{center}
    \includegraphics[width=0.45\textwidth]{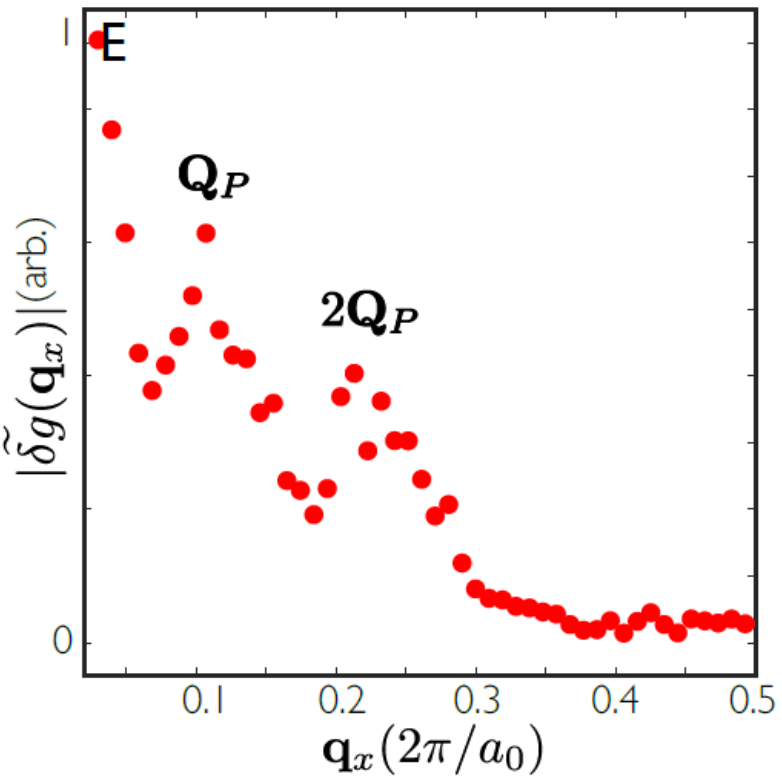}
  \end{center}
  \caption{Measured field-induced changes of the local density of states, $\tilde \delta g(q,E=30\,$ meV$)$. The field induced $N(q)$ modulations leads to clear peaks at $Q_p = 0.117 \pm 0.01$ and $2Q_p = 0.231 \pm 0.01$. Figure reproduced from (Edkins et al. 2019).}
  \label{fig:stm5}
\end{figure}

In a more recent study, Edkins and coworkers have reverted to the use of the single particle tunnelling microscopy but now in the presence of a strong (8.25\,T) magnetic field~\cite{edkins2019magnetic}. The key to this study is the recognition that within the vortex cores, reflecting the presence of the magnetic field, the mixing of superconductivity and the pair density wave state should result in modulations of the density of electronic states, $N(r,E)$, with wave vectors $Q_p$ and $2Q_p$. The actual observation is made by comparing the local measured density of states with and without the magnetic field to give the field induced changes, $\tilde \delta g(r,E,B)$. Figure~\ref{fig:stm5} shows a plot of the Fourier transform $\delta g(q)$  in momentum space of the field induced changes in real space, $\tilde\delta g(r)$, the latter being measured at an energy of $30$ meV.

Norman and Davis have taken the experimental observations from the PDW studies and calculated the associated Fermi surface~\cite{norman2018quantum}. They find that the calculated Fermi surface contains electron pockets that are consistent with quantum oscillation studies, and hole pockets in the single-particle Fermi surface that are similar to those of YRZ-type models. While such studies are encouraging, it seems reasonable to make the observation that the modulations in electron density with wave vector $2Q_p$, associated with the mixing of superconductivity and pair density waves have only ever been seen in the presence of a large magnetic field~\cite{wu2011magneticfieldinduced,wu2013emergence,wu2015incipient,gerber2015threedimensional,edkins2019magnetic}.

\subsection[Carrier density change in the Hall effect]{Carrier density change at the onset of the pseudogap observed in the Hall Effect}
\label{sec:carrierdens}

The Hall effect is a standard method to measure the density of carriers in a material~\cite{ashcroft1976solid}. The interpretation of Hall effect measurements in the cuprates, however, has several serious complications. First, the high temperature of the transition to superconductivity at $T_\text{c}$ does not allow measurements in the superconducting state at low magnetic field strengths and temperatures. A second complication is the appearance of static CDW states at magnetic fields $H > 17$\,T in NMR measurements in much of the pseudogap region~\cite{wu2011magneticfieldinduced,wu2013emergence}. At these high fields a reconstruction of the Fermi surface into small pockets is observed in quantum oscillation experiments~\cite{sebastian2011quantum,sebastian2012towards,vignolle2013from,sebastian2015quantum}. ARPES and related experiments at zero magnetic field find a Fermi surface broken up into four non-overlapping nodal arcs (see the previous two sections), which complicates the theory of possible closed magnetic orbits in the pseudogap phase~\cite{chakravarty2008fermi,elfimov2008theory,senthil2009synthesis,yao2011fermisurface,banerjee2013theory}. We shall not discuss this particular issue further here, and instead we refer the reader elsewhere for more details on this topic~\cite{sebastian2012towards}.

\begin{figure}
\begin{tabular}{ll}
(a) & (b) \\
\includegraphics[width=0.45\textwidth]{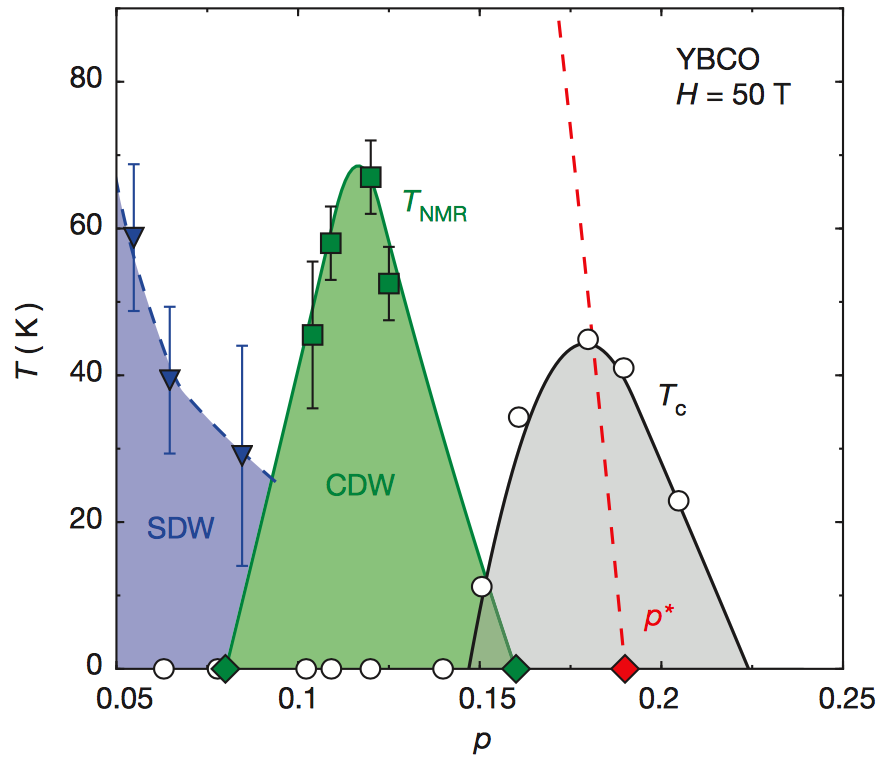} & 
\includegraphics[width=0.44\textwidth]{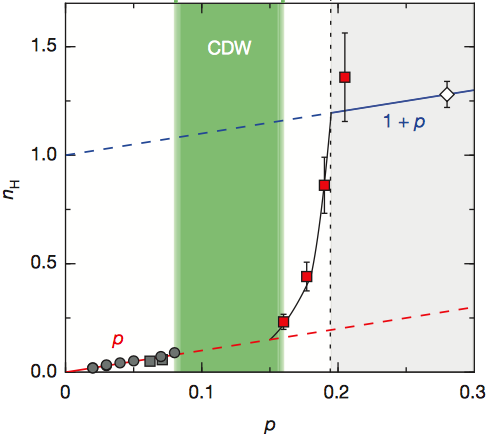}
\end{tabular}
\caption{(a) The high field $H=50$\,T phase diagram of YBa$_2$Cu$_3$O$_y$ as a function of hole doping $p$ and temperature $T$. Such a large field $H$ is applied to destroy superconductivity at low temperatures, and hence allowing the Hall effect down to be measured down to low temperatures. The low doping (blue shaded) region displays static spin density wave (SDW) order. At intermediate hole dopings (green shaded region), YBa$_2$Cu$_3$O$_y$ exhibits magnetic field induced charge density wave (CDW) order, as seen in NMR experiments. At large dopings (grey region) there is a uniform superconducting state. The red dashed line shows the pseudogap transition temperature $T^\ast(p)$. (b) The evolution of the Hall number, $n_\text{H} = 1/(eR_\text{H})$ with doping $p$ at $T=50$\,K in YBa$_2$Cu$_3$O$_y$ (red/grey squares). The green shaded region in (b) corresponds to that in (a) at $T=50$\,K. Figures adapted from (Badoux et al. 2016).}
\label{fig:badoux2016}
\end{figure}

In recent years Taillefer and coworkers have reported an interesting series of experiments on several cuprate materials~\cite{badoux2016change}. Here we focus our attention on the type B cuprate, YBa$_2$Cu$_3$O$_y$, whose high field phase diagram ($H =50$\,T) is presented in figure~\ref{fig:badoux2016}(a)~\footnote{Note that for the purposes of our discussions only $p>0.15$ of figure~\ref{fig:badoux2016}(a) is of interest.}. This phase diagram consists of three distinct regions. Firstly, at very low hole densities, $p < 0.1$, static SDW and CDW order are seen in scattering experiments. This region is outside the scope of this review and we will not discuss it further. The intermediate hole density region, $0.1 < p < 0.15$, displays long range CDW order that is also clearly observed in NMR at high field, $H = 50$\,T \cite{wu2011magneticfieldinduced,wu2013emergence}. The transition temperature for the magnetic field induced CDW drops to zero at $p =0.15$ and is replaced by uniform superconductivity in the reduced hole density range  $0.15 < p < 0.2125$. This range spans the onset of the pseudogap at $p^\ast = 0.19 \pm 0.01$, with the pseudogap transition temperature $T^\ast(p)$ represented by the red dashed line. 

The Hall number $n_\text{H}$ serves as a direct probe of the carrier density, and is defined as 
\begin{equation}
n_\text{H} = \frac{1}{eR_\text{H}},
\label{nH}
\end{equation} 
where $e$ is the electron charge and $R_\text{H}$ is the Hall coefficient, which may be measured directly from the high-field Hall resistivity $\rho_{xy}(H) = R_\text{H} H$. Figure~\ref{fig:badoux2016}(b) shows the evolution of the Hall number with hole density $p$ in the normal state at $T = 50$\,K, with data for YBa$_2$Cu$_3$O$_{7-x}$ shown as squares (see figure~4b of~\cite{badoux2016change} for further details). This shows a continuous drop from a value of $n_\text{H}= 1+p$ at $p = 0.2$ to the much smaller $n_\text{H} = p$ at $p < 0.15$. The former value of the Hall number is the hole density at overdoping from the full band structure Fermi surface. The latter is the Mott value which is reduced by the commensurate change of one carrier per site~\footnote{We note that a similar drop in $n_\text{H}$ with decreasing doping has been observed in a number of other cuprate materials, see for example~\cite{collignon2017fermisurface,laliberte2016origin}.}. Such a change in the Hall number would occur if the Brillouin zone was exactly halved, as in the case of long range commensurate $(\pi,\pi)$ antiferromagnetic order. However, long range commensurate antiferromagnetic order is ruled out by NMR~\cite{wu2011magneticfieldinduced,wu2013emergence} and other experiments~\cite{haug2009magneticfieldenhanced,haug2010neutron}. 

Storey calculated the evolution of the Hall number at the onset of the pseudogap~\cite{storey2016hall} using the YRZ phenomenological ansatz~\cite{yang2006phenomenological,rice2012phenomenological}. This theory is based on umklapp scattering arising from short range commensurate antiferromagnetic order in the pseudogap phase (see section~\ref{sec:yrzprop}). By introducing a continuous increase in the gap parameter with decreasing hole density, additional small electron pockets appear in the YRZ picture near the antinodal regions at intermediate densities. These small electron pockets improve the fit to the data as shown in~\cite{storey2016hall} figure~4b. However, such a detailed interpretation must be treated with some caution. Nevertheless, the conclusion that short range commensurate $(\pi,\pi)$ antiferromagnetic correlations appear at the entrance to the pseudogap state is clearly supported by the Hall effect data reported in~\cite{badoux2016change}.

We shall return to the Hall effect later in the transport section and examine the surprising experimental results which show that it also changes as the temperature is raised through the pseudogap temperature $T^\ast$.

\subsection{Antiferromagnetic correlations in the pseudogap phase}
\label{sec:afmcorr}

Let us now elaborate further on the antiferromagnetic correlations in the pseudogap phase of the type B cuprates. As we saw in the previous section, the behaviour of the Hall effect seen in \cite{badoux2016change} can, possibly, be explained via commensurate antiferromagnetic $(\pi,\pi)$ order that leads to a restructuring of the Fermi surface in the pseudogap phase. This is a completely different scenario to that we discussed in the theory section, where the pseudogap arises from the short range order of two-leg ladders generalised to two spatial dimensions. We will now briefly discuss the difference between these two scenarios, and supporting experimental evidence, in more detail. 

In the theoretical picture of type B cuprates, at half-filling the ground state is an isolated singlet and the lowest energy magnetic excitation is the gapped (finite energy) $S=1$ state. This excitation is centred on the wave vector $(\pi,\pi)$, which is associated with short range antiferromagnetic correlations. Using Ossadnik's approximation~\cite{ossadnik2016wave} to describe a two-dimensional material via a superlattice of supercells, each with a $S=1$ excitation about $(\pi,\pi)$, we realise a clear difference between the two types of cuprates. In type A cuprates there is a \textit{gapless} $S=1$ excitation due to long range antiferromagnetic order, while in type B cuprates the corresponding $S=1$ excitation has a finite energy gap. This important difference between the type A and type B cuprates shows up clearly in experiment. 

\begin{figure}
\begin{center}
\includegraphics[width=0.9\textwidth]{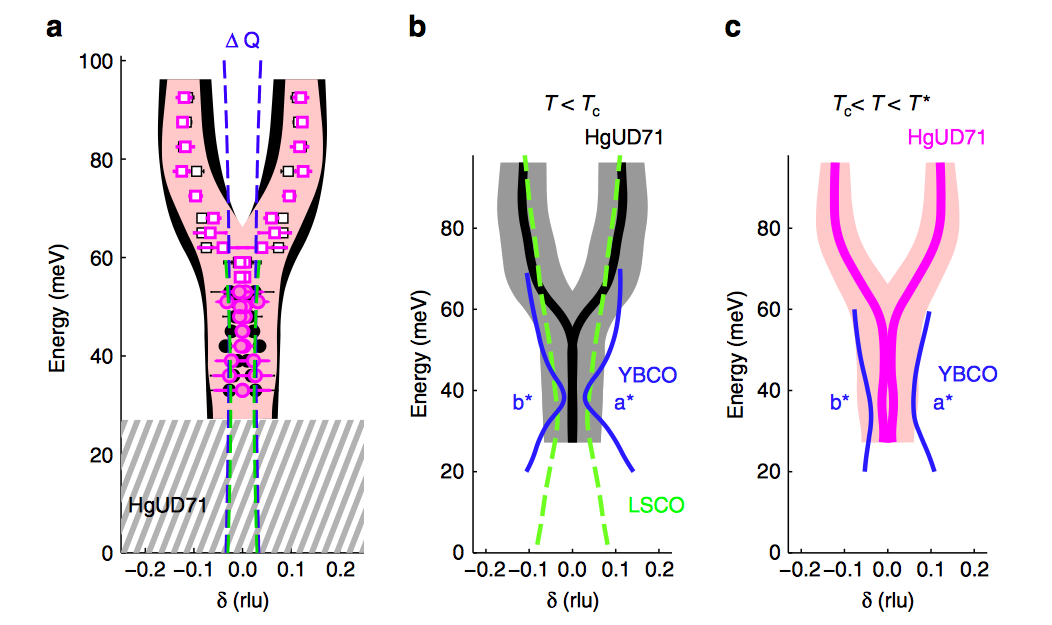}
\end{center}
\vspace{-4mm}
\caption{A comparison of the Y-shaped magnon spectrum as measured in underdoped Hg1201 with $T_\text{c} = 71$\,K (HgUD71) and that in YBa$_2$Cu$_3$O$_{6.6}$ (YBCO6.6) and LSCO at similar dopings. (a) The magnon spectrum as a function of wave vector $\delta$ measured relative to $(\pi,\pi)$ at $T=5$\,K (black points) and $T=85$\,K (magenta points). Green and blue dashed lines ($\Delta Q$) denote the full width half maximum (FWHM) experimental resolution. The shaded black and pink regions show the FWHM (in the momentum direction) of the magnon response. The shaded grey region shows the energy gap, where no magnon response is seen. (b) Comparison of dispersion of the magnon in HgUD71 with YBCO6.6 and LSCO deep in the superconducting state. (c) Comparison of the dispersion of the magnon in HgUD71 with YBCO6.6 above $T_\text{c}$. YBCO6.6 is orthorhombic, leading to anisotropic response -- shown is both $\delta$ along $a^\ast$ (right) and $b^\ast$ (left) directions. As in (a), the shaded regions in both (b) and (c) denote the FWHM of the magnon line in HgUD71. Figure reproduced from (Chan et al. 2016).}
\label{fig:chan2016}
\end{figure}

\cite{chan2016commensurate} reported a detailed set of neutron scattering experiments on the commensurate antiferromagnetic $(\pi,\pi)$ spin excitations in an underdoped type B cuprate, Hg1201, with $T_\text{c} = 71$\,K and pseudogap onset temperature $T^\ast = 305$\,K. They found a set of magnons with a Y-shape, centred at $(\pi,\pi)$, with an energy gap of 25\,meV. Although the wave vector $(\pi,\pi)$ is constant along the pseudogap region of the Fermi surface, the magnitude of the energy gap varies along this surface, due to changes in the parameters (such as the effective mass, etc.) of the supercells. This leads to good qualitative agreement with the theory for type B cuprates. In figure~\ref{fig:chan2016} we reproduce a figure from~\cite{chan2016commensurate} that shows a comparison between their results for the magnon spectrum in Hg1201 and earlier results on another type B cuprate, YBa$_2$Cu$_3$O$_{6.6}$ ($T_\text{c} = 61$\,K) from~\cite{hinkov2007spin}. We see that the results are in general agreement at high energies. The situation at low energies is more confused, but the magnetic excitations remain gapped in both materials in contrast to type A cuprates. To demonstrate this we also show results from a similar experiment on an underdoped type A cuprate, LSCO, which shows the standard hourglass spectrum with legs extending down to zero energy. The recent extensive NMR investigations of~\cite{imai2017revisiting} on LSCO found both CDW and SDW order, with strong local disorder. We see that the expected differences in the magnetic spectra between type A and type B cuprates show up clearly in the discussed experiments. 

We note that recent theory work using dual fermion methods, cluster dynamical mean field theory, and quantum Monte Carlo have also see the gapped Y-shaped magnon spectrum~\cite{leblanc2019magnetic}. With the agreement between the analytical approaches of section~\ref{sec:approaches} and cluster dynamical mean field theory for transport properties, it is worth examining the magnetic properties of the analytical models in future work.

Hg1201 and YBa$_2$Cu$_3$O$_{6.6}$ are the cleanest underdoped cuprates, as attested by observations of quasiparticle orbits through the de Haas-van Alphen effect~\cite{sebastian2011quantum,basiric2013universal}. The major difference between the two materials is the number of copper layers per unit cell: Hg1201 has a single layer, whilst YBa$_2$Cu$_3$O$_{6.6}$ has two. The neutron scattering experiments of~\cite{chan2016commensurate} on Hg1201 did not observe a substantial difference between the dynamical magnetic susceptibility above and below $T_c$. On the other hand, measurements on  YBa$_2$Cu$_3$O$_{6.6}$ show rather profound differences. In general, one can expect such differences to occur on the grounds of change in the contribution of quasiparticles to the magnetic response function in the superconducting state.

At high energies the dynamical susceptibility can be well approximated by a damped harmonic oscillator: 
\begin{equation}
\chi^{(R)}({\bi q},\omega) = \frac{A}{(\omega + \rmi\Gamma)^2 - \Omega_{\bi q}^2},
\end{equation}
with constant damping $\Gamma$. As can be seen from figure~3(c) of \cite{hinkov2007spin} and figure~2 of~\cite{chan2016commensurate}, the dispersion $\Omega_q$ is rather anisotropic. This is consistent with the hypothesis that this collective mode is formed from the electronic states located on small portions of the underlying Fermi surface.

\subsection{Optical properties of the pseudogap phase}
\label{sec:optical}
The layered nature of the cuprates leads to important differences between the optical response to light polarised parallel and perpendicular to the CuO$_2$ layers and we discuss these two cases separately. We also concentrate our discussion to specific recent data sets which illustrate key results and only reference more comprehensive reviews~\cite{timusk1999pseudogap,basov2005electrodynamics,tajima2016optical}.

\subsubsection{In-layer optical conductivity.}

\begin{figure}
\begin{center}
\includegraphics[width=0.55\textwidth]{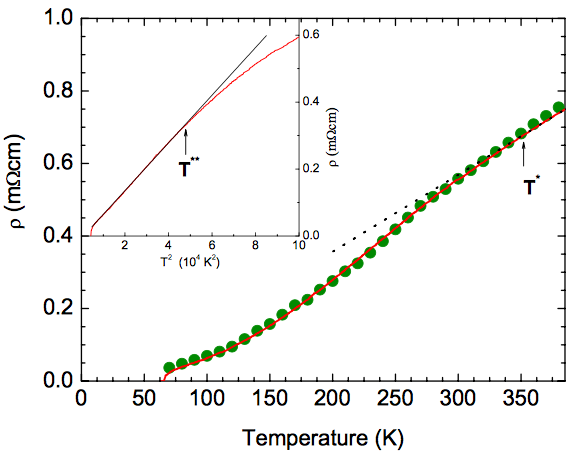}
\end{center}
\vspace{-3mm}
\caption{The d.c. resistivity, $\rho(T)$, of HgBa$_2$CuO$_{4+\delta}$ (Hg1201, $T_\text{c} = 67$\,K) obtained by extrapolation of the inverse optical conductivity (green points) as compared to transport measurements (red line). The dashed black line shows a linear fit. (Inset) The same data as the main figure plotted as a function of $T^2$, emphasising the low temperature $T_\text{c} < T \ll T^\ast$ behaviour. Figures adapted from (Mirzaei et al. 2013).}
\label{fig:mirzaei2013}
\end{figure}

We begin by considering the optical response parallel to the CuO$_2$ layers. An ideal material for examining this is the single layer underdoped cuprate Hg1201, with $T_\text{c}=67$\,K and a hole density estimated to be $x=0.1$, as studied in~\cite{mirzaei2013spectroscopic}. The d.c. resistivity $\rho(T)$, as extrapolated from the inverse optical conductivity, is in excellent agreement with transport measurements and is shown in figure~\ref{fig:mirzaei2013}. Above $T_\text{c}$, the resistivity shows first a $T^2$ form (see the inset of figure~\ref{fig:mirzaei2013}), which holds up until $T = 220$\,K, and then crosses over to a linear form above $T^\ast = 350$\,K. These transport properties will be discussed further in section~\ref{sec:transport}.

\begin{figure}
\begin{center}
\includegraphics[width=0.75\textwidth]{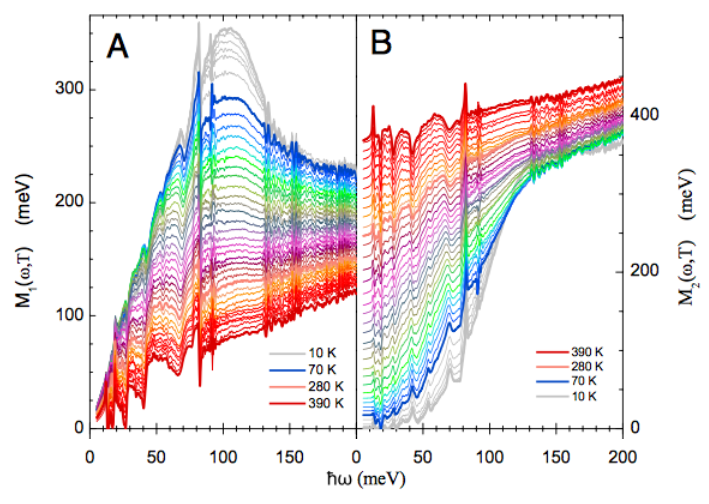}
\end{center}
\vspace{-5mm}
\caption{The (A) real and (B) imaginary parts of the memory function $M(\omega,T)$ in HgBa$_2$CuO$_{4+\delta}$ (Hg1201, $T_\text{c} = 67$\,K). Results are shown in 10\,K steps from $T=10$\, K (grey) to $T = 390$\,K (red). The memory function $M(\omega,T)$ is related to the related to the in-layer optical conductivity through~\eqref{opticalcond}. Figure adapted from (Mirzaei et al. 2013).}
\label{fig:mirzaei2013fig4}
\end{figure}

The in-layer optical conductivity $G(\omega,T)$ can be characterised in terms of the ``memory function'' $M(\omega,T) = M_1(\omega,T)+\rmi M_2(\omega,T)$ through
\begin{equation}
G(\omega,T) = \frac{\rmi \pi K}{\hbar \omega + M(\omega,T)} G_0, 
\label{opticalcond}
\end{equation}
where $K$ is the spectral weight and $G_0 = 2e^2/h$ is the conductance quantum. The memory function is shown in figure~\ref{fig:mirzaei2013fig4}; at high temperatures $T > T^\ast$, the optical conductivity shows typical Drude behaviour, while below $T^\ast$ the weight from high energies is transferred to a low frequency peak due to the onset of the pseudogap. Below $T_\text{c}$ a second transfer of weight occurs, into a $\delta(\omega=0)$ feature in the superconducting state. This is reflected in the fact that below $T_\text{c}$ both the real and imaginary parts of the memory function become very small as $\omega \to 0$. This illustrates how the low-temperature optical conductivity is governed by superconductivity confined to the nodal pockets. 

At temperature above the formation of the pseudogap, $T > T^\ast$, the antinodal regions contribute to the low frequency conductivity in accordance the theoretical analysis discussed in sections~\ref{sec:yrzprop}--\ref{sec:twofluid}. \cite{mirzaei2013spectroscopic} also reported the total weight of the low frequency coherent conductivity, $K^\ast$, for underdoped samples and found that this scales proportional to the doping. This behaviour agrees with the assignment of the low frequency weight to the nodal pockets. 

\subsubsection{c-axis optical conductivity.}
\label{sec:caxiscond}

\begin{figure}
\begin{center}
\begin{tabular}{ll}
\includegraphics[width=0.45\textwidth]{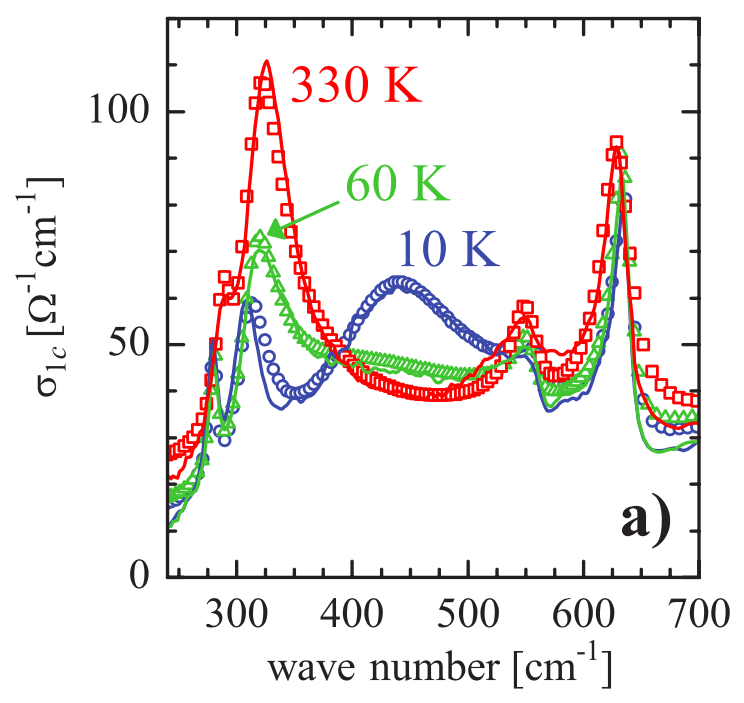} &
\includegraphics[width=0.5\textwidth]{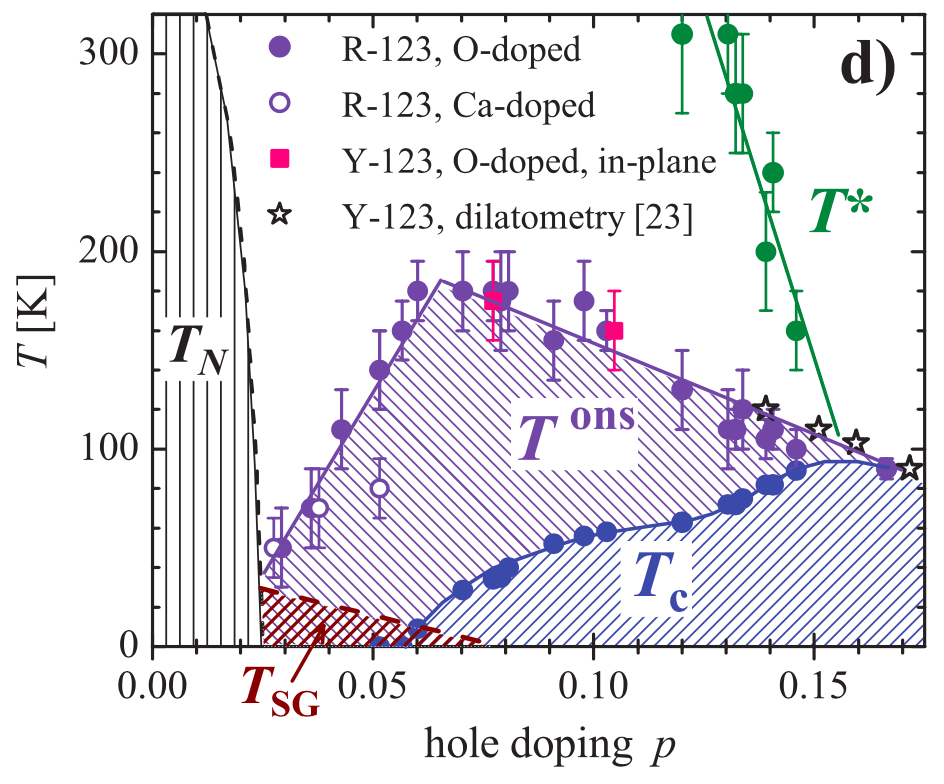}
\end{tabular}
\end{center}
\caption{(Left) The $c$-axis response of YBa$_2$Cu$_3$O$_{7-x}$ with $T_\text{c} = 58$\,K, showing the real part of the optical  conductivity $\sigma_{1c}$. The measurements are shown as solid lines, whilst symbols show a multilayer-model analysis.  (Right) The resulting doping phase diagram of $T^\ast$, $T^\text{ons}$, and $T_\text{c}$. Figures reproduced from~(Dubroka et al. 2011).}
\label{fig:dubroka}
\end{figure}

The behaviour of the optical response perpendicular to the superconducting layers is different, with the $c$-axis optical conductivity $\s_c(\omega,T)$ being much smaller than the in-layer conductivity. It also contains response from a series of infrared optical phonons, as well as the electronic transitions. Dubroka and collaborators studied the infrared response as a function of temperature in a series of YBa$_2$Cu$_3$O$_{7-x}$ samples within the pseudogap phase \cite{dubroka2011evidence}. There, the onset of the pseudogap is also associated with a transfer of weight in the $c$-axis conductivity from low energies to high energies, see figure~\ref{fig:dubroka}(a). The temperature at which this weight transfer starts was used by~\cite{dubroka2011evidence} to study the evolution of $T^\ast$ upon change of the doping $x$, as shown in figure~\ref{fig:dubroka}(b).  

The pseudogap temperature, $T^\ast(x)$, rises quite rapidly as the doping $x$ decreases. This is presumably due to a substantial increase in the strength of the renormalised low-energy Coulomb repulsion, as would be expected with screening from particle-hole transitions becoming weaker with decreasing $x$. Note, however, that the superconducting transition temperature $T_\text{c}$ decreases with doping, as expected from the shrinking of the nodal pockets. 

The data of~\cite{dubroka2011evidence} also identifies another temperature scale, denoted $T^\text{ons}(x)$ (see figure~\ref{fig:dubroka}(b)), which initially increases slowly at the onset of the pseudogap and goes through a maximum at $x=0.07$. This temperature is defined via the appearance of an additional peak in the midst of the infrared $c$-axis phonon spectra. It persists into the superconducting state \cite{homes1993optical,schutzmann1995doping,berhard2000farinfrared}, and has been interpreted as a signature of a transverse plasmon mode, arising from the intra-bilayer transfer of Cooper pairs in the superconducting state. Such a state was initially proposed for bilayer superconductors by van der Marel and Tsvetkov~\cite{vandermarel1996transverse}, and was later used explain the appearance of a new mode in underdoped YBa$_2$Cu$_3$O$_{7-x}$~\cite{munzar1999new,gruninger2000midinfrared}. Note, however, that $T^\text{ons}(x)$ is much higher than $T_\text{c}(x)$, for example $T^\text{ons}(x=0.1) = 150$\,K while $T_\text{c}(x=0.1) = 50$\,K! Such an extended range of superconducting fluctuations above the transition temperature $T_\text{c}$ is exceptional and not observed in the low frequency in-layer conductivity. The interpretation of this behaviour shall be discussed in the next section, when we cover the giant phonon anomalies observed in YBa$_2$Cu$_3$O$_{7-x}$ samples at the same hole density~\cite{letacon2013inelastic}.  

\subsection{Inelastic and resonant inelastic X-ray scattering}
\label{sec:xray}

Let us move away from optical measurements, and consider higher energy probes. Two types of X-ray scattering experiments have been performed on underdoped cuprates, resonant inelastic X-ray scattering and inelastic X-ray scattering. An important differences between these two experiments is the energy resolution -- this is poor for resonant inelastic X-ray scattering and much better in inelastic X-ray scattering experiments. In the former, scattering response is integrated over a range of energies or order 10 meV or more, whereas in the latter experiments energy resolution of order 1 meV or better can be achieved. In the earlier review by Keimer \textit{et al.} \cite{keimer2015quantum} resonant inelastic X-ray scattering experiments were interpreted as evidence for charge order, which is competing with the superconductivity. For a detailed review of resonant X-ray scattering in the cuprates, we refer the reader to~\cite{comin2016resonant}.

\subsubsection{Giant phonon anomalies in type B cuprates.}

\begin{figure}
\begin{center}
\includegraphics[width=0.9\textwidth]{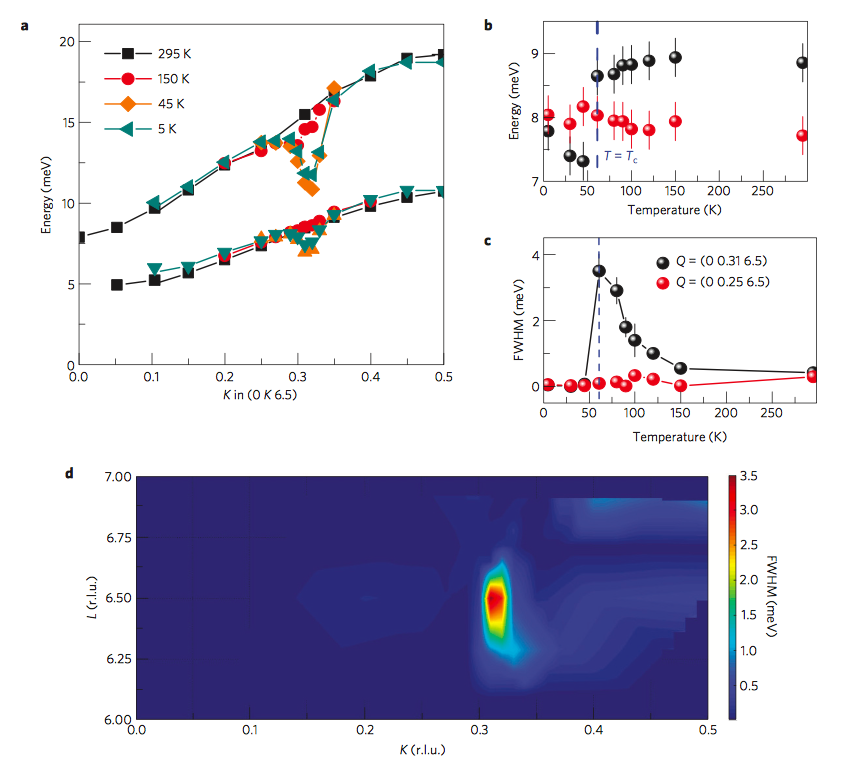}
\end{center}
\vspace{-7mm}
\caption{The temperature dependence of the phonon dispersion. (a) The dispersion of two of the low-energy phonons along the (0 K 6.5) direction for four values of the temperature. A clear feature is observed centred on $K \sim 0.31$. The temperature dependence of the (a) energy of the acoustic phonon and (b) the full-width half-maximum of the phonon line for $Q$ = (0 0.25 6.5) (red dots) and $Q$ = (0 0.31 6.5) (black dots). Notice the broadening of the $Q =$ (0 0.31 6.5) phonon line, which occurs at the same momentum as the feature in (a). Error bars show the uncertainty of the fit. (d) The momentum dependence of the full-width half-maximum of the acoustic phonon at $T=T_\text{c}$. Figure reproduced from (Le~Tacon et al. 2013).}
\label{fig:letacon}
\end{figure}

Le Tacon and coworkers~\cite{letacon2013inelastic} studied the phonon spectra of YBa$_2$Cu$_3$O$_{6.6}$ using high resolution inelastic X-ray scattering. They found an unusual set of giant phonon anomalies that appeared as the temperature was lowered below $T = 150$\,K and which persisted down to the superconducting transition at $T_\text{c} =55$\,K. Giant phonon anomalies are an enhancement of the phonon damping at a certain position in $\bi{k}$-space. In the case at hand, the giant phonon anomaly is localised at $Q=$(0 0.31 6.5), and becomes stronger at the temperature is lowered towards the superconducting transition temperature, as shown in figure~\ref{fig:letacon}. The onset of long-range superconducting order at $T_\text{c} = 55$\,K leads to an abrupt change in the nature of the anomaly, changing from an increase in damping to a dip in the phonon energy (see figure~\ref{fig:letacon}). It is also worth noting that, as discussed in section~\ref{sec:caxiscond}, YBCO samples with the same transition temperature $T_\text{c} = 55$\,K display a transverse plasmon mode in the $c$-axis optical conductivity as the temperature is lowered through the same value of the temperature, $T = 150$\,K. 

\begin{figure}
\begin{center}
\begin{tabular}{ll}
(a) & (b) \\
\includegraphics[width=0.33\textwidth]{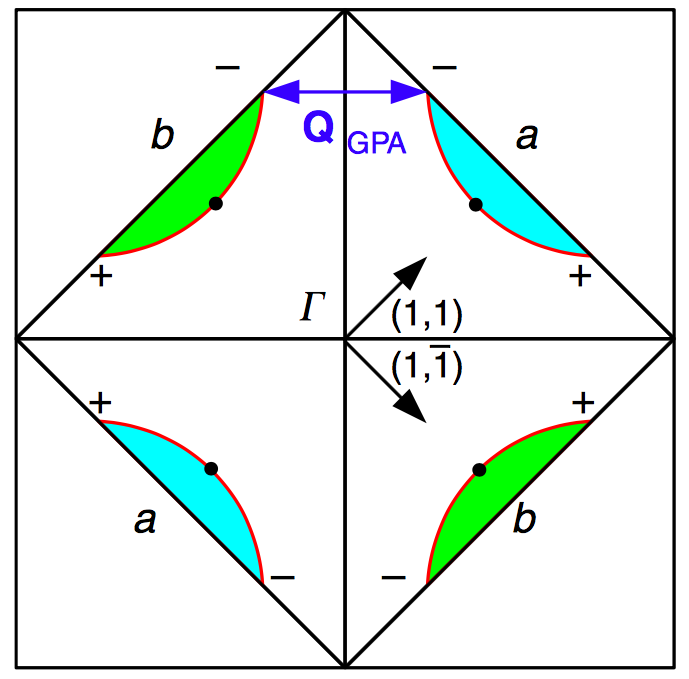} &
\includegraphics[width=0.45\textwidth]{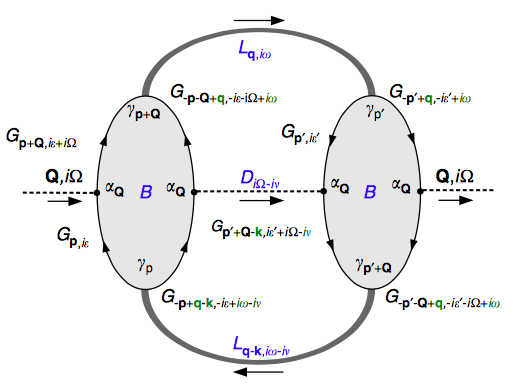}
\end{tabular}
\end{center}
\vspace{-0.3cm}
\caption{(a) The Fermi arcs in the Brillouin zone are separated into sub-bands, labelled ${\bi a}$ and ${\bi b}$, in the $(1,1)$ and $(1,\bar 1)$ directions. Labelled is $Q_\text{GPA}$, the wave vector of the giant phonon anomaly, which connects the two sub-bands. (b) The Feynman diagram depicting the phonon self-energy. Phonon propagators (D, dashed lines), Leggett mode propagators (L, thick grey lines) are coupled by the effective interaction vertex (B), which is an electronic bubble (electron Green's function, G). The phonon is forward scattered by absorption/emission of a Leggett mode. Adapted from (Liu et al. 2016) figures~1 and 2. }
\label{fig:liu2016}
\end{figure}

The intimate relationship between these anomalies and the presence of pairing fluctuations led Liu and collaborators to speculate that these results might be interpreted via a coupling of phonons with momentum $Q$ with pairing fluctuations, as shown in figure~\ref{fig:liu2016}. Note that the observed $Q$ vector of the giant phonon anomalies connects the ends of the superconducting arcs in the $(1,1)$ and $(1,\bar 1)$ directions, see figure~\ref{fig:liu2016}. The d-wave characteristic of pairing in the cuprates leads to a sign change in the phase of pairs in these two nodal directions, which Liu et al argued frustrated the coupling between Cooper pairs in the two nodal directions. This helps to explain the unusually large temperature range over which pairing fluctuations are observed in these materials, up to temperatures $\sim 3T_\text{c}$. Liu and coworkers also argued for the existence of Leggett modes~\cite{leggett1966numberphase} due to this frustrated weak coupling between the two sets of Cooper pair arcs. Their proposal radically differs from the usual Kohn anomalies~\cite{kohn1959image} that precede standard charge density wave instabilities, and offers an alternative explanation of the observed anomalies (which are observed with X-ray scattering that integrates response over a wide energy range due to the low energy resolution). As we noted earlier, NMR/NQR spectra in underdoped YBCO do not show evidence of static charge order unless large magnetic fields are applied~\cite{wu2011magneticfieldinduced,wu2013emergence}. For further details of the Leggett mode proposal and their coupling to phonons, we refer the reader to~\cite{liu2016giant}.

\subsubsection[High resolution inelastic X-ray scattering.]{High resolution inelastic X-ray scattering on type A cuprates and comparison to type B cuprates.}
Recently high resolution inelastic X-ray scattering measurements have been reported for the type A cuprate LBCO at $x=1/8$~\cite{miao2018incommensurate}. Measurements on LBCO show similar behaviour to the type B cuprates at room temperature, but evolve differently as the temperature is lowered. The ground state of type A cuprates possesses complex intertwined order, with hole stripes of wave vector $(\pm 0.25\times 2\pi/a_0,0)$, SDW antiferromagnetic order with wave vector $(\pm 0.125 \times 2\pi/a_0,0)$, and PDW order with the same wave vector as the SDWs. Measurements by~\cite{miao2018incommensurate} showed that the CDW order associated with the hole stripes is influenced by the onset of the SDW as the temperature is lowered towards $T = 55$\,K. Below this temperature the patterns of the CDW and SDW are locked together, with the unit cell of the CDW half that of the SDW. This modification of the wave vector of the CDW, together with the onset of SDW correlations as the temperature is lowered, is a clear difference between type A and type B cuprates in the low temperature limit. On the other hand, these inelastic X-ray scattering experiments point towards \textit{universal behaviour} at room temperature, with the split into two types emerging only upon cooling. We shall return to this point later, in section~\ref{sec:open}, with the discussion of open issues. 

\subsubsection[Short-range CDW correlations in YBCO and Hg1201.]{Short-range CDW correlations in YBCO and Hg1201 as observed in X-ray scattering.}
X-ray scattering has also played a pivotal role, over the last six or so years, in works that have shown CDW \textit{correlations} are ubiquitous in the cuprates, being found in all major families including those of type B, such as YBCO and Hg1201~\cite{ghiringhelli2012longrange,chang2012direct,achkar2012distinct,tabis2014charge,comin2015broken,campi2015inhomogeneity,forgan2015microscopic,tabis2017synchrotron}. This often leads to the statement that CDW \textit{order} is present in all cuprates. It is, however, worth emphasising that in the type B cuprates the observed CDW \textit{correlations} are short-ranged in the absence of a strong magnetic field, and hence do not correspond to what one would traditionally call (long-range) order. This leads to some confusion in the literature, and it is worth stating that these X-ray scattering experiments are consistent with NMR/NQR, which shows an absence of order, as discussed in section~\ref{sec:nmr}. NMR/NQR does not observe splitting of peaks associated with the presence of static charge order, evidencing that long-range charge order is absent in the type B cuprates. 

\subsection{Transport properties}
\label{sec:transport}

Let us now return to transport properties (briefly mentioned in section~\ref{sec:carrierdens}). It was recently emphasised by Anderson that the temperature dependence of the transport properties of the cuprates was one of the earliest signs of unusual physics, and is one of the quintessential features that theory must capture~\cite{anderson2018four}. The d.c. resistivity as a function of temperature is linear in $T$ at high temperatures, crossing over to $T^2$ within the pseudogap phase. On its own this is already surprising, with the opening of the pseudogap leading to a drop in carrier density, but an \textit{increase} in conductivity! The situation only gets stranger when considering Hall transport, where there are no signs of the anomalous linear-in-$T$ transport and instead behaviour consistent with Fermi liquid theory is observed. We now discuss each of these surprising results in more detail.

\subsubsection{In-plane d.c. resistivity as a function of temperature.}
\label{sec:linear}

\begin{figure}
\begin{center}
\includegraphics[width=0.6\textwidth]{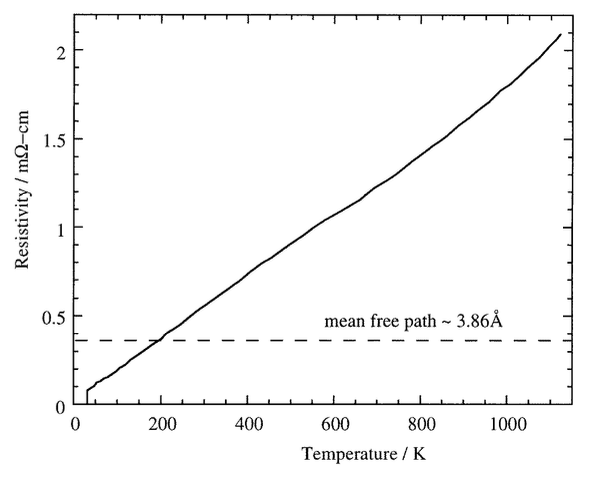}
\end{center}
\vspace{-5mm}
\caption{The temperature dependence of the d.c. resistivity in the type A cuprate, La$_{1.85}$Sr$_{0.15}$CuO$_{4-\delta}$, showing linear-in-$T$ behaviour to well beyond the Mott-Ioffe-Regel, denoted by the dashed line. Figure reproduced from (Liang 1998).}
\label{fig:liang}
\end{figure}

Perhaps the best known anomalous property of the normal states is the unusual temperature dependence of the in-plane d.c. resistivity~\cite{gurvitch1987resistivity,orenstein1992frequency,takagi1992systematic,mandrus1992resitivity,wang1996observation,watanabe1997anisotropic,ando2001mobility,calandra2003violation,ando2004electronic,daou2009linear,barisic2013universal,proust2016fermi}. At temperatures above the formation of the pseudogap, $T>T^\ast$, it is linear in temperature $\rho(T) \propto T$, in striking contrast to the quadratic-in-$T$ behaviour of Fermi liquid theory. We have already seen an example of this for Hg1201 in figure~\ref{fig:mirzaei2013} (extracted from optical conductivity data), and figure~\ref{fig:liang} shows this for LSCO. 

This unusual non-Fermi liquid temperature dependence has lead to the high temperature phase being known as the ``strange metal''. The linear-in-$T$ resistivity can persist to very high temperatures, and to values far beyond the Mott-Ioffe-Regel bound, where conventional scattering theory yields a mean free path of less than the atomic spacing~\cite{ioffe1960noncrystalline,mott1972conduction,gurvitch1981iofferegel,gurvitch1983experimental}. The Mott-Ioffe-Regel limit is shown in figure~\ref{fig:liang} by the horizontal dashed line. Despite strongly violating this bound, some samples have exhibited saturation of the resistivity at high temperatures, see e.g.~\cite{wang1996observation,calandra2003violation}. 

Upon cooling to temperatures below the formation of the pseudogap, $T<T^\ast$, the resistivity of underdoped cuprates is consistent with conventional Fermi liquid behaviour $\rho(T)\propto T^2$, see figure~\ref{fig:mirzaei2013} and \cite{barisic2013universal,proust2016fermi}. This appears to be universal, being seen in both type A and type B cuprates~\cite{barisic2013universal}~\footnote{BSCCO does not show the crossover to $T^2$ at low temperatures, which has been attributed to disorder effects~\cite{barisic2013universal}.}. It is worth emphasising that this behaviour is, in fact, very surprising (perhaps more so than the $\rho(T)\propto T$ at high temperatures alone). Upon cooling into the pseudogap phase, the number of charge carriers is suppressed with parts of the Fermi surface become gapped. Nevertheless, the resistivity \textit{decreases}, in complete contradiction with usual expectations! Not only this, but it changes from the unconventional $\rho(T)\propto T$ behaviour of the strange metal back to the standard Fermi liquid form, $\rho(T) \propto T^2$.

As a brief aside, it is worth mentioning the behaviour of overdoped cuprates. In the normal phase, with a full Fermi surface $T>T_\text{c}$, they do not revert to the Fermi liquid behaviour $\rho(T) \propto T^2$. This lead Anderson and coworkers to propose that strong coupling, with separated lower and upper Mott Hubbard bands, persists into the overdoped region~\cite{anderson2008hidden,jain2009beyond,casey2011hidden}. This may not, however, be the case. Umklapp scattering processes still occur at overdoping but are not dominant as at underdoping. \cite{buhmann2013numerical,buhmann2013unconventional} analysed their contribution to the temperature dependence of the resistivity in the normal phase under a weak coupling approach, based on the numerical analysis of a Landau-Boltzmann transport equation. A more complete analysis of the frequency and as well as the temperature dependence of the in plane conductivity would be useful to settle the relative role of strong coupling and umklapp processes in overdoped cuprates~\cite{hussey2008phenomenology}.

\subsubsection{Hall effect and Hall angle as a function of temperature.}
\label{sec:hallT}
We have seen that the in-plane d.c. resistivity in the normal states of the cuprates is anomalous, showing behaviour outside of the Landau-Fermi liquid paradigm in the strange metal. It is then natural to expect that \textit{all} transport is anomalous, including upon the addition of a magnetic field. Here we will discuss such a scenario, and see that there are some surprises here too~\cite{harris1995violation,terasaki1995normalstate,kimura1996inplane,barisic2015hidden}. 

\begin{figure}
\begin{center}
\begin{tabular}{ll}
(a) & (b) \\
\includegraphics[width=0.42\textwidth]{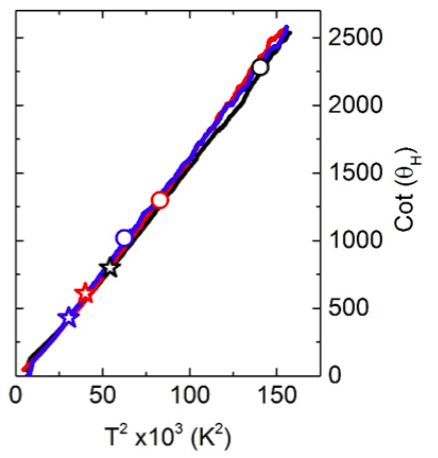} &
\includegraphics[width=0.4\textwidth]{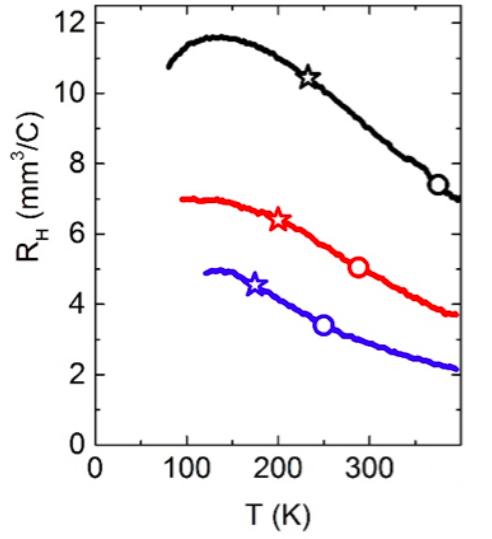}
\end{tabular}
\end{center}
\vspace{-4mm}
\caption{(a) The cotangent of the Hall angle, $\cot(\theta_\text{H})$, as a function of the square of the temperature $T^2$;
(b) the Hall coefficient $R_\text{H}$ as a function of temperature $T$.  Data is presented for three underdoped samples of HgBa$_2$CuO$_{4+\delta}$ ($T_\text{c} = 55$\,K in black, $T_\text{c} = 71$\,K in red, $T_\text{c} = 95$\,K in blue). Circles show the pseudogap temperature, $T^\ast$, for each of the samples. Stars denote the temperature below which the in-plane d.c. longitudinal resistivity is purely $\rho(T) \propto T^2$. Figures adapted from (Bari\v{s}i\'c et al. 2015).}
\label{fig:hall}
\end{figure}

With the application of a magnetic field, the Hall effect leads to finite transverse transport and thus finite transverse resistivity. This Hall resistivity, $\rho_\text{Hall} = R_HH$, is proportional to the Hall constant $R_H$ $(=1/n_Hec)$, where $n_H$ is the carrier density (see the discussion of section~\ref{sec:carrierdens}). At low temperatures only the hole carriers in the nodal pockets contribute to transport, so that $n_H$ is the hole density referenced to the Mott insulating state. When the temperature is raised through $T^\ast$, the pseudogap on the Fermi surface vanishes, which naively leads one to expect a drastic increase in the carrier density to a value consistent with the  hole density referenced to the Bloch band structure. The measurements in figure~\ref{fig:hall}, however, show only a gradual increase in $n_H$  at $T^\ast$. 

In a recent paper~\cite{rice2017umklapp} we put forward an explanation for this surprising behaviour. This was based on assuming a special straight form of the dispersion in the antinodal regions of the Fermi surface, away from the nodal pockets. Along these straight pieces of the Fermi surface, the direction of the Fermi velocity is constant and this prevents quasiparticles at the Fermi level from moving around a closed path. Such a closed path is required to enclose a magnetic flux and give rise to a contribution to the Hall effect. As a result, raising $T$ through $T^\ast$ leads only to a small and continuous increase in the value of $n_H$ linear in $T$, which in turn leads to the slowly decreasing value of $R_H$ in figure~\ref{fig:hall}.

In a standard metal, described by Landau-Fermi liquid theory, the Hall angle $\theta_H$ is defined by
\begin{equation}
\cot\theta_\text{H} = \frac{\rho}{\rho_\text{Hall}} \equiv \frac{\rho}{R_H H},
\end{equation}
which exhibits a temperature dependence $\sim T^2$ due to the constant carrier density combined with a scattering rate increasing in the resistivity $\rho(T) \sim T^2$. Curiously, similar behaviour is found in the cuprates~\cite{harris1995violation,terasaki1995normalstate,kimura1996inplane,barisic2015hidden}. This surprising behaviour need not solely imply Landau-Fermi liquid behaviour, associated with a constant carrier density that is independent of doping. Instead, this behaviour can also arise from the linear-in-$T$ behaviour of both $\rho(T)$ and $R_H(T)$, as illustrated for the case of Hg1201 in figure~\ref{fig:hall}. At finite temperatures $T$, the separation between the Fermi-liquid-like nodal quasiparticles and the antinodal region should not be regarded as strict, and both regions could contribute to the Hall angle in the strange metal.

It is also worth noting that the high magnetic field formation of CDW order, as discussed in section~\ref{sec:nmr}, can have clear signatures in transport measurements. If the CDW order leads to a reconstruction of the Fermi surface with the associated formation of electron pockets, Hall coefficient measurements will show this: cooling through $T_\text{CDW}$ the Hall coefficient will become negative. An example of such behaviour in YBCO can be found in, e.g., \cite{leboeuf2007electron,badoux2016change}. For more detailed discussion of the phenomenology of transport in the normal states of cuprates, we refer the reader to the review~\cite{hussey2008phenomenology}.

\section{Overview, open issues, and future directions}
\label{sec:open}

\subsection{Overview}

In this review, we have presented what we believe to be compelling evidence for dividing the high temperature cuprate superconductors into two types, which we denote as `type A' and `type B'. Our motivation for doing so is strongly influenced by \textit{microscopic theoretical studies} of the two-dimensional Hubbard and $t$-$J$ models, where the presence of quasi-degenerate ground states appears almost universally [see section~\ref{sec:competinggs} and, e.g.,~\cite{zheng2017stripe}]. Remarkably, these quasi-degenerate ground states have markedly different physical properties, with one exhibiting intertwined long-range orders and associated translational symmetry breaking, while the other possesses uniform $d$-wave superconductivity. This separation into two types is borne out by experiment, with each type having distinctive features in the ``normal'' states, while also sharing many common features. 

The differences between type A and type B cuprates is perhaps best exhibited in NMR/NQR experiments. The critical slowing down that accompanies the onset of the long-range antiferromagnetic order leads to a wipe out of the NMR signal in LBCO, while the onset of long-range CDW order leads to extra peaks in NQR, due to differences in the local quadrupolar fields in the presence of static CDW order. LBCO, and other material with similar NMR/NQR features, are what we have called type A cuprates. On the other hand, type B materials behave very different, with NMR/NQR spectra showing well-resolved narrow lines that show no signs of static CDW order (away from a transition to CDW order in high magnetic fields). NMR/NQR also shows significant differences in the behaviour of the spin relaxation rate in the two types of cuprates. Other experimental examples of this dichotomy in behaviour have been presented in the previous section. We summarize the type A, type B dichotomy for some of the most-studied cuprates in table~\ref{tab:summary}.

Despite more than three decades of intense theoretical and experimental studies, with hundreds of thousands of papers being published, there still remain many open issues for research in the high-$T_\text{c}$ cuprates. We now finish with a discussion of some of these open issues and future directions, both for experiment and theory.

\begin{table}[ht]
  \begin{center}
    \begin{tabular}{c|c|c|c}
      \hline\hline
      Cuprate & ~Type A~ & ~Type B~ & ~Experimental evidence~ \\
      \hline\hline
      La$_{2-x}$Ba$_x$CuO$_4$ & $\times$ & & \begin{tabular}{c}
                                               Neutron scattering \\
                                               \cite{tranquada1995evidence} \\
                                               c-axis transport \\
                                               \cite{li2007twodimensional}\\
                                               Inelastic X-ray scattering \\
                                               \cite{miao2017hightemperature,miao2018incommensurate}
                                             \end{tabular} \\
      \hline
      La$_{2-x}$Sr$_x$CuO$_4$ & $\times$ & & \begin{tabular}{c}
                                               Nuclear magnetic resonance \\
                                               \cite{imai2017revisiting}\\
                                               X-ray diffraction \\
                                               \cite{thampy2014rotated,croft2014charge} \\
                                               Non-linear optical response \\
                                               \cite{rajasekaran2018probing} \\
                                               Hourglass magnon spectrum \\
                                               \cite{chan2016commensurate}
                                             \end{tabular}  \\
      \hline
      YBa$_x$Cu$_y$O$_{7-x}$ & & $\times$ &  \begin{tabular}{c}
                                               Nuclear magnetic resonance \\
                                               \cite{tomeno1994nmr,wu2015incipient} \\
                                               Magnon spectral gap \\
                                               \cite{hinkov2007spin}
                                             \end{tabular} \\
      \hline
      HgBa$_2$CuO$_{4+\delta}$ & & $\times$ & \begin{tabular}{c}
                                                Nuclear magnetic resonance \\
                                                \cite{itoh1998pseudo,mukuda2012high} \\
                                                Magnon spectrum gap \\
                                                \cite{chan2016commensurate}
                                              \end{tabular} \\
      \hline
      \begin{tabular}{c}
        Bi$_2$Sr$_2$CaCu$_2$O$_{8+x}$ \\
        Bi$_2$Sr$_2$CaCuO$_{2+\delta}$
      \end{tabular}
              & $\sim$ & $\sim$ & \begin{tabular}{c}
                                                          Nuclear magnetic resonance \\
                                                          \cite{ishida1998pseudogap,crocker2011nmr} \\
                                                          Scanning tunneling microscopy \\
                                                          (Hoffman, Hudson, Lang, Madhavan, \\
                                                          Eisaki, Uchida \& Davis 2002)
                                                          
                                                 \end{tabular} \\
      \hline\hline
    \end{tabular}
  \end{center}
  \caption{A summary of the type A / type B dichotomy for some of the most commonly studied cuprate superconductors. The fourth column, `Significant experimental evidence', lists the experiments (with example references) that point most strongly towards ground states with (or without) translational symmetry breaking charge order for type A (type B) materials. For Bi$_2$Sr$_2$CaCu$_2$O$_{8+x}$ and Bi$_2$Sr$_2$CaCuO$_{2+\delta}$ the experimental evidence is unclear; nuclear magnetic resonance suggests that it could be a type B ground state with significant levels of disorder, whilst scanning tunneling microscopy measurements find low-amplitude charge modulations on the surface.}
  \label{tab:summary}
\end{table}

\subsection{Experiment}

With the dichotomy presented within this review, an interesting question is raised: what is the relation between the two very different types of ground states, and can one tune between them in an experimentally achievable manner? One interesting and suggestive recent study is that of~\cite{miao2018incommensurate},  where X-ray scattering on a type A cuprate observes that the wave vector of CDW order changes from $q\sim0.31$ at room temperature to $q \sim 0.25$ at low-temperatures. The first value should now be familiar to the reader, coinciding with features in both the $c$-axis optical conductivity and giant phonon anomaly measurements on type B cuprates. Could this constitute an example where at high temperatures one probes properties of a close-in-energy type B ground state, while at low-temperatures that system freezes into the low-energy type A state? 

The possibility that the onset of the pseudogap is accompanied by a thermodynamic phase transition has been discussed for some years. It has, for example, been suggested that the pseudogap phase breaks time-reversal, four-fold rotation, and mirror symmetries. Recent experiments by Hsieh and coworkers~\cite{zhao2017global} on underdoped YBCO have reported detailed studies of the changes in the underlying lattice symmetry at the onset of the pseudogap at $T=T^\ast$. They used high-sensitivity linear and second-harmonic optical anisotropy measurements to determine the point group of YBa$_2$Cu$_3$O$_y$ in the pseudogap phase, finding that both the spatial inversion and two-fold rotational symmetries are broken below $T^\ast$. On the other hand,  mirror symmetries perpendicular to the Cu--O plane are absent at all temperatures. They studied a wide doping range and came to the conclusion that the pseudogap region coincides with an odd-parity order that does not arise from a competing Fermi surface instability. The authors have not, however, established any connection between this order and those associated with CDWs or superconductivity. Another recent study addressing similar issues is that of Matsuda and coworkers, where high precision torque magnetometry measurements of the anisotropic susceptibility within the $ab$ planes in orthorhombic YBa$_2$Cu$_3$O$_y$ were reported~\cite{sato2017thermodynamic}. The in-plane anisotropy displays a significant increase with a distinct kink at $T = T^\ast$, and scaling behaviour with respect to $T/T^\ast$ over a wide range of doping. The rotational symmetry breaking sets in at $T^\ast$, leading the authors to conclude that the pseudogap onset is associated with a second order nematic phase transition. At present it remains to be clarified if the symmetry breaking shown in these studies is associated with umklapp scattering in some manner.

The nature of the crossover at the end of the nodal arcs/pockets to the antinodal pseudogap is another open issue.. In the superconducting phase, the nodal region has a single-particle gap due to Cooper pairing, with the lowest energy excitations being part of the two-particle continuum. The antinodal region at underdoping, on the other hand, possesses the larger pseudogap. The reason for the large insulating pseudogap is clear, being due to the additional particle-hole $(\pi,\pi)$ antiferromagnetic short-range order opening a two-particle gap as one moves towards underdoping. Nevertheless, details of the crossover between these two gaps at the ends of the pockets observed in ARPES is not clear, and e.g., the relation to the YRZ ansatz needs further study.

\subsection{Theory}
From the perspective of theory, there are also a number of open issues and future directions to explore. With our discussion, we have seen that the free energy landscape of two-dimensional Hubbard models (plus various perturbations) plays a vital role in the physics of the cuprates. This leads to the obvious question: is it possible to map out this free energy landscape as a function of the model parameters and perturbations? From the viewpoint of microscopic calculations, this seems like a huge challenge, requiring large scale and high accuracy simulations across a wide range of Hamiltonians. More focused numerical studies could address issues such as the role of extended hopping and beyond Hubbard interactions, which are undoubtedly important. One suggested route, which may significantly help the mapping out of the free energy landscape, is to focus on better understanding the mapping between microscopic theories and Landau-Ginzburg functionals. New techniques for understanding the ground states of such functionals are also being developed~\cite{coser2014truncated}. 

The lack of understanding of the free-energy landscape also obscures many other issues. How do external tuning parameters, such as applied pressure or magnetic field, modify it? Understanding this may help guide experimental efforts as to how to tune between the two different types of behaviour in the same material. Are recent results that show the vanishing of the giant phonon anomalies under applied pressure~\cite{souliou2018rapid} a signature of the crossover from type B to type A behaviour in YBCO?

Related questions that also requires detailed understanding of the free energy landscape are: What is the role that finite temperature plays in determining the physics? Can one can see signatures of both type A and type B behaviour in the same material at reasonable temperature? With energy separations being as small as $\sim 0.01t$ per site, one may expect that even relatively small temperatures can lead to blending of the type A and type B behaviours, especially in thermodynamic measurements that (theoretically) trace over the thermal density matrix $\rho \sim e^{-\beta H}$. For finite temperatures, it is not clear that these questions can simply be addressed by pushing state-of-the-art tensor network approaches, and other computational approaches are often limited to reasonably high temperatures. This remains a challenging forefront for theory. 

Yet another interesting theoretical issue is where is the type A ground state in one-loop FRG studies of the two-dimensional Hubbard model~\cite{metzner2012functional,li2014competing}? This is perhaps best exemplified by the singular mode formulation of the functional RG, developed by Huseman and Salmhofer~\cite{husemann2009efficient}; it is an unbiased formulation of the functional RG that allows a comparison between different instabilities. Explicitly, it resolves the renormalised wave vector dependent irreducible interaction vertices as a function of the running energy (or temperature) scale. Thus a scale-dependent comparison between effective interactions in all relevant particle-particle or particle-hole channels can be made. Wang and coworkers~\cite{li2014competing} have applied this method to the cuprates, examining not only the simplest one-band models but also various extensions to multi-orbital models with additional interaction terms. They were able to find  instabilities in the CDW and SDW channels that competed with d-wave superconductivity, which depended on the strength of additional interaction terms in the bare model. However, it should be noted that the form of these CDW and SDW instabilities, i.e. their $\bi{Q}$-vectors, are different to those shown in the type A cuprates. This implies that either: (i) the analyses so far are missing terms that are important for the type A behaviour; or (ii) higher order, i.e. two-loop, terms are important. This is certainly an issue that should be addressed in future studies, with the relationship between the two ground states observed in experiments remaining open. 

There is also the question of how to `derive' the YRZ ansatz for the single electron propagator within the pseudogap phase. While we have presented some approaches to doing this in the appendix, each method uses the random phase approximation (predicated on the presence of the long-range interactions or sufficiently weak next-neighbour hopping) and one may doubt its validity in a realistic cuprate setting. One recent interesting and suggestive result is that of~\cite{quinn2018splitting}, whose single particle propagator is quite reminiscent of the YRZ one. Alternatively, a full numerical implementation of the ideas of~\cite{ossadnik2016wave} for the two-dimensional Hubbard model may lead to some insights for this issue.   

\subsection{Insights from other directions}

There is also the potential for insights to arise from other directions that, at first glance, appear unconnected to the high-$T_\text{c}$ cuprates.

The Fe-based superconductors, discovered a coupled decades later than the cuprates, have attracted more interest than the cuprate superconductors over the past few years. These two sets of unconventional superconductors have, by and large, been treated separately due to the large differences in electronic structure, Fermi surface, etc. It is likely, however, that this will soon change, due to a very recent paper by Maier, Berlijn and Scalapino~\cite{maier2018dwave}. They reported an analysis of the electronic structures of Ba$_2$CuO$_{3+\delta}$, which belongs to a group of highly overdoped planar cuprates that was recently found to exhibit bulk superconductivity with $T_\text{c} \approx 80$\,K~\cite{li2018new}. Interest in the highly overdoped cuprates goes back some years, see for example the review of Geballe and Marezio~\cite{geballe2009enhanced}, and in this case the Fermi surface for Ba$_2$CuO$_{3+\delta}$ at $\delta =0.5$ computed in~\cite{maier2018dwave} is very similar to that of many Fe-pnictides, such as Ba$_{0.6}$K$_{0.4}$Fe$_2$As$_2$. Both consist of an electron pocket at the zone centre and a hole pocket in the corners of the unit cell. For the Fe-pnictides, microscopic modelling finds strong pairing instabilities in two different channels, namely $d$-wave and $s_\pm$ pairing. Maier and coworkers report similar instability for the overdoped cuprates~\cite{maier2018dwave}. These results promise interesting close comparisons between this pair of unconventional superconductors. We note, however, that these highly overdoped cuprates are synthesised at high temperatures and oxidation pressures, which are not easy to handle and limit crystal growth. As a result, single crystal experiments to determine the order parameter symmetry, etc, will likely be challenging. 

Another intriguing direction, currently under intense investigation, is so-called ``magic angle'' bilayer graphene. By stacking undoped single sheets of graphene, rotated in-plane through a ``magic angle'' of~$\theta\sim 1.1^\circ$ relative to one-another, one obtains a strongly-correlated Mott insulator~\cite{cao2018correlated}. Addition of a back gate allows one to tune the carrier density of the system~\cite{cao2018unconventional}, revealing a temperature-doping phase diagram~\cite{cao2018unconventional,yankowitz2018tuning} that is remarkably reminiscent of the high-temperature cuprate superconductors. It contains a low-doping Mott insulator, a superconducting dome with unconventional superconductivity, and a large doping metallic phase. Whether the analogy with the cuprates persists further is an interesting avenue for future studies, with the hope that additional understanding of the superconductivity in this highly tunable system will shed light on the cuprates. 

Finally, we finish by highlighting interesting experimental results emerging from a different community, which have the potential to provide deep insights into the physics of the high temperature superconductors. Recent advances in the field of ultra-cold atomic gases have enabled the analogue quantum simulation of the two-dimensional Hubbard model to an unprecedented degree of accuracy~\cite{lewenstein2007ultracold,bloch2008manybody,esslinger2010fermihubbard,bloch2012quantum}. Experiments in such optical lattice systems are able to directly measure transport phenomena~\cite{ott2004collisionally,strohmaier2007interactioncontrolled,schneider2012fermionic,xu2016bad,anderson2017optical}, including recently the temperature dependence of the resistivity of the two-dimensional Hubbard model~\cite{xu2016bad,brown2018bad}. Whilst these experiments are currently limited to the high temperature regime (with temperatures being a significant fraction of the Fermi temperature, a huge temperature by condensed matter standards), advances over the coming years will undoubtedly lead to lower temperatures and access to the full phase diagram of the two-dimensional Hubbard model. This will allow one to experimentally probe the physics of the two-dimensional Hubbard model and its extensions in a setting where one has fine control of many parameters (such as the hopping amplitudes, filling, interactions, temperature, and so forth). This may help isolate whether ``beyond Hubbard model'' physics plays a crucial role in high temperature superconductivity. 

\ack

We thank Philip W. Anderson, Laura Classen, Andrew James, Robert Konik, Ye-Hua Liu, Masao Ogata, Matthias Ossadnik, Manfred Sigrist, Russ Walstedt, and Fuchun Zhang for useful discussions and collaborations. N.J.R. would like to thank the members of the Quantum Materials group at the University of Amsterdam, particularly Erik van Heumen and Mark Golden, for many interesting and educational group meetings about the `normal states' of the cuprates. We also thank Philippe Corboz and Sangwoo Choi for useful comments and suggestions relating to recent progress in numerical simulations of the 2D Hubbard model.   

N.J.R. is supported by the European Union's Horizon 2020 research and innovation programme under grant agreement No 745944. Work at Brookhaven National Laboratory was supported by the Condensed Matter Physics and Materials Science Division, in turn funded by the U.S. Department of Energy, Office of Basic Energy Sciences, under Contract DE-SC0012704.

\section*{Glossary of acronyms}

\begin{tabular}{lcl}
  {\bf ARPES} & \qquad & angle resolved photoemission spectroscopy \\
  {\bf Bi2212} & & Bi$_2$Sr$_2$CaCuO$_{2+\delta}$ \\  
  {\bf BSCCO} & & Bi$_2$Sr$_2$CaCu$_2$O$_{8+x}$ \\
  {\bf CDW} &  & charge density wave \\
  {\bf Hg1201} & & HgBa$_2$CuO$_{4+\delta}$ \\
  {\bf irreps} & & irreducible representations \\
  {\bf LBCO} & & La$_{2-x}$Ba$_x$CuO$_4$ \\
  {\bf LSCO} & & La$_{2-x}$Sr$_x$CuO$_4$ \\
  {\bf NMR} & & nuclear magnetic resonance \\
  {\bf NQR} & & nuclear quadrupolar resonance \\
  {\bf PDW} & & pair density wave \\
  {\bf RG} & & renormalization group \\
  {\bf RPA} & & random phase approximations \\
  {\bf SDW} & & spin density wave \\
  {\bf SISTM} & & spectroscopic intensity scanning tunnelling microscopy \\
  {\bf STM} & & scanning tunnelling microscopy \\
  {\bf Y248} & & YBa$_2$Cu$_4$O$_8$ \\
  {\bf YBCO} & &  YBa$_x$Cu$_y$O$_{7-z}$ \\
  {\bf YRZ} & & Yang-Rice-Zhang (ansatz)
\end{tabular}

\appendix

\section{Efforts towards a microscopic derivation of the YRZ propagator}

In this appendix, we discuss some of the efforts made towards a microscopic derivation of the YRZ ansatz for the single electron propagator. In particular, we first consider the random phase approximation (RPA) analysis of coupled two-leg ladders from~\cite{konik2006doped}. This used exact information from the SO(8) formulation of the two-leg ladder $d$-Mott state (see section~\ref{sec:dmott}) as a starting point for a perturbative RPA treatment. In the second approach of~\cite{james2012magnetic}, we recount an approach based upon a slave boson treatment of the $t$-$J$ model, which treats next-nearest-neighbour hopping under an RPA-like approximation. 

\subsection[\qquad\qquad Random phase approximation for coupled two-leg ladders.]{Random phase approximation for coupled two-leg ladders.}

The initial inspiration for the YRZ ansatz for the electron Green's function was provided by coupled two-leg ladders treated under the random phase approximation (RPA)~\cite{konik2006doped}. In particular, Konik and coworkers considered coupling together half-filled two-leg ladders, described by an effective SO(8) Gross-Neveu field theory (as discussed in section~\ref{sec:dmott}), via long-range hopping
\begin{equation}
H = \sum_{i} H^\text{SO(8)}_i + \sum_{i,i',\ell,\ell',\s} t^\perp_{ii'\ell\ell'} \int\rmd x \Big[ c\dg_{\ell i \s}(x) c_{\ell' i' \s}(x) + \text{H.c.} \Big]. 
\label{coupledladders}
\end{equation} 
Here $c\dg_{\ell i \s}(x)$ creates an electron at position $x$ on the $\ell$th leg ($\ell = 1,2$) of the $i$th ladder with spin $\s=\up,\dn$, and $H_i^\text{SO(8)}$ is the low-energy SO(8) description of the half-filled two leg Hubbard ladder~\cite{lin1998exact}. 

Within the SO(8) theory, the electron Green's function is given by~\cite{konik2000twoleg,konik2001exact,essler2005application}
\begin{equation}
G^\text{SO(8)}_{\alpha,\s}(\omega,k_x) = Z_{\alpha,\s} \frac{\omega + \epsilon_{\alpha}(k_x)}{\omega^2 - \epsilon^2_{\alpha}(k_x) - \Delta^2} + G_{\alpha,\s}^\text{incoh},
\end{equation}
Here $\alpha=\pm$ is a band index, $\epsilon_{\alpha}(k_x)$ is the bare dispersion of the bands, and $\Delta$ is the charge gap. The quasiparticle weight satisfies $Z_{\alpha,\s} \approx 1$ as the incoherent part of the Green's function, $G_{\alpha,\s}^\text{incoh}$, makes negligible contribution to the spectral weight. When the coupling between ladders, $t_{ii'\ell\ell'}^\perp$, is long ranged there is an emergent small parameter~\cite{essler2002weakly,essler2005theory} and the coupled problem,~\eqref{coupledladders}, can be treated under the RPA. This gives the electron Green's function~\cite{konik2006doped}
\begin{equation}
G^\text{RPA}_{\alpha,\s}(k_x,\bi{k}_\perp) = \bigg\{ \Big[ G^\text{SO(8)}_{\alpha,\s}(k_x) \Big]^{-1} - t^\perp_{\alpha}(\bi{k}_\perp) \bigg\}^{-1}.
\label{rpa}
\end{equation}
The resulting quasiparticle spectrum $E_{\alpha,\s}(k_x,\bi{k}_\perp)$ can be obtained from the poles in the RPA Green's function, i.e.
\begin{equation}
\omega - \epsilon_{\alpha}(k_x) - \frac{\Delta^2}{\omega + \epsilon_{\alpha}(k_x)} - t^\perp_{\alpha}(\bi{k}_\perp) = 0,
\end{equation}
yielding
\begin{equation}
  \begin{split}
  E_{\alpha,\s,\pm}(\bi{k}) &\equiv E_{\alpha,\s,\pm}(k_x,\bi{k}_\perp) \\
  &= \frac{t^\perp_{\alpha}(\bi{k}_\perp)}{2} \pm \sqrt{\Big( \epsilon_\alpha(k_x) + t^\perp_\alpha(\bi{k}_\perp) \Big)^2 + \Delta^2}. 
\end{split}
\label{qpspectrum}
\end{equation}

At this point we highlight that the RPA Green's function~\eqref{rpa} and the associated quasiparticle spectrum~\eqref{qpspectrum} have a remarkably similar form to the YRZ Green's function (see section~\ref{sec:yrzprop}) and the known phenomenological form for the self-energy necessary to capture ARPES observations~\cite{norman1998phenomenology}.

\begin{figure}
\begin{center}
\includegraphics[width=0.4\textwidth]{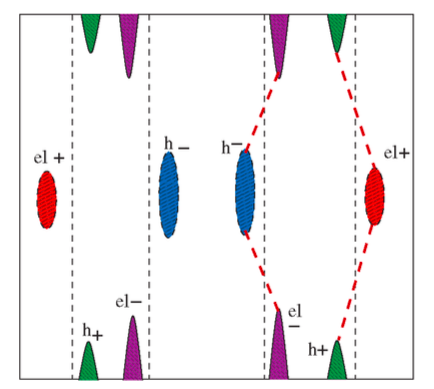} 
\end{center}
\vspace{-4mm}
\caption{The Fermi surface, with electron (red/magenta) and hole (green/blue) pockets, from the RPA treatment of coupled ladders. Pockets of different size, formed from the bonding (+) and anti-bonding (-) bands, originates from different hopping parameters $t^\perp_\pm$. Dashed lines (vertical) denote where the one-dimensional bands are zero, $\epsilon_\pm(k_x)=0$, and thick dashed lines show the gap minima $\epsilon_\alpha(k) = t^\perp_\alpha(k)/2$. Figure reproduced from (Konik et al. 2006).}
\label{fig:konik2006}
\end{figure}

Taking the interchain hopping amplitude $t^\perp_\alpha(\bi{k}_\perp)$ to be strongly peaked near $\bi{k}_\perp = 0, \bi{G}/2$ (where $\bi{G}$ is the reciprocal lattice vector in the direction perpendicular to the chain) with amplitude $t_{\alpha,0}$, the Fermi surface is defined through the equation $E_{\alpha,\s,\pm}(\bi{k}_F) = 0$, i.e.
\begin{equation}
\Big[ 2(k_x - k_{\text{F},\alpha}) v_\text{F} \mp t_{\alpha,0} \Big]^2 + \frac{2 t_{\alpha,0}^2}{\kappa_0^2} \bi{k}_\perp \cdot \bi{k}_\perp  = t_{a,0}^2 - 4\Delta^2.
\label{eq:fermisurface}
\end{equation}
Here $\kappa_0$ is a coefficient in the following expansion of the interchain hopping amplitude:
\begin{equation}   
t_\alpha \Big(\bi{k}_\perp + (1\mp1)\bi{G}/4\Big) = \mp t_{a,0} \bigg( 1 - \frac{\bi{k}_\perp \cdot \bi{k}_\perp}{\kappa_0^2} + \ldots \bigg).
\end{equation}
From~\eqref{eq:fermisurface}, it is apparent that there are only gapless excitations (i.e., a Fermi surface) when $t_{a,0}^2 - 4\Delta^2 > 0$. The Fermi surface separates into electron and hole pockets (depending on the sign of $t_{\alpha,0}$), shown schematically in figure~\ref{fig:konik2006}, in a similar manner to dynamical mean field theory calculations on doped Mott insulators~\cite{civelii2005dynamical,stanescu2006fermi}.

Beyond the Fermi surface discussed above, Konik and coworkers showed that the RPA analysis of coupled ladders reproduces various features of the cuprates and theories describing them~\cite{konik2006doped}. At half-filling, their model possesses an SO(6) symmetry, and hence exhibits the same phenomenology as well-known SO(5) theories of superconductivity~\cite{zhang1997unified,rabello1998microscopic}. With finite doping, the model favours superconductivity and calculations are well-controlled, allowing one to estimate $T_\text{c}$. 

In this section, we have seen that one can obtain an electron Green's function with striking similarity to the YRZ form through a RPA treatment of coupled ladders. In the following section, we will see an alternative construction, based upon a slave boson treatment of doped Mott insulators~\cite{james2012magnetic}.

\subsection[\qquad\qquad Slave bosons treatment of a doped Mott insulator.]{Slave bosons treatment of a doped Mott insulator.}

We will now review a different microscopic approach to obtaining a Green's function of the YRZ form, following~\cite{james2012magnetic}. Here we begin with the $t$-$J$ model on the two-dimensional square lattice, 
\begin{equation}
  \begin{split}
  H_{t-J} =& - \sum_{\la i,j\ra,\s} t \Big( c\dg_{i,\s} c_{j,\s} + \text{H.c.} \Big) + \sum_{\la i,j\ra} J_\text{H} \bi{S}_i \cdot \bi{S}_j \\
  & - \sum_{\la\la i,j \ra\ra,\s} t'\Big( c\dg_{i,\s} c_{j,\s} + \text{H.c.} \Big),
  \end{split}
\end{equation}
where $t$ ($t'$) is the nearest-neighbour (next-nearest neighbour) hopping amplitude and $J_\text{H}$ is the spin exchange interaction. We first treat the nearest-neighbour terms, setting $t'=0$, through slave boson mean field theory. This involves factorising the electron operator $c\dg_{i,\s}$ in terms of spinon $f\dg_{i,\s}$ and holon $b_i$ degrees of freedom, through the identity $c\dg_{i,\s} = f\dg_{i,\s} b_i$. These two new degrees of freedom satisfy a constraint: $\sum_\s f\dg_{i,\s}f_{i,s} + b\dg_i b_i = 1$ at the operatorial level, although in the mean field treatment this is relaxed to being true at the level of a thermal expectation value. 

Following the standard calculation, see~\cite{brinckmann2001renormalized}, one arrives at the (slave boson mean field theory) spinon Green's function
\begin{equation}
G^f_\s(\bi{k},\omega) = \frac{1}{\omega - \xi_0(\bi{k}) - \Sigma_\text{R}(\bi{k},\omega)}, 
\end{equation}
where $\xi_0(\bi{k}) = -2t_0(x) (\cos k_x + \cos k_y)$ is the non-interacting single electron dispersion and $\Sigma_\text{R}(\bi{k},\omega) = |\Delta_R(\bi{k})|^2/(\omega + \xi_0(\bi{k}))$ with $\Delta_R(\bi{k}) = \Delta_0(x) (\cos k_x - \cos k_y)$. Here with $t_0(x), \Delta_0(x)$ are doping $x$ dependent. Under the approximation that the holons are `almost condensed' the electron Green's function is proportional to the spinon one:
\begin{equation}
G_\s(\bi{k},\omega) = g_t(x) G^f_\s(\bi{k},\omega), \label{Gs}
\end{equation}
where $g_t(x) = x$ in the slave boson mean field theory~\footnote{Under the Gutzwiller approximation, the condensed holon Green's function is $g_t(x) = 2x/(1+x^3)$.}. 

\begin{figure}
\begin{center}
\includegraphics[width = 0.65\textwidth]{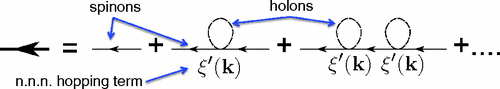}
\end{center}
\vspace{-4mm}
\caption{The random phase approximation form of the YRZ spinon propagator (left hand side) in terms of the spinon (thin lines) and holon (dashed lines) degrees of freedom of the slave boson mean field theory treatment. Figure reproduced from (James et al. 2012).}
\label{fig:james2012}
\end{figure}

The Green's function~\eqref{Gs} is not exactly of the YRZ form, with the dispersion in the denominator being solely due to nearest-neighbour hopping. This can be remedied by treatment of the next-nearest-neighbour hopping terms under a RPA-like approximation (shown in figure~\ref{fig:james2012}), yielding~\footnote{A similar result for the spinon Green's function can be obtained whilst treating both nearest- and next-nearest neighbour hopping on the same footing, see the supplementary material of~\cite{james2012magnetic}.}
\begin{equation}
G^f_\s(\bi{k},\omega) = \frac{1}{\omega - \xi_0(\bi{k}) - \xi'(\bi{k}) - \Sigma_\text{R}(\bi{k},\omega)}, 
\label{rpannn}
\end{equation}
with $\xi'(\bi{k}) = -4t'(x) \cos k_x \cos k_y - 2t''(x) ( \cos 2k_x + \cos 2k_y)$, where we also include next-next-neighbour hopping. The next-nearest-neighbour term, under the RPA treatment, serves to bind together the spinons and holons, a physical feature of the slave boson factorisation that is absent in the standard mean field treatment. Thus, with~\eqref{rpannn} we arrive at a Green's function of the YRZ form. 

\section*{References}

\bibliographystyle{jphysicsB_withTitles}
\bibliography{Anomalies_Pseudogap_Review}

\end{document}